\documentclass[10pt,conference]{IEEEtran}

\usepackage{xspace}
\usepackage{xcolor}
\usepackage{graphicx}
\usepackage[normalem]{ulem} %
\usepackage{amsmath}
\usepackage{amssymb} %
\usepackage{amsfonts} %
\usepackage{enumitem}
\usepackage[autolanguage]{numprint}
\nprounddigits{0}

\usepackage[linkcolor=black,citecolor=black,anchorcolor=black,filecolor=black,menucolor=black,runcolor=black,urlcolor=black,hidelinks]{hyperref} %
\usepackage{url} %
\usepackage{breakurl}
\usepackage{cite} %

\usepackage{booktabs}
\usepackage{multirow}
\usepackage{makecell}
\usepackage{ragged2e}
\usepackage{array}

\usepackage{caption}
\usepackage{subcaption}

\usepackage{listings}

\usepackage{algorithm}
\usepackage{algpseudocode}
\algrenewcommand\algorithmicindent{.6em}
\algrenewcommand\algorithmicrequire{\textbf{Input:}}
\algrenewcommand\algorithmicensure{\textbf{Output:}}

\usepackage{tikz}
\usetikzlibrary{calc}
\usetikzlibrary{shapes.geometric}
\usetikzlibrary{decorations.pathreplacing}
\usetikzlibrary{positioning}

\usepackage{balance}
\usepackage{datetime} %
\usepackage{pgffor}
\usepackage{tcolorbox}

\usepackage{verbatim}

\newcommand{\XSpace}[1]{}
\newcommand{\XComment}[1]{}
\makeatletter
\def\gnewcommand{\g@star@or@long\new@command}
\def\grenewcommand{\g@star@or@long\renew@command}
\def\g@star@or@long#1{%
  \@ifstar{\let\l@ngrel@x\global#1}{\def\l@ngrel@x{\long\global}#1}}
\makeatother
\newcommand{\DefMacro}[2]{\expandafter\gnewcommand\csname rmk-#1\endcsname{#2}}
\newcommand{\UseMacro}[1]{\csname rmk-#1\endcsname}

\newcommand{\MyPara}[1]{\vspace{1pt}\noindent\textbf{#1}.}
\newcommand{\MyParaOnly}[1]{\vspace{1pt}\noindent\textbf{#1}}

\newcommand{\InputWithSpace}[1]{\bgroup\def\arraystretch{1.2}\input{#1}\egroup}
\newcommand{\Code}[1]{{\ifmmode{\mathtt{#1}}\else$\mathtt{#1}$\fi}}
\newcommand{\CodeIn}[1]{{\ifmmode{\mathtt{#1}}\else$\mathtt{#1}$\fi}}

\newcolumntype{R}[1]{>{\RaggedLeft\arraybackslash}p{#1}}
\newcolumntype{L}[1]{>{\RaggedRight\arraybackslash}p{#1}}

\definecolor{gray}{RGB}{211,211,211}
\newcommand{\jbasicstyle}{\small\sffamily} %

\newcommand{\jnumberstyle}{\scriptsize}

\lstdefinelanguage{pseudo}
{
  morekeywords={},
  keywordstyle=\bfseries,
  lineskip=-0.1em,
  numbers=left, %
  numberstyle=\jnumberstyle,
  numbersep=4pt,
  basicstyle=\jbasicstyle,
  breaklines=true,
  breakautoindent=true,
  tabsize=2,
  columns=fullflexible,
  morecomment=*[l][\textsl]{//},
  mathescape=true,
  xleftmargin=10pt,
}

\lstdefinelanguage{todo-comment}
{
  morekeywords={},
  keywordstyle=\bfseries,
  lineskip=-0.1em,
  numbers=none,
  basicstyle=\jbasicstyle,
  breaklines=true,
  breakautoindent=true,
  tabsize=2,
  columns=fullflexible,
  morecomment=*[l][\textsl]{//},
  mathescape=true,
  xleftmargin=-10pt,
}

\lstdefinelanguage{java-pretty}
{
  language=java,
  numbers=left,
  numbersep=4pt,
  basicstyle=\scriptsize\ttfamily,
  numberstyle=\scriptsize,
  breaklines=true,
  columns=fullflexible,
  xleftmargin=16pt,
  showstringspaces=false,
}

\lstdefinelanguage{java-pretty-no-number}
{
  language=java,
  numbers=none,
  basicstyle=\scriptsize\ttfamily,
  numberstyle=\scriptsize,
  breaklines=true,
  columns=fullflexible,
  xleftmargin=0,
  showstringspaces=false,
}
\newcommand{\TecoTool}{\textsc{TeCo}\xspace}
\newcommand{\Title}{Learning Deep Semantics for Test Completion}
\newcommand{\TecoURL}{\url{https://github.com/EngineeringSoftware/teco}}

\newcommand{\testcomp}{test completion\xspace}

\newcommand{\codeut}{code under test\xspace}
\newcommand{\mut}{method under test\xspace}

\newcommand{\clzut}{class under test\xspace}
\newcommand{\testm}{test method\xspace}

\newcommand{\testms}{test methods\xspace}

\newcommand{\testc}{test class\xspace}
\newcommand{\testsign}{\testm signature\xspace}
\newcommand{\sufdata}{syntax-level data\xspace}
\newcommand{\semdata}{code semantics\xspace}
\newcommand{\Semdata}{Code semantics\xspace}
\newcommand{\SemData}{Code Semantics\xspace}

\newcommand{\stmt}{statement\xspace}
\newcommand{\stmts}{statements\xspace}

\newcommand{\priorstmts}{prior statements\xspace}

\newcommand{\sanalysis}{static analysis\xspace}
\newcommand{\Sanalysis}{Static analysis\xspace}

\newcommand{\exectx}{execution context\xspace}

\newcommand{\exeres}{execution result\xspace}
\newcommand{\Exeres}{Execution result\xspace}

\newcommand{\oraclegen}{test oracle generation\xspace}
\newcommand{\Oraclegen}{Test oracle generation\xspace}
\newcommand{\OracleGen}{Test Oracle Generation\xspace}
\newcommand{\runnablesubset}{runnable subset\xspace}
\newcommand{\oraclesubset}{oracle subset\xspace}
\newcommand{\oraclerunnablesubset}{oracle-runnable subset\xspace}

\DefMacro{no-types-local}{S1\xspace}  %
\DefMacro{name-types-local}{local var types\xspace}
\DefMacro{no-types-absent}{S2\xspace}  %
\DefMacro{name-types-absent}{absent types\xspace}
\DefMacro{no-fields-notset}{S3\xspace}  %
\DefMacro{name-fields-notset}{unset fields\xspace}
\DefMacro{no-setup-teardown}{S4\xspace}  %
\DefMacro{name-setup-teardown}{setup teardown\xspace}
\DefMacro{no-last-called-method}{S5\xspace}  %
\DefMacro{name-last-called-method}{last called method\xspace}
\DefMacro{no-prev-insns}{\Fix{UNUSED}\xspace}  %
\DefMacro{name-prev-insns}{\Fix{UNUSED bytecode}\xspace}
\DefMacro{no-similar-stmt}{S6\xspace}  %
\DefMacro{name-similar-stmt}{similar statement\xspace}
\DefMacro{no-runtime-types}{S7\xspace}  %
\DefMacro{name-runtime-types}{runtime types\xspace}
\DefMacro{no-runtime-values}{S8\xspace}  %
\DefMacro{name-runtime-values}{runtime values\xspace}
\DefMacro{no-runtime-values-depth1}{S9\xspace}  %
\DefMacro{name-runtime-values-depth1}{depth-1 runtime values\xspace}

\DefMacro{model-teco-icse23}{\TecoTool{}\xspace}
\DefMacro{model-teco-icse23-norr}{\TecoTool{}-noRr\xspace}
\DefMacro{model-teco-id}{\TecoTool-ID\xspace}
\DefMacro{model-teco-s1}{\TecoTool-\UseMacro{no-types-local}{}\xspace}
\DefMacro{model-teco-s2}{\TecoTool-\UseMacro{no-types-absent}{}\xspace}
\DefMacro{model-teco-s3}{\TecoTool-\UseMacro{no-fields-notset}{}\xspace}
\DefMacro{model-teco-s4}{\TecoTool-\UseMacro{no-setup-teardown}{}\xspace}
\DefMacro{model-teco-s5}{\TecoTool-\UseMacro{no-last-called-method}{}\xspace}
\DefMacro{model-teco-s6}{\TecoTool-\UseMacro{no-similar-stmt}{}\xspace}
\DefMacro{model-codet5-noft}{CodeT5-noFt\xspace}
\DefMacro{model-codet5}{CodeT5\xspace}
\DefMacro{model-codegpt-noft}{CodeGPT-noFt\xspace}
\DefMacro{model-codegpt}{CodeGPT\xspace}
\DefMacro{model-codex}{Codex\xspace}
\DefMacro{model-plbart}{PLBART\xspace}
\DefMacro{model-atlas}{ATLAS\xspace}
\DefMacro{model-toga}{TOGA\xspace}

\newcommand{\subtoken}{subtoken\xspace}
\newcommand{\subtokens}{subtokens\xspace}

\newcommand{\subtokenizing}{subtokenizing\xspace}

\newcommand{\train}{training\xspace}
\newcommand{\val}{validation\xspace}
\newcommand{\test}{evaluation\xspace}
\newcommand{\Test}{Evaluation\xspace}

\newcommand{\pretraining}{pre-training\xspace}
\newcommand{\pretrained}{pre-trained\xspace}
\newcommand{\finetune}{fine-tune\xspace}
\newcommand{\finetuned}{fine-tuned\xspace}

\newcommand{\finetuning}{fine-tuning\xspace}
\newcommand{\Finetuning}{Fine-tuning\xspace}

\newcommand{\BLEU}{BLEU\xspace}
\newcommand{\CodeBLEU}{CodeBLEU\xspace}
\newcommand{\editsim}{edit similarity\xspace}
\newcommand{\Editsim}{Edit similarity\xspace}
\newcommand{\editsimAcro}{EditSim\xspace}
\newcommand{\ROUGEL}{ROUGE\xspace}  %

\newcommand{\xmatch}{exact-match accuracy\xspace}
\newcommand{\Xmatch}{Exact-match accuracy\xspace}
\newcommand{\xmatchAcro}{XM\xspace}
\newcommand{\xmatchk}{top-10 accuracy\xspace}
\newcommand{\Xmatchk}{Top-10 accuracy\xspace}
\newcommand{\xmatchkAcro}{Acc@10\xspace}
\newcommand{\PctCompilable}{\%Compile\xspace}
\newcommand{\PctRunnable}{\%Run\xspace}

\newcommand{\Leditdis}{Levenshtein edit distance\xspace}

\newcommand{\aTest}{\ensuremath{T}\xspace}
\newcommand{\aCodeUT}{\ensuremath{C}\xspace}
\newcommand{\aInputPiece}{\ensuremath{x}\xspace}
\newcommand{\aInput}{\ensuremath{x}\xspace}
\newcommand{\aOutput}{\ensuremath{y}\xspace}
\newcommand{\aMUT}{\ensuremath{\aInputPiece_{mut}}\xspace}
\newcommand{\aSign}{\ensuremath{\aInputPiece_{sign}}\xspace}
\newcommand{\aPriorStmts}{\ensuremath{\aInputPiece_{prior}}\xspace}
\DefMacro{a-types-local}{\ensuremath{\aInputPiece_{\UseMacro{no-types-local}}}\xspace}
\DefMacro{a-types-absent}{\ensuremath{\aInputPiece_{\UseMacro{no-types-absent}}}\xspace}
\DefMacro{a-fields-notset}{\ensuremath{\aInputPiece_{\UseMacro{no-fields-notset}}}\xspace}
\DefMacro{a-setup-teardown}{\ensuremath{\aInputPiece_{\UseMacro{no-setup-teardown}}}\xspace}
\DefMacro{a-last-called-method}{\ensuremath{\aInputPiece_{\UseMacro{no-last-called-method}}}\xspace}
\DefMacro{a-prev-insns}{\ensuremath{\aInputPiece_{\UseMacro{no-prev-insns}}}\xspace}
\DefMacro{a-similar-stmt}{\ensuremath{\aInputPiece_{\UseMacro{no-similar-stmt}}}\xspace}
\DefMacro{a-runtime-types}{\ensuremath{\aInputPiece_{\UseMacro{no-runtime-types}}}\xspace}
\DefMacro{a-runtime-values}{\ensuremath{\aInputPiece_{\UseMacro{no-runtime-values}}}\xspace}
\DefMacro{a-runtime-values-depth1}{\ensuremath{\aInputPiece_{\UseMacro{no-runtime-values-depth1}}}\xspace}
\newcommand{\aSanalysis}{\ensuremath{\mathtt{analysis}}\xspace}

\newcommand{\aSep}{\ensuremath{\mathtt{\langle sep\rangle}}\xspace}
\newcommand{\aBos}{\ensuremath{\mathtt{\langle s\rangle}}\xspace}
\newcommand{\aEos}{\ensuremath{\mathtt{\langle /s\rangle}}\xspace}
\newcommand{\aEncoder}{\ensuremath{\mathtt{encoder}}\xspace}
\newcommand{\aEncH}{\ensuremath{h}\xspace}
\newcommand{\aDecoder}{\ensuremath{\mathtt{decoder}}\xspace}
\newcommand{\aLoss}{\ensuremath{\mathtt{loss}}\xspace}
\newcommand{\aCorpus}{\ensuremath{\mathcal{TC}}\xspace}
\newcommand{\aOutputs}{\ensuremath{\mathcal{Y}}\xspace}

\DefMacro{TH-model}{Model\xspace}
\DefMacro{TH-xmatch}{\xmatchAcro\xspace}
\DefMacro{TH-xmatch-top10}{\xmatchkAcro\xspace}
\DefMacro{TH-bleu}{\BLEU\xspace}
\DefMacro{TH-code-bleu}{\CodeBLEU\xspace}
\DefMacro{TH-edit-sim}{\editsimAcro\xspace}
\DefMacro{TH-rouge}{\ROUGEL\xspace}
\DefMacro{TH-rouge-p}{P\xspace}
\DefMacro{TH-rouge-r}{R\xspace}
\DefMacro{TH-rouge-f}{F1\xspace}
\DefMacro{TH-compilable}{\PctCompilable\xspace}
\DefMacro{TH-runnable}{\PctRunnable\xspace}

\newcommand{\THNA}{N/A\xspace}
\DefMacro{TH-SUM}{$\Sigma$\xspace}
\DefMacro{TH-AVG}{Avg.\xspace}
\DefMacro{TH-corpus-all}{all\xspace}
\DefMacro{TH-corpus-train}{\train\xspace}
\DefMacro{TH-corpus-val}{\val\xspace}
\DefMacro{TH-corpus-test}{\test\xspace}
\DefMacro{TH-num_proj}{\#proj\xspace}
\DefMacro{TH-num_test}{\#test\xspace}
\DefMacro{TH-num_stmt}{\#stmt\xspace}
\DefMacro{TH-tok_mut}{len(MUT)\xspace}
\DefMacro{TH-tok_test}{len(test)\xspace}

\DefMacro{TC-notice-bold}{The best number for each metric is
bolded.\xspace} 
\DefMacro{TC-notice-allsign}{The differences between models for each
metric are statistical significant.\xspace}
\DefMacro{TC-notice-sign}{Numbers marked with the same greek letter
prefix are not statistically significantly different.\xspace}
\DefMacro{TC-notice-sign-subtable}{In each table, numbers marked with
the same greek letter prefix are not statistically significantly
different.\xspace}

\DefMacro{corpus-filter-min_test_stmt}{5}
\DefMacro{corpus-filter-max_test_stmt}{8,222}
\DefMacro{corpus-filter-min_focalm_tok}{0}
\DefMacro{corpus-filter-max_focalm_tok}{9,787}
\DefMacro{corpus-filter-max_test_focalm_tok}{1,288}
\DefMacro{corpus-filter-max_tok_per_stmt}{1,726}
\DefMacro{corpus-filter-max_data_per_proj}{0}
\DefMacro{corpus-filter-min_data_per_proj}{0}
\DefMacro{corpus-filter-min_star}{0}
\DefMacro{corpus-filter-remove_badly_named_tests}{2,490}
\DefMacro{corpus-filter-remove_irregular_tests}{1,908}
\DefMacro{corpus-filter-cannot_locate_focalm}{36,818}
\DefMacro{corpus-filter-no_data}{265}
\DefMacro{corpus-filter-code_mapping_error}{633}
\DefMacro{corpus-filter-insn_simplify_warning}{0}
\DefMacro{corpus-filter-has_control_flow}{22,435}
\DefMacro{corpus-filter-has_invokedynamic}{5,420}
\DefMacro{corpus-initial-num_test}{221,666}
\DefMacro{corpus-limit-min_test_stmt}{1}
\DefMacro{corpus-limit-max_test_stmt}{20}
\DefMacro{corpus-limit-min_focalm_tok}{1}
\DefMacro{corpus-limit-max_focalm_tok}{200}
\DefMacro{corpus-limit-max_test_focalm_tok}{400}
\DefMacro{corpus-limit-max_tok_per_stmt}{100}
\DefMacro{corpus-limit-max_data_per_proj}{10,000}
\DefMacro{corpus-limit-min_data_per_proj}{1}
\DefMacro{corpus-limit-min_star}{0}
\DefMacro{corpus-limit-remove_badly_named_tests}{1}
\DefMacro{corpus-limit-remove_irregular_tests}{1}

\DefMacro{corpus-all-num_proj}{1,270}
\DefMacro{corpus-all-num_test}{130,934}
\DefMacro{corpus-all-num_stmt}{645,633}
\DefMacro{corpus-all-tok_mut}{40.88}
\DefMacro{corpus-all-tok_test}{79.57}
\DefMacro{corpus-all-seq_len-AVG}{243.88}
\DefMacro{corpus-all-seq_len-SUM}{157,453,971}
\DefMacro{corpus-all-seq_len-MAX}{2,426}
\DefMacro{corpus-all-seq_len-MIN}{23}
\DefMacro{corpus-all-seq_len-MEDIAN}{203.00}
\DefMacro{corpus-all-seq_len-STDEV}{176.47}
\DefMacro{corpus-all-seq_len-MODE}{139}
\DefMacro{corpus-all-seq_len-CNT}{645,633}
\DefMacro{corpus-all-seq_len_gt_512}{6}
\DefMacro{corpus-train-num_proj}{1,163}
\DefMacro{corpus-train-num_test}{120,521}
\DefMacro{corpus-train-num_stmt}{584,924}
\DefMacro{corpus-train-tok_mut}{40.61}
\DefMacro{corpus-train-tok_test}{79.58}
\DefMacro{corpus-train-seq_len-AVG}{239.07}
\DefMacro{corpus-train-seq_len-SUM}{139,838,682}
\DefMacro{corpus-train-seq_len-MAX}{2,426}
\DefMacro{corpus-train-seq_len-MIN}{23}
\DefMacro{corpus-train-seq_len-MEDIAN}{199.00}
\DefMacro{corpus-train-seq_len-STDEV}{171.57}
\DefMacro{corpus-train-seq_len-MODE}{140}
\DefMacro{corpus-train-seq_len-CNT}{584,924}
\DefMacro{corpus-train-seq_len_gt_512}{6}
\DefMacro{corpus-val-num_proj}{43}
\DefMacro{corpus-val-num_test}{5,413}
\DefMacro{corpus-val-num_stmt}{30,515}
\DefMacro{corpus-val-tok_mut}{45.09}
\DefMacro{corpus-val-tok_test}{73.24}
\DefMacro{corpus-val-seq_len-AVG}{267.59}
\DefMacro{corpus-val-seq_len-SUM}{8,165,425}
\DefMacro{corpus-val-seq_len-MAX}{1,227}
\DefMacro{corpus-val-seq_len-MIN}{24}
\DefMacro{corpus-val-seq_len-MEDIAN}{240.00}
\DefMacro{corpus-val-seq_len-STDEV}{154.81}
\DefMacro{corpus-val-seq_len-MODE}{122}
\DefMacro{corpus-val-seq_len-CNT}{30,515}
\DefMacro{corpus-val-seq_len_gt_512}{8}
\DefMacro{corpus-test-num_proj}{64}
\DefMacro{corpus-test-num_test}{5,000}
\DefMacro{corpus-test-num_stmt}{30,194}
\DefMacro{corpus-test-tok_mut}{42.85}
\DefMacro{corpus-test-tok_test}{86.26}
\DefMacro{corpus-test-seq_len-AVG}{312.97}
\DefMacro{corpus-test-seq_len-SUM}{9,449,864}
\DefMacro{corpus-test-seq_len-MAX}{2,198}
\DefMacro{corpus-test-seq_len-MIN}{26}
\DefMacro{corpus-test-seq_len-MEDIAN}{254.00}
\DefMacro{corpus-test-seq_len-STDEV}{256.15}
\DefMacro{corpus-test-seq_len-MODE}{198}
\DefMacro{corpus-test-seq_len-CNT}{30,194}
\DefMacro{corpus-test-seq_len_gt_512}{13}

\DefMacro{corpus-test-runnable-any-num_stmt}{25,074}
\DefMacro{corpus-test-runnable-any-pct}{83.04}
\DefMacro{corpus-test-runnable-any-num_test}{4,223}
\DefMacro{corpus-test-runnable-any-num_proj}{55}
\DefMacro{corpus-test-assert-num_stmt}{4,212}
\DefMacro{corpus-test-assert-pct}{13.95}
\DefMacro{corpus-test-assert-num_test}{4,212}
\DefMacro{corpus-test-assert-num_proj}{61}
\DefMacro{corpus-test-runnable-assert-num_stmt}{3,540}
\DefMacro{corpus-test-runnable-assert-pct}{11.72}
\DefMacro{corpus-test-runnable-assert-num_test}{3,540}
\DefMacro{corpus-test-runnable-assert-num_proj}{51}
\DefMacro{corpus-all-runnable-any-num_test}{101,965}
\DefMacro{corpus-all-runnable-any-pct}{77.88}
\DefMacro{corpus-train-assert-num_stmt}{92,567}
\DefMacro{corpus-train-assert-num_test}{92,531}
\DefMacro{corpus-train-assert-num_proj}{1,118}
\DefMacro{corpus-val-assert-num_stmt}{3,050}
\DefMacro{corpus-val-assert-num_test}{3,050}
\DefMacro{corpus-val-assert-num_proj}{41}

\DefMacro{res-atlas-CSNm-eval-any-stmt/test-bs10-last-xmatch}{-1.00}
\DefMacro{res-atlas-CSNm-eval-any-stmt/test-bs10-last-xmatch-top10}{-1.00}
\DefMacro{res-atlas-CSNm-eval-any-stmt/test-bs10-last-bleu}{-1.00}
\DefMacro{res-atlas-CSNm-eval-any-stmt/test-bs10-last-code-bleu}{-1.00}
\DefMacro{res-atlas-CSNm-eval-any-stmt/test-bs10-last-edit-sim}{-1.00}
\DefMacro{res-atlas-CSNm-eval-any-stmt/test-bs10-last-rouge-f}{-1.00}
\DefMacro{res-atlas-CSNm-eval-any-stmt/test-bs10-last-compilable}{-1.00}
\DefMacro{res-atlas-CSNm-eval-any-stmt/test-bs10-last-runnable}{-1.00}
\DefMacro{res-atlas-CSNm-eval-assert-stmt/test-bs10-last-xmatch}{0.21}
\DefMacro{res-atlas-CSNm-eval-assert-stmt/test-bs10-last-xmatch-top3}{0.36}
\DefMacro{res-atlas-CSNm-eval-assert-stmt/test-bs10-last-xmatch-top5}{0.47}
\DefMacro{res-atlas-CSNm-eval-assert-stmt/test-bs10-last-xmatch-top10}{0.66}
\DefMacro{res-atlas-CSNm-eval-assert-stmt/test-bs10-last-xmatch-top100}{0.66}
\DefMacro{res-atlas-CSNm-eval-assert-stmt/test-bs10-last-bleu}{21.55}
\DefMacro{res-atlas-CSNm-eval-assert-stmt/test-bs10-last-bleu-max}{28.56}
\DefMacro{res-atlas-CSNm-eval-assert-stmt/test-bs10-last-bleu-min}{12.25}
\DefMacro{res-atlas-CSNm-eval-assert-stmt/test-bs10-last-bleu-avg}{19.43}
\DefMacro{res-atlas-CSNm-eval-assert-stmt/test-bs10-last-code-bleu}{13.39}
\DefMacro{res-atlas-CSNm-eval-assert-stmt/test-bs10-last-code-bleu-max}{22.54}
\DefMacro{res-atlas-CSNm-eval-assert-stmt/test-bs10-last-code-bleu-min}{5.36}
\DefMacro{res-atlas-CSNm-eval-assert-stmt/test-bs10-last-code-bleu-avg}{12.25}
\DefMacro{res-atlas-CSNm-eval-assert-stmt/test-bs10-last-edit-sim}{54.06}
\DefMacro{res-atlas-CSNm-eval-assert-stmt/test-bs10-last-edit-sim-max}{60.90}
\DefMacro{res-atlas-CSNm-eval-assert-stmt/test-bs10-last-edit-sim-min}{44.08}
\DefMacro{res-atlas-CSNm-eval-assert-stmt/test-bs10-last-edit-sim-avg}{52.36}
\DefMacro{res-atlas-CSNm-eval-assert-stmt/test-bs10-last-rouge-p}{54.07}
\DefMacro{res-atlas-CSNm-eval-assert-stmt/test-bs10-last-rouge-p-max}{59.63}
\DefMacro{res-atlas-CSNm-eval-assert-stmt/test-bs10-last-rouge-p-min}{37.14}
\DefMacro{res-atlas-CSNm-eval-assert-stmt/test-bs10-last-rouge-p-avg}{47.64}
\DefMacro{res-atlas-CSNm-eval-assert-stmt/test-bs10-last-rouge-r}{49.95}
\DefMacro{res-atlas-CSNm-eval-assert-stmt/test-bs10-last-rouge-r-max}{57.37}
\DefMacro{res-atlas-CSNm-eval-assert-stmt/test-bs10-last-rouge-r-min}{45.67}
\DefMacro{res-atlas-CSNm-eval-assert-stmt/test-bs10-last-rouge-r-avg}{51.30}
\DefMacro{res-atlas-CSNm-eval-assert-stmt/test-bs10-last-rouge-f}{50.70}
\DefMacro{res-atlas-CSNm-eval-assert-stmt/test-bs10-last-rouge-f-max}{56.70}
\DefMacro{res-atlas-CSNm-eval-assert-stmt/test-bs10-last-rouge-f-min}{40.19}
\DefMacro{res-atlas-CSNm-eval-assert-stmt/test-bs10-last-rouge-f-avg}{48.03}
\DefMacro{res-atlas-CSNm-eval-assert-stmt/test-bs10-last-compilable}{-1.00}
\DefMacro{res-atlas-CSNm-eval-assert-stmt/test-bs10-last-runnable}{-1.00}
\DefMacro{res-atlas-CSNm-eval-runnable-any-stmt/test-bs10-last-xmatch}{-1.00}
\DefMacro{res-atlas-CSNm-eval-runnable-any-stmt/test-bs10-last-xmatch-top10}{-1.00}
\DefMacro{res-atlas-CSNm-eval-runnable-any-stmt/test-bs10-last-bleu}{-1.00}
\DefMacro{res-atlas-CSNm-eval-runnable-any-stmt/test-bs10-last-code-bleu}{-1.00}
\DefMacro{res-atlas-CSNm-eval-runnable-any-stmt/test-bs10-last-edit-sim}{-1.00}
\DefMacro{res-atlas-CSNm-eval-runnable-any-stmt/test-bs10-last-rouge-f}{-1.00}
\DefMacro{res-atlas-CSNm-eval-runnable-any-stmt/test-bs10-last-compilable}{-1.00}
\DefMacro{res-atlas-CSNm-eval-runnable-any-stmt/test-bs10-last-runnable}{-1.00}
\DefMacro{res-atlas-CSNm-eval-runnable-assert-stmt/test-bs10-last-xmatch}{0.23}
\DefMacro{res-atlas-CSNm-eval-runnable-assert-stmt/test-bs10-last-xmatch-top3}{0.39}
\DefMacro{res-atlas-CSNm-eval-runnable-assert-stmt/test-bs10-last-xmatch-top5}{0.51}
\DefMacro{res-atlas-CSNm-eval-runnable-assert-stmt/test-bs10-last-xmatch-top10}{0.68}
\DefMacro{res-atlas-CSNm-eval-runnable-assert-stmt/test-bs10-last-xmatch-top100}{0.68}
\DefMacro{res-atlas-CSNm-eval-runnable-assert-stmt/test-bs10-last-bleu}{21.81}
\DefMacro{res-atlas-CSNm-eval-runnable-assert-stmt/test-bs10-last-bleu-max}{28.97}
\DefMacro{res-atlas-CSNm-eval-runnable-assert-stmt/test-bs10-last-bleu-min}{12.40}
\DefMacro{res-atlas-CSNm-eval-runnable-assert-stmt/test-bs10-last-bleu-avg}{19.61}
\DefMacro{res-atlas-CSNm-eval-runnable-assert-stmt/test-bs10-last-code-bleu}{13.56}
\DefMacro{res-atlas-CSNm-eval-runnable-assert-stmt/test-bs10-last-code-bleu-max}{22.96}
\DefMacro{res-atlas-CSNm-eval-runnable-assert-stmt/test-bs10-last-code-bleu-min}{5.42}
\DefMacro{res-atlas-CSNm-eval-runnable-assert-stmt/test-bs10-last-code-bleu-avg}{12.42}
\DefMacro{res-atlas-CSNm-eval-runnable-assert-stmt/test-bs10-last-edit-sim}{54.19}
\DefMacro{res-atlas-CSNm-eval-runnable-assert-stmt/test-bs10-last-edit-sim-max}{61.02}
\DefMacro{res-atlas-CSNm-eval-runnable-assert-stmt/test-bs10-last-edit-sim-min}{44.23}
\DefMacro{res-atlas-CSNm-eval-runnable-assert-stmt/test-bs10-last-edit-sim-avg}{52.46}
\DefMacro{res-atlas-CSNm-eval-runnable-assert-stmt/test-bs10-last-rouge-p}{54.06}
\DefMacro{res-atlas-CSNm-eval-runnable-assert-stmt/test-bs10-last-rouge-p-max}{59.75}
\DefMacro{res-atlas-CSNm-eval-runnable-assert-stmt/test-bs10-last-rouge-p-min}{37.21}
\DefMacro{res-atlas-CSNm-eval-runnable-assert-stmt/test-bs10-last-rouge-p-avg}{47.66}
\DefMacro{res-atlas-CSNm-eval-runnable-assert-stmt/test-bs10-last-rouge-r}{50.25}
\DefMacro{res-atlas-CSNm-eval-runnable-assert-stmt/test-bs10-last-rouge-r-max}{57.80}
\DefMacro{res-atlas-CSNm-eval-runnable-assert-stmt/test-bs10-last-rouge-r-min}{45.92}
\DefMacro{res-atlas-CSNm-eval-runnable-assert-stmt/test-bs10-last-rouge-r-avg}{51.70}
\DefMacro{res-atlas-CSNm-eval-runnable-assert-stmt/test-bs10-last-rouge-f}{50.87}
\DefMacro{res-atlas-CSNm-eval-runnable-assert-stmt/test-bs10-last-rouge-f-max}{56.99}
\DefMacro{res-atlas-CSNm-eval-runnable-assert-stmt/test-bs10-last-rouge-f-min}{40.34}
\DefMacro{res-atlas-CSNm-eval-runnable-assert-stmt/test-bs10-last-rouge-f-avg}{48.24}
\DefMacro{res-atlas-CSNm-eval-runnable-assert-stmt/test-bs10-last-compilable}{3.62}
\DefMacro{res-atlas-CSNm-eval-runnable-assert-stmt/test-bs10-last-runnable}{1.45}
\DefMacro{res-codegpt-CSNm-eval-any-stmt/test-bs10-last-xmatch}{12.20}
\DefMacro{res-codegpt-CSNm-eval-any-stmt/test-bs10-last-xmatch-top3}{17.99}
\DefMacro{res-codegpt-CSNm-eval-any-stmt/test-bs10-last-xmatch-top5}{20.13}
\DefMacro{res-codegpt-CSNm-eval-any-stmt/test-bs10-last-xmatch-top10}{22.67}
\DefMacro{res-codegpt-CSNm-eval-any-stmt/test-bs10-last-xmatch-top100}{22.67}
\DefMacro{res-codegpt-CSNm-eval-any-stmt/test-bs10-last-bleu}{36.30}
\DefMacro{res-codegpt-CSNm-eval-any-stmt/test-bs10-last-bleu-max}{49.82}
\DefMacro{res-codegpt-CSNm-eval-any-stmt/test-bs10-last-bleu-min}{16.13}
\DefMacro{res-codegpt-CSNm-eval-any-stmt/test-bs10-last-bleu-avg}{28.42}
\DefMacro{res-codegpt-CSNm-eval-any-stmt/test-bs10-last-code-bleu}{31.84}
\DefMacro{res-codegpt-CSNm-eval-any-stmt/test-bs10-last-code-bleu-max}{47.47}
\DefMacro{res-codegpt-CSNm-eval-any-stmt/test-bs10-last-code-bleu-min}{10.33}
\DefMacro{res-codegpt-CSNm-eval-any-stmt/test-bs10-last-code-bleu-avg}{24.09}
\DefMacro{res-codegpt-CSNm-eval-any-stmt/test-bs10-last-edit-sim}{59.09}
\DefMacro{res-codegpt-CSNm-eval-any-stmt/test-bs10-last-edit-sim-max}{69.66}
\DefMacro{res-codegpt-CSNm-eval-any-stmt/test-bs10-last-edit-sim-min}{42.60}
\DefMacro{res-codegpt-CSNm-eval-any-stmt/test-bs10-last-edit-sim-avg}{54.28}
\DefMacro{res-codegpt-CSNm-eval-any-stmt/test-bs10-last-rouge-p}{62.16}
\DefMacro{res-codegpt-CSNm-eval-any-stmt/test-bs10-last-rouge-p-max}{74.25}
\DefMacro{res-codegpt-CSNm-eval-any-stmt/test-bs10-last-rouge-p-min}{42.52}
\DefMacro{res-codegpt-CSNm-eval-any-stmt/test-bs10-last-rouge-p-avg}{55.65}
\DefMacro{res-codegpt-CSNm-eval-any-stmt/test-bs10-last-rouge-r}{62.87}
\DefMacro{res-codegpt-CSNm-eval-any-stmt/test-bs10-last-rouge-r-max}{72.95}
\DefMacro{res-codegpt-CSNm-eval-any-stmt/test-bs10-last-rouge-r-min}{49.06}
\DefMacro{res-codegpt-CSNm-eval-any-stmt/test-bs10-last-rouge-r-avg}{60.58}
\DefMacro{res-codegpt-CSNm-eval-any-stmt/test-bs10-last-rouge-f}{61.10}
\DefMacro{res-codegpt-CSNm-eval-any-stmt/test-bs10-last-rouge-f-max}{71.84}
\DefMacro{res-codegpt-CSNm-eval-any-stmt/test-bs10-last-rouge-f-min}{44.40}
\DefMacro{res-codegpt-CSNm-eval-any-stmt/test-bs10-last-rouge-f-avg}{56.37}
\DefMacro{res-codegpt-CSNm-eval-any-stmt/test-bs10-last-num-seq}{10.00}
\DefMacro{res-codegpt-CSNm-eval-any-stmt/test-bs10-last-compilable}{-1.00}
\DefMacro{res-codegpt-CSNm-eval-any-stmt/test-bs10-last-runnable}{-1.00}
\DefMacro{res-codegpt-CSNm-eval-assert-stmt/test-bs10-last-xmatch}{10.56}
\DefMacro{res-codegpt-CSNm-eval-assert-stmt/test-bs10-last-xmatch-top3}{18.47}
\DefMacro{res-codegpt-CSNm-eval-assert-stmt/test-bs10-last-xmatch-top5}{22.10}
\DefMacro{res-codegpt-CSNm-eval-assert-stmt/test-bs10-last-xmatch-top10}{27.19}
\DefMacro{res-codegpt-CSNm-eval-assert-stmt/test-bs10-last-xmatch-top100}{27.19}
\DefMacro{res-codegpt-CSNm-eval-assert-stmt/test-bs10-last-bleu}{40.91}
\DefMacro{res-codegpt-CSNm-eval-assert-stmt/test-bs10-last-bleu-max}{60.19}
\DefMacro{res-codegpt-CSNm-eval-assert-stmt/test-bs10-last-bleu-min}{17.74}
\DefMacro{res-codegpt-CSNm-eval-assert-stmt/test-bs10-last-bleu-avg}{33.55}
\DefMacro{res-codegpt-CSNm-eval-assert-stmt/test-bs10-last-code-bleu}{33.33}
\DefMacro{res-codegpt-CSNm-eval-assert-stmt/test-bs10-last-code-bleu-max}{54.85}
\DefMacro{res-codegpt-CSNm-eval-assert-stmt/test-bs10-last-code-bleu-min}{10.67}
\DefMacro{res-codegpt-CSNm-eval-assert-stmt/test-bs10-last-code-bleu-avg}{26.64}
\DefMacro{res-codegpt-CSNm-eval-assert-stmt/test-bs10-last-edit-sim}{67.63}
\DefMacro{res-codegpt-CSNm-eval-assert-stmt/test-bs10-last-edit-sim-max}{80.50}
\DefMacro{res-codegpt-CSNm-eval-assert-stmt/test-bs10-last-edit-sim-min}{49.16}
\DefMacro{res-codegpt-CSNm-eval-assert-stmt/test-bs10-last-edit-sim-avg}{63.22}
\DefMacro{res-codegpt-CSNm-eval-assert-stmt/test-bs10-last-rouge-p}{70.11}
\DefMacro{res-codegpt-CSNm-eval-assert-stmt/test-bs10-last-rouge-p-max}{84.24}
\DefMacro{res-codegpt-CSNm-eval-assert-stmt/test-bs10-last-rouge-p-min}{47.39}
\DefMacro{res-codegpt-CSNm-eval-assert-stmt/test-bs10-last-rouge-p-avg}{63.09}
\DefMacro{res-codegpt-CSNm-eval-assert-stmt/test-bs10-last-rouge-r}{67.95}
\DefMacro{res-codegpt-CSNm-eval-assert-stmt/test-bs10-last-rouge-r-max}{80.56}
\DefMacro{res-codegpt-CSNm-eval-assert-stmt/test-bs10-last-rouge-r-min}{53.29}
\DefMacro{res-codegpt-CSNm-eval-assert-stmt/test-bs10-last-rouge-r-avg}{66.14}
\DefMacro{res-codegpt-CSNm-eval-assert-stmt/test-bs10-last-rouge-f}{67.94}
\DefMacro{res-codegpt-CSNm-eval-assert-stmt/test-bs10-last-rouge-f-max}{81.02}
\DefMacro{res-codegpt-CSNm-eval-assert-stmt/test-bs10-last-rouge-f-min}{49.46}
\DefMacro{res-codegpt-CSNm-eval-assert-stmt/test-bs10-last-rouge-f-avg}{63.36}
\DefMacro{res-codegpt-CSNm-eval-assert-stmt/test-bs10-last-num-seq}{10.00}
\DefMacro{res-codegpt-CSNm-eval-assert-stmt/test-bs10-last-compilable}{-1.00}
\DefMacro{res-codegpt-CSNm-eval-assert-stmt/test-bs10-last-runnable}{-1.00}
\DefMacro{res-codegpt-CSNm-eval-runnable-any-stmt/test-bs10-last-xmatch}{12.95}
\DefMacro{res-codegpt-CSNm-eval-runnable-any-stmt/test-bs10-last-xmatch-top3}{19.13}
\DefMacro{res-codegpt-CSNm-eval-runnable-any-stmt/test-bs10-last-xmatch-top5}{21.39}
\DefMacro{res-codegpt-CSNm-eval-runnable-any-stmt/test-bs10-last-xmatch-top10}{24.03}
\DefMacro{res-codegpt-CSNm-eval-runnable-any-stmt/test-bs10-last-xmatch-top100}{24.03}
\DefMacro{res-codegpt-CSNm-eval-runnable-any-stmt/test-bs10-last-bleu}{37.19}
\DefMacro{res-codegpt-CSNm-eval-runnable-any-stmt/test-bs10-last-bleu-max}{51.14}
\DefMacro{res-codegpt-CSNm-eval-runnable-any-stmt/test-bs10-last-bleu-min}{16.32}
\DefMacro{res-codegpt-CSNm-eval-runnable-any-stmt/test-bs10-last-bleu-avg}{28.96}
\DefMacro{res-codegpt-CSNm-eval-runnable-any-stmt/test-bs10-last-code-bleu}{32.46}
\DefMacro{res-codegpt-CSNm-eval-runnable-any-stmt/test-bs10-last-code-bleu-max}{48.65}
\DefMacro{res-codegpt-CSNm-eval-runnable-any-stmt/test-bs10-last-code-bleu-min}{10.19}
\DefMacro{res-codegpt-CSNm-eval-runnable-any-stmt/test-bs10-last-code-bleu-avg}{24.35}
\DefMacro{res-codegpt-CSNm-eval-runnable-any-stmt/test-bs10-last-edit-sim}{59.75}
\DefMacro{res-codegpt-CSNm-eval-runnable-any-stmt/test-bs10-last-edit-sim-max}{70.58}
\DefMacro{res-codegpt-CSNm-eval-runnable-any-stmt/test-bs10-last-edit-sim-min}{42.81}
\DefMacro{res-codegpt-CSNm-eval-runnable-any-stmt/test-bs10-last-edit-sim-avg}{54.74}
\DefMacro{res-codegpt-CSNm-eval-runnable-any-stmt/test-bs10-last-rouge-p}{62.89}
\DefMacro{res-codegpt-CSNm-eval-runnable-any-stmt/test-bs10-last-rouge-p-max}{75.11}
\DefMacro{res-codegpt-CSNm-eval-runnable-any-stmt/test-bs10-last-rouge-p-min}{42.76}
\DefMacro{res-codegpt-CSNm-eval-runnable-any-stmt/test-bs10-last-rouge-p-avg}{56.16}
\DefMacro{res-codegpt-CSNm-eval-runnable-any-stmt/test-bs10-last-rouge-r}{63.66}
\DefMacro{res-codegpt-CSNm-eval-runnable-any-stmt/test-bs10-last-rouge-r-max}{73.95}
\DefMacro{res-codegpt-CSNm-eval-runnable-any-stmt/test-bs10-last-rouge-r-min}{49.36}
\DefMacro{res-codegpt-CSNm-eval-runnable-any-stmt/test-bs10-last-rouge-r-avg}{61.20}
\DefMacro{res-codegpt-CSNm-eval-runnable-any-stmt/test-bs10-last-rouge-f}{61.90}
\DefMacro{res-codegpt-CSNm-eval-runnable-any-stmt/test-bs10-last-rouge-f-max}{72.85}
\DefMacro{res-codegpt-CSNm-eval-runnable-any-stmt/test-bs10-last-rouge-f-min}{44.71}
\DefMacro{res-codegpt-CSNm-eval-runnable-any-stmt/test-bs10-last-rouge-f-avg}{56.97}
\DefMacro{res-codegpt-CSNm-eval-runnable-any-stmt/test-bs10-last-num-seq}{10.00}
\DefMacro{res-codegpt-CSNm-eval-runnable-any-stmt/test-bs10-last-compilable}{53.77}
\DefMacro{res-codegpt-CSNm-eval-runnable-any-stmt/test-bs10-last-runnable}{15.13}
\DefMacro{res-codegpt-CSNm-eval-runnable-assert-stmt/test-bs10-last-xmatch}{10.41}
\DefMacro{res-codegpt-CSNm-eval-runnable-assert-stmt/test-bs10-last-xmatch-top3}{18.99}
\DefMacro{res-codegpt-CSNm-eval-runnable-assert-stmt/test-bs10-last-xmatch-top5}{22.93}
\DefMacro{res-codegpt-CSNm-eval-runnable-assert-stmt/test-bs10-last-xmatch-top10}{28.17}
\DefMacro{res-codegpt-CSNm-eval-runnable-assert-stmt/test-bs10-last-xmatch-top100}{28.17}
\DefMacro{res-codegpt-CSNm-eval-runnable-assert-stmt/test-bs10-last-bleu}{41.56}
\DefMacro{res-codegpt-CSNm-eval-runnable-assert-stmt/test-bs10-last-bleu-max}{60.67}
\DefMacro{res-codegpt-CSNm-eval-runnable-assert-stmt/test-bs10-last-bleu-min}{18.32}
\DefMacro{res-codegpt-CSNm-eval-runnable-assert-stmt/test-bs10-last-bleu-avg}{34.20}
\DefMacro{res-codegpt-CSNm-eval-runnable-assert-stmt/test-bs10-last-code-bleu}{33.79}
\DefMacro{res-codegpt-CSNm-eval-runnable-assert-stmt/test-bs10-last-code-bleu-max}{55.22}
\DefMacro{res-codegpt-CSNm-eval-runnable-assert-stmt/test-bs10-last-code-bleu-min}{11.18}
\DefMacro{res-codegpt-CSNm-eval-runnable-assert-stmt/test-bs10-last-code-bleu-avg}{27.28}
\DefMacro{res-codegpt-CSNm-eval-runnable-assert-stmt/test-bs10-last-edit-sim}{67.78}
\DefMacro{res-codegpt-CSNm-eval-runnable-assert-stmt/test-bs10-last-edit-sim-max}{80.64}
\DefMacro{res-codegpt-CSNm-eval-runnable-assert-stmt/test-bs10-last-edit-sim-min}{49.56}
\DefMacro{res-codegpt-CSNm-eval-runnable-assert-stmt/test-bs10-last-edit-sim-avg}{63.45}
\DefMacro{res-codegpt-CSNm-eval-runnable-assert-stmt/test-bs10-last-rouge-p}{70.34}
\DefMacro{res-codegpt-CSNm-eval-runnable-assert-stmt/test-bs10-last-rouge-p-max}{84.27}
\DefMacro{res-codegpt-CSNm-eval-runnable-assert-stmt/test-bs10-last-rouge-p-min}{47.53}
\DefMacro{res-codegpt-CSNm-eval-runnable-assert-stmt/test-bs10-last-rouge-p-avg}{63.24}
\DefMacro{res-codegpt-CSNm-eval-runnable-assert-stmt/test-bs10-last-rouge-r}{68.68}
\DefMacro{res-codegpt-CSNm-eval-runnable-assert-stmt/test-bs10-last-rouge-r-max}{80.96}
\DefMacro{res-codegpt-CSNm-eval-runnable-assert-stmt/test-bs10-last-rouge-r-min}{53.73}
\DefMacro{res-codegpt-CSNm-eval-runnable-assert-stmt/test-bs10-last-rouge-r-avg}{66.65}
\DefMacro{res-codegpt-CSNm-eval-runnable-assert-stmt/test-bs10-last-rouge-f}{68.43}
\DefMacro{res-codegpt-CSNm-eval-runnable-assert-stmt/test-bs10-last-rouge-f-max}{81.33}
\DefMacro{res-codegpt-CSNm-eval-runnable-assert-stmt/test-bs10-last-rouge-f-min}{49.75}
\DefMacro{res-codegpt-CSNm-eval-runnable-assert-stmt/test-bs10-last-rouge-f-avg}{63.70}
\DefMacro{res-codegpt-CSNm-eval-runnable-assert-stmt/test-bs10-last-num-seq}{10.00}
\DefMacro{res-codegpt-CSNm-eval-runnable-assert-stmt/test-bs10-last-compilable}{47.39}
\DefMacro{res-codegpt-CSNm-eval-runnable-assert-stmt/test-bs10-last-runnable}{16.13}
\DefMacro{res-codet5-CSNm-eval-any-stmt/test-bs10-last-xmatch}{13.57}
\DefMacro{res-codet5-CSNm-eval-any-stmt/test-bs10-last-xmatch-top3}{19.19}
\DefMacro{res-codet5-CSNm-eval-any-stmt/test-bs10-last-xmatch-top5}{21.32}
\DefMacro{res-codet5-CSNm-eval-any-stmt/test-bs10-last-xmatch-top10}{24.11}
\DefMacro{res-codet5-CSNm-eval-any-stmt/test-bs10-last-xmatch-top100}{24.11}
\DefMacro{res-codet5-CSNm-eval-any-stmt/test-bs10-last-bleu}{38.33}
\DefMacro{res-codet5-CSNm-eval-any-stmt/test-bs10-last-bleu-max}{52.28}
\DefMacro{res-codet5-CSNm-eval-any-stmt/test-bs10-last-bleu-min}{17.38}
\DefMacro{res-codet5-CSNm-eval-any-stmt/test-bs10-last-bleu-avg}{30.58}
\DefMacro{res-codet5-CSNm-eval-any-stmt/test-bs10-last-code-bleu}{33.88}
\DefMacro{res-codet5-CSNm-eval-any-stmt/test-bs10-last-code-bleu-max}{50.19}
\DefMacro{res-codet5-CSNm-eval-any-stmt/test-bs10-last-code-bleu-min}{11.35}
\DefMacro{res-codet5-CSNm-eval-any-stmt/test-bs10-last-code-bleu-avg}{26.25}
\DefMacro{res-codet5-CSNm-eval-any-stmt/test-bs10-last-edit-sim}{60.81}
\DefMacro{res-codet5-CSNm-eval-any-stmt/test-bs10-last-edit-sim-max}{71.95}
\DefMacro{res-codet5-CSNm-eval-any-stmt/test-bs10-last-edit-sim-min}{43.73}
\DefMacro{res-codet5-CSNm-eval-any-stmt/test-bs10-last-edit-sim-avg}{56.36}
\DefMacro{res-codet5-CSNm-eval-any-stmt/test-bs10-last-rouge-p}{63.55}
\DefMacro{res-codet5-CSNm-eval-any-stmt/test-bs10-last-rouge-p-max}{75.73}
\DefMacro{res-codet5-CSNm-eval-any-stmt/test-bs10-last-rouge-p-min}{43.33}
\DefMacro{res-codet5-CSNm-eval-any-stmt/test-bs10-last-rouge-p-avg}{57.04}
\DefMacro{res-codet5-CSNm-eval-any-stmt/test-bs10-last-rouge-r}{63.97}
\DefMacro{res-codet5-CSNm-eval-any-stmt/test-bs10-last-rouge-r-max}{74.41}
\DefMacro{res-codet5-CSNm-eval-any-stmt/test-bs10-last-rouge-r-min}{49.34}
\DefMacro{res-codet5-CSNm-eval-any-stmt/test-bs10-last-rouge-r-avg}{61.60}
\DefMacro{res-codet5-CSNm-eval-any-stmt/test-bs10-last-rouge-f}{62.40}
\DefMacro{res-codet5-CSNm-eval-any-stmt/test-bs10-last-rouge-f-max}{73.40}
\DefMacro{res-codet5-CSNm-eval-any-stmt/test-bs10-last-rouge-f-min}{45.09}
\DefMacro{res-codet5-CSNm-eval-any-stmt/test-bs10-last-rouge-f-avg}{57.66}
\DefMacro{res-codet5-CSNm-eval-any-stmt/test-bs10-last-num-seq}{10.00}
\DefMacro{res-codet5-CSNm-eval-any-stmt/test-bs10-last-compilable}{-1.00}
\DefMacro{res-codet5-CSNm-eval-any-stmt/test-bs10-last-runnable}{-1.00}
\DefMacro{res-codet5-CSNm-eval-assert-stmt/test-bs10-last-xmatch}{8.45}
\DefMacro{res-codet5-CSNm-eval-assert-stmt/test-bs10-last-xmatch-top3}{14.63}
\DefMacro{res-codet5-CSNm-eval-assert-stmt/test-bs10-last-xmatch-top5}{18.29}
\DefMacro{res-codet5-CSNm-eval-assert-stmt/test-bs10-last-xmatch-top10}{24.04}
\DefMacro{res-codet5-CSNm-eval-assert-stmt/test-bs10-last-xmatch-top100}{24.04}
\DefMacro{res-codet5-CSNm-eval-assert-stmt/test-bs10-last-bleu}{39.03}
\DefMacro{res-codet5-CSNm-eval-assert-stmt/test-bs10-last-bleu-max}{58.22}
\DefMacro{res-codet5-CSNm-eval-assert-stmt/test-bs10-last-bleu-min}{18.62}
\DefMacro{res-codet5-CSNm-eval-assert-stmt/test-bs10-last-bleu-avg}{33.91}
\DefMacro{res-codet5-CSNm-eval-assert-stmt/test-bs10-last-code-bleu}{31.03}
\DefMacro{res-codet5-CSNm-eval-assert-stmt/test-bs10-last-code-bleu-max}{52.85}
\DefMacro{res-codet5-CSNm-eval-assert-stmt/test-bs10-last-code-bleu-min}{11.21}
\DefMacro{res-codet5-CSNm-eval-assert-stmt/test-bs10-last-code-bleu-avg}{26.67}
\DefMacro{res-codet5-CSNm-eval-assert-stmt/test-bs10-last-edit-sim}{66.50}
\DefMacro{res-codet5-CSNm-eval-assert-stmt/test-bs10-last-edit-sim-max}{79.79}
\DefMacro{res-codet5-CSNm-eval-assert-stmt/test-bs10-last-edit-sim-min}{49.58}
\DefMacro{res-codet5-CSNm-eval-assert-stmt/test-bs10-last-edit-sim-avg}{63.71}
\DefMacro{res-codet5-CSNm-eval-assert-stmt/test-bs10-last-rouge-p}{67.78}
\DefMacro{res-codet5-CSNm-eval-assert-stmt/test-bs10-last-rouge-p-max}{81.84}
\DefMacro{res-codet5-CSNm-eval-assert-stmt/test-bs10-last-rouge-p-min}{46.94}
\DefMacro{res-codet5-CSNm-eval-assert-stmt/test-bs10-last-rouge-p-avg}{62.14}
\DefMacro{res-codet5-CSNm-eval-assert-stmt/test-bs10-last-rouge-r}{67.66}
\DefMacro{res-codet5-CSNm-eval-assert-stmt/test-bs10-last-rouge-r-max}{80.34}
\DefMacro{res-codet5-CSNm-eval-assert-stmt/test-bs10-last-rouge-r-min}{53.25}
\DefMacro{res-codet5-CSNm-eval-assert-stmt/test-bs10-last-rouge-r-avg}{66.25}
\DefMacro{res-codet5-CSNm-eval-assert-stmt/test-bs10-last-rouge-f}{66.63}
\DefMacro{res-codet5-CSNm-eval-assert-stmt/test-bs10-last-rouge-f-max}{79.85}
\DefMacro{res-codet5-CSNm-eval-assert-stmt/test-bs10-last-rouge-f-min}{49.06}
\DefMacro{res-codet5-CSNm-eval-assert-stmt/test-bs10-last-rouge-f-avg}{63.00}
\DefMacro{res-codet5-CSNm-eval-assert-stmt/test-bs10-last-num-seq}{10.00}
\DefMacro{res-codet5-CSNm-eval-assert-stmt/test-bs10-last-compilable}{-1.00}
\DefMacro{res-codet5-CSNm-eval-assert-stmt/test-bs10-last-runnable}{-1.00}
\DefMacro{res-codet5-CSNm-eval-runnable-any-stmt/test-bs10-last-xmatch}{14.38}
\DefMacro{res-codet5-CSNm-eval-runnable-any-stmt/test-bs10-last-xmatch-top3}{20.26}
\DefMacro{res-codet5-CSNm-eval-runnable-any-stmt/test-bs10-last-xmatch-top5}{22.50}
\DefMacro{res-codet5-CSNm-eval-runnable-any-stmt/test-bs10-last-xmatch-top10}{25.39}
\DefMacro{res-codet5-CSNm-eval-runnable-any-stmt/test-bs10-last-xmatch-top100}{25.39}
\DefMacro{res-codet5-CSNm-eval-runnable-any-stmt/test-bs10-last-bleu}{39.26}
\DefMacro{res-codet5-CSNm-eval-runnable-any-stmt/test-bs10-last-bleu-max}{53.48}
\DefMacro{res-codet5-CSNm-eval-runnable-any-stmt/test-bs10-last-bleu-min}{17.45}
\DefMacro{res-codet5-CSNm-eval-runnable-any-stmt/test-bs10-last-bleu-avg}{31.08}
\DefMacro{res-codet5-CSNm-eval-runnable-any-stmt/test-bs10-last-code-bleu}{34.55}
\DefMacro{res-codet5-CSNm-eval-runnable-any-stmt/test-bs10-last-code-bleu-max}{51.19}
\DefMacro{res-codet5-CSNm-eval-runnable-any-stmt/test-bs10-last-code-bleu-min}{11.08}
\DefMacro{res-codet5-CSNm-eval-runnable-any-stmt/test-bs10-last-code-bleu-avg}{26.46}
\DefMacro{res-codet5-CSNm-eval-runnable-any-stmt/test-bs10-last-edit-sim}{61.36}
\DefMacro{res-codet5-CSNm-eval-runnable-any-stmt/test-bs10-last-edit-sim-max}{72.70}
\DefMacro{res-codet5-CSNm-eval-runnable-any-stmt/test-bs10-last-edit-sim-min}{43.71}
\DefMacro{res-codet5-CSNm-eval-runnable-any-stmt/test-bs10-last-edit-sim-avg}{56.68}
\DefMacro{res-codet5-CSNm-eval-runnable-any-stmt/test-bs10-last-rouge-p}{64.28}
\DefMacro{res-codet5-CSNm-eval-runnable-any-stmt/test-bs10-last-rouge-p-max}{76.57}
\DefMacro{res-codet5-CSNm-eval-runnable-any-stmt/test-bs10-last-rouge-p-min}{43.51}
\DefMacro{res-codet5-CSNm-eval-runnable-any-stmt/test-bs10-last-rouge-p-avg}{57.54}
\DefMacro{res-codet5-CSNm-eval-runnable-any-stmt/test-bs10-last-rouge-r}{64.64}
\DefMacro{res-codet5-CSNm-eval-runnable-any-stmt/test-bs10-last-rouge-r-max}{75.25}
\DefMacro{res-codet5-CSNm-eval-runnable-any-stmt/test-bs10-last-rouge-r-min}{49.50}
\DefMacro{res-codet5-CSNm-eval-runnable-any-stmt/test-bs10-last-rouge-r-avg}{62.13}
\DefMacro{res-codet5-CSNm-eval-runnable-any-stmt/test-bs10-last-rouge-f}{63.15}
\DefMacro{res-codet5-CSNm-eval-runnable-any-stmt/test-bs10-last-rouge-f-max}{74.30}
\DefMacro{res-codet5-CSNm-eval-runnable-any-stmt/test-bs10-last-rouge-f-min}{45.31}
\DefMacro{res-codet5-CSNm-eval-runnable-any-stmt/test-bs10-last-rouge-f-avg}{58.20}
\DefMacro{res-codet5-CSNm-eval-runnable-any-stmt/test-bs10-last-num-seq}{10.00}
\DefMacro{res-codet5-CSNm-eval-runnable-any-stmt/test-bs10-last-compilable}{54.84}
\DefMacro{res-codet5-CSNm-eval-runnable-any-stmt/test-bs10-last-runnable}{17.62}
\DefMacro{res-codet5-CSNm-eval-runnable-assert-stmt/test-bs10-last-xmatch}{8.79}
\DefMacro{res-codet5-CSNm-eval-runnable-assert-stmt/test-bs10-last-xmatch-top3}{14.93}
\DefMacro{res-codet5-CSNm-eval-runnable-assert-stmt/test-bs10-last-xmatch-top5}{18.68}
\DefMacro{res-codet5-CSNm-eval-runnable-assert-stmt/test-bs10-last-xmatch-top10}{24.37}
\DefMacro{res-codet5-CSNm-eval-runnable-assert-stmt/test-bs10-last-xmatch-top100}{24.37}
\DefMacro{res-codet5-CSNm-eval-runnable-assert-stmt/test-bs10-last-bleu}{40.26}
\DefMacro{res-codet5-CSNm-eval-runnable-assert-stmt/test-bs10-last-bleu-max}{58.37}
\DefMacro{res-codet5-CSNm-eval-runnable-assert-stmt/test-bs10-last-bleu-min}{19.02}
\DefMacro{res-codet5-CSNm-eval-runnable-assert-stmt/test-bs10-last-bleu-avg}{34.38}
\DefMacro{res-codet5-CSNm-eval-runnable-assert-stmt/test-bs10-last-code-bleu}{32.23}
\DefMacro{res-codet5-CSNm-eval-runnable-assert-stmt/test-bs10-last-code-bleu-max}{52.85}
\DefMacro{res-codet5-CSNm-eval-runnable-assert-stmt/test-bs10-last-code-bleu-min}{11.64}
\DefMacro{res-codet5-CSNm-eval-runnable-assert-stmt/test-bs10-last-code-bleu-avg}{27.16}
\DefMacro{res-codet5-CSNm-eval-runnable-assert-stmt/test-bs10-last-edit-sim}{67.08}
\DefMacro{res-codet5-CSNm-eval-runnable-assert-stmt/test-bs10-last-edit-sim-max}{79.87}
\DefMacro{res-codet5-CSNm-eval-runnable-assert-stmt/test-bs10-last-edit-sim-min}{49.72}
\DefMacro{res-codet5-CSNm-eval-runnable-assert-stmt/test-bs10-last-edit-sim-avg}{63.85}
\DefMacro{res-codet5-CSNm-eval-runnable-assert-stmt/test-bs10-last-rouge-p}{68.38}
\DefMacro{res-codet5-CSNm-eval-runnable-assert-stmt/test-bs10-last-rouge-p-max}{81.69}
\DefMacro{res-codet5-CSNm-eval-runnable-assert-stmt/test-bs10-last-rouge-p-min}{47.03}
\DefMacro{res-codet5-CSNm-eval-runnable-assert-stmt/test-bs10-last-rouge-p-avg}{62.23}
\DefMacro{res-codet5-CSNm-eval-runnable-assert-stmt/test-bs10-last-rouge-r}{68.55}
\DefMacro{res-codet5-CSNm-eval-runnable-assert-stmt/test-bs10-last-rouge-r-max}{80.61}
\DefMacro{res-codet5-CSNm-eval-runnable-assert-stmt/test-bs10-last-rouge-r-min}{53.64}
\DefMacro{res-codet5-CSNm-eval-runnable-assert-stmt/test-bs10-last-rouge-r-avg}{66.68}
\DefMacro{res-codet5-CSNm-eval-runnable-assert-stmt/test-bs10-last-rouge-f}{67.42}
\DefMacro{res-codet5-CSNm-eval-runnable-assert-stmt/test-bs10-last-rouge-f-max}{79.98}
\DefMacro{res-codet5-CSNm-eval-runnable-assert-stmt/test-bs10-last-rouge-f-min}{49.30}
\DefMacro{res-codet5-CSNm-eval-runnable-assert-stmt/test-bs10-last-rouge-f-avg}{63.27}
\DefMacro{res-codet5-CSNm-eval-runnable-assert-stmt/test-bs10-last-num-seq}{10.00}
\DefMacro{res-codet5-CSNm-eval-runnable-assert-stmt/test-bs10-last-compilable}{44.45}
\DefMacro{res-codet5-CSNm-eval-runnable-assert-stmt/test-bs10-last-runnable}{16.87}
\DefMacro{res-codet5-noft-CSNm-eval-any-stmt/test-bs10-last-xmatch}{0.00}
\DefMacro{res-codet5-noft-CSNm-eval-any-stmt/test-bs10-last-xmatch-top3}{0.00}
\DefMacro{res-codet5-noft-CSNm-eval-any-stmt/test-bs10-last-xmatch-top5}{0.00}
\DefMacro{res-codet5-noft-CSNm-eval-any-stmt/test-bs10-last-xmatch-top10}{0.00}
\DefMacro{res-codet5-noft-CSNm-eval-any-stmt/test-bs10-last-xmatch-top100}{0.00}
\DefMacro{res-codet5-noft-CSNm-eval-any-stmt/test-bs10-last-bleu}{0.00}
\DefMacro{res-codet5-noft-CSNm-eval-any-stmt/test-bs10-last-bleu-max}{1.23}
\DefMacro{res-codet5-noft-CSNm-eval-any-stmt/test-bs10-last-bleu-min}{0.00}
\DefMacro{res-codet5-noft-CSNm-eval-any-stmt/test-bs10-last-bleu-avg}{0.27}
\DefMacro{res-codet5-noft-CSNm-eval-any-stmt/test-bs10-last-code-bleu}{3.36}
\DefMacro{res-codet5-noft-CSNm-eval-any-stmt/test-bs10-last-code-bleu-max}{3.78}
\DefMacro{res-codet5-noft-CSNm-eval-any-stmt/test-bs10-last-code-bleu-min}{3.26}
\DefMacro{res-codet5-noft-CSNm-eval-any-stmt/test-bs10-last-code-bleu-avg}{3.44}
\DefMacro{res-codet5-noft-CSNm-eval-any-stmt/test-bs10-last-edit-sim}{1.68}
\DefMacro{res-codet5-noft-CSNm-eval-any-stmt/test-bs10-last-edit-sim-max}{2.38}
\DefMacro{res-codet5-noft-CSNm-eval-any-stmt/test-bs10-last-edit-sim-min}{1.55}
\DefMacro{res-codet5-noft-CSNm-eval-any-stmt/test-bs10-last-edit-sim-avg}{1.96}
\DefMacro{res-codet5-noft-CSNm-eval-any-stmt/test-bs10-last-rouge-p}{0.08}
\DefMacro{res-codet5-noft-CSNm-eval-any-stmt/test-bs10-last-rouge-p-max}{8.03}
\DefMacro{res-codet5-noft-CSNm-eval-any-stmt/test-bs10-last-rouge-p-min}{0.02}
\DefMacro{res-codet5-noft-CSNm-eval-any-stmt/test-bs10-last-rouge-p-avg}{1.84}
\DefMacro{res-codet5-noft-CSNm-eval-any-stmt/test-bs10-last-rouge-r}{0.01}
\DefMacro{res-codet5-noft-CSNm-eval-any-stmt/test-bs10-last-rouge-r-max}{3.04}
\DefMacro{res-codet5-noft-CSNm-eval-any-stmt/test-bs10-last-rouge-r-min}{0.01}
\DefMacro{res-codet5-noft-CSNm-eval-any-stmt/test-bs10-last-rouge-r-avg}{0.68}
\DefMacro{res-codet5-noft-CSNm-eval-any-stmt/test-bs10-last-rouge-f}{0.02}
\DefMacro{res-codet5-noft-CSNm-eval-any-stmt/test-bs10-last-rouge-f-max}{4.28}
\DefMacro{res-codet5-noft-CSNm-eval-any-stmt/test-bs10-last-rouge-f-min}{0.01}
\DefMacro{res-codet5-noft-CSNm-eval-any-stmt/test-bs10-last-rouge-f-avg}{0.95}
\DefMacro{res-codet5-noft-CSNm-eval-any-stmt/test-bs10-last-num-seq}{10.00}
\DefMacro{res-codet5-noft-CSNm-eval-any-stmt/test-bs10-last-compilable}{-1.00}
\DefMacro{res-codet5-noft-CSNm-eval-any-stmt/test-bs10-last-runnable}{-1.00}
\DefMacro{res-codet5-noft-CSNm-eval-assert-stmt/test-bs10-last-xmatch}{0.00}
\DefMacro{res-codet5-noft-CSNm-eval-assert-stmt/test-bs10-last-xmatch-top3}{0.00}
\DefMacro{res-codet5-noft-CSNm-eval-assert-stmt/test-bs10-last-xmatch-top5}{0.00}
\DefMacro{res-codet5-noft-CSNm-eval-assert-stmt/test-bs10-last-xmatch-top10}{0.00}
\DefMacro{res-codet5-noft-CSNm-eval-assert-stmt/test-bs10-last-xmatch-top100}{0.00}
\DefMacro{res-codet5-noft-CSNm-eval-assert-stmt/test-bs10-last-bleu}{0.00}
\DefMacro{res-codet5-noft-CSNm-eval-assert-stmt/test-bs10-last-bleu-max}{1.01}
\DefMacro{res-codet5-noft-CSNm-eval-assert-stmt/test-bs10-last-bleu-min}{0.00}
\DefMacro{res-codet5-noft-CSNm-eval-assert-stmt/test-bs10-last-bleu-avg}{0.20}
\DefMacro{res-codet5-noft-CSNm-eval-assert-stmt/test-bs10-last-code-bleu}{1.18}
\DefMacro{res-codet5-noft-CSNm-eval-assert-stmt/test-bs10-last-code-bleu-max}{1.43}
\DefMacro{res-codet5-noft-CSNm-eval-assert-stmt/test-bs10-last-code-bleu-min}{1.17}
\DefMacro{res-codet5-noft-CSNm-eval-assert-stmt/test-bs10-last-code-bleu-avg}{1.23}
\DefMacro{res-codet5-noft-CSNm-eval-assert-stmt/test-bs10-last-edit-sim}{1.86}
\DefMacro{res-codet5-noft-CSNm-eval-assert-stmt/test-bs10-last-edit-sim-max}{2.34}
\DefMacro{res-codet5-noft-CSNm-eval-assert-stmt/test-bs10-last-edit-sim-min}{1.77}
\DefMacro{res-codet5-noft-CSNm-eval-assert-stmt/test-bs10-last-edit-sim-avg}{2.04}
\DefMacro{res-codet5-noft-CSNm-eval-assert-stmt/test-bs10-last-rouge-p}{0.02}
\DefMacro{res-codet5-noft-CSNm-eval-assert-stmt/test-bs10-last-rouge-p-max}{6.73}
\DefMacro{res-codet5-noft-CSNm-eval-assert-stmt/test-bs10-last-rouge-p-min}{0.01}
\DefMacro{res-codet5-noft-CSNm-eval-assert-stmt/test-bs10-last-rouge-p-avg}{1.44}
\DefMacro{res-codet5-noft-CSNm-eval-assert-stmt/test-bs10-last-rouge-r}{0.00}
\DefMacro{res-codet5-noft-CSNm-eval-assert-stmt/test-bs10-last-rouge-r-max}{2.58}
\DefMacro{res-codet5-noft-CSNm-eval-assert-stmt/test-bs10-last-rouge-r-min}{0.00}
\DefMacro{res-codet5-noft-CSNm-eval-assert-stmt/test-bs10-last-rouge-r-avg}{0.54}
\DefMacro{res-codet5-noft-CSNm-eval-assert-stmt/test-bs10-last-rouge-f}{0.01}
\DefMacro{res-codet5-noft-CSNm-eval-assert-stmt/test-bs10-last-rouge-f-max}{3.65}
\DefMacro{res-codet5-noft-CSNm-eval-assert-stmt/test-bs10-last-rouge-f-min}{0.00}
\DefMacro{res-codet5-noft-CSNm-eval-assert-stmt/test-bs10-last-rouge-f-avg}{0.77}
\DefMacro{res-codet5-noft-CSNm-eval-assert-stmt/test-bs10-last-num-seq}{10.00}
\DefMacro{res-codet5-noft-CSNm-eval-assert-stmt/test-bs10-last-compilable}{-1.00}
\DefMacro{res-codet5-noft-CSNm-eval-assert-stmt/test-bs10-last-runnable}{-1.00}
\DefMacro{res-codet5-noft-CSNm-eval-runnable-any-stmt/test-bs10-last-xmatch}{0.00}
\DefMacro{res-codet5-noft-CSNm-eval-runnable-any-stmt/test-bs10-last-xmatch-top3}{0.00}
\DefMacro{res-codet5-noft-CSNm-eval-runnable-any-stmt/test-bs10-last-xmatch-top5}{0.00}
\DefMacro{res-codet5-noft-CSNm-eval-runnable-any-stmt/test-bs10-last-xmatch-top10}{0.00}
\DefMacro{res-codet5-noft-CSNm-eval-runnable-any-stmt/test-bs10-last-xmatch-top100}{0.00}
\DefMacro{res-codet5-noft-CSNm-eval-runnable-any-stmt/test-bs10-last-bleu}{0.00}
\DefMacro{res-codet5-noft-CSNm-eval-runnable-any-stmt/test-bs10-last-bleu-max}{1.25}
\DefMacro{res-codet5-noft-CSNm-eval-runnable-any-stmt/test-bs10-last-bleu-min}{0.00}
\DefMacro{res-codet5-noft-CSNm-eval-runnable-any-stmt/test-bs10-last-bleu-avg}{0.27}
\DefMacro{res-codet5-noft-CSNm-eval-runnable-any-stmt/test-bs10-last-code-bleu}{3.35}
\DefMacro{res-codet5-noft-CSNm-eval-runnable-any-stmt/test-bs10-last-code-bleu-max}{3.77}
\DefMacro{res-codet5-noft-CSNm-eval-runnable-any-stmt/test-bs10-last-code-bleu-min}{3.26}
\DefMacro{res-codet5-noft-CSNm-eval-runnable-any-stmt/test-bs10-last-code-bleu-avg}{3.43}
\DefMacro{res-codet5-noft-CSNm-eval-runnable-any-stmt/test-bs10-last-edit-sim}{1.65}
\DefMacro{res-codet5-noft-CSNm-eval-runnable-any-stmt/test-bs10-last-edit-sim-max}{2.33}
\DefMacro{res-codet5-noft-CSNm-eval-runnable-any-stmt/test-bs10-last-edit-sim-min}{1.51}
\DefMacro{res-codet5-noft-CSNm-eval-runnable-any-stmt/test-bs10-last-edit-sim-avg}{1.92}
\DefMacro{res-codet5-noft-CSNm-eval-runnable-any-stmt/test-bs10-last-rouge-p}{0.09}
\DefMacro{res-codet5-noft-CSNm-eval-runnable-any-stmt/test-bs10-last-rouge-p-max}{8.04}
\DefMacro{res-codet5-noft-CSNm-eval-runnable-any-stmt/test-bs10-last-rouge-p-min}{0.02}
\DefMacro{res-codet5-noft-CSNm-eval-runnable-any-stmt/test-bs10-last-rouge-p-avg}{1.87}
\DefMacro{res-codet5-noft-CSNm-eval-runnable-any-stmt/test-bs10-last-rouge-r}{0.02}
\DefMacro{res-codet5-noft-CSNm-eval-runnable-any-stmt/test-bs10-last-rouge-r-max}{3.07}
\DefMacro{res-codet5-noft-CSNm-eval-runnable-any-stmt/test-bs10-last-rouge-r-min}{0.01}
\DefMacro{res-codet5-noft-CSNm-eval-runnable-any-stmt/test-bs10-last-rouge-r-avg}{0.69}
\DefMacro{res-codet5-noft-CSNm-eval-runnable-any-stmt/test-bs10-last-rouge-f}{0.03}
\DefMacro{res-codet5-noft-CSNm-eval-runnable-any-stmt/test-bs10-last-rouge-f-max}{4.30}
\DefMacro{res-codet5-noft-CSNm-eval-runnable-any-stmt/test-bs10-last-rouge-f-min}{0.01}
\DefMacro{res-codet5-noft-CSNm-eval-runnable-any-stmt/test-bs10-last-rouge-f-avg}{0.97}
\DefMacro{res-codet5-noft-CSNm-eval-runnable-any-stmt/test-bs10-last-num-seq}{10.00}
\DefMacro{res-codet5-noft-CSNm-eval-runnable-any-stmt/test-bs10-last-compilable}{0.00}
\DefMacro{res-codet5-noft-CSNm-eval-runnable-any-stmt/test-bs10-last-runnable}{0.00}
\DefMacro{res-codet5-noft-CSNm-eval-runnable-assert-stmt/test-bs10-last-xmatch}{0.00}
\DefMacro{res-codet5-noft-CSNm-eval-runnable-assert-stmt/test-bs10-last-xmatch-top3}{0.00}
\DefMacro{res-codet5-noft-CSNm-eval-runnable-assert-stmt/test-bs10-last-xmatch-top5}{0.00}
\DefMacro{res-codet5-noft-CSNm-eval-runnable-assert-stmt/test-bs10-last-xmatch-top10}{0.00}
\DefMacro{res-codet5-noft-CSNm-eval-runnable-assert-stmt/test-bs10-last-xmatch-top100}{0.00}
\DefMacro{res-codet5-noft-CSNm-eval-runnable-assert-stmt/test-bs10-last-bleu}{0.00}
\DefMacro{res-codet5-noft-CSNm-eval-runnable-assert-stmt/test-bs10-last-bleu-max}{0.98}
\DefMacro{res-codet5-noft-CSNm-eval-runnable-assert-stmt/test-bs10-last-bleu-min}{0.00}
\DefMacro{res-codet5-noft-CSNm-eval-runnable-assert-stmt/test-bs10-last-bleu-avg}{0.19}
\DefMacro{res-codet5-noft-CSNm-eval-runnable-assert-stmt/test-bs10-last-code-bleu}{1.26}
\DefMacro{res-codet5-noft-CSNm-eval-runnable-assert-stmt/test-bs10-last-code-bleu-max}{1.50}
\DefMacro{res-codet5-noft-CSNm-eval-runnable-assert-stmt/test-bs10-last-code-bleu-min}{1.25}
\DefMacro{res-codet5-noft-CSNm-eval-runnable-assert-stmt/test-bs10-last-code-bleu-avg}{1.31}
\DefMacro{res-codet5-noft-CSNm-eval-runnable-assert-stmt/test-bs10-last-edit-sim}{1.87}
\DefMacro{res-codet5-noft-CSNm-eval-runnable-assert-stmt/test-bs10-last-edit-sim-max}{2.31}
\DefMacro{res-codet5-noft-CSNm-eval-runnable-assert-stmt/test-bs10-last-edit-sim-min}{1.77}
\DefMacro{res-codet5-noft-CSNm-eval-runnable-assert-stmt/test-bs10-last-edit-sim-avg}{2.02}
\DefMacro{res-codet5-noft-CSNm-eval-runnable-assert-stmt/test-bs10-last-rouge-p}{0.02}
\DefMacro{res-codet5-noft-CSNm-eval-runnable-assert-stmt/test-bs10-last-rouge-p-max}{6.50}
\DefMacro{res-codet5-noft-CSNm-eval-runnable-assert-stmt/test-bs10-last-rouge-p-min}{0.01}
\DefMacro{res-codet5-noft-CSNm-eval-runnable-assert-stmt/test-bs10-last-rouge-p-avg}{1.41}
\DefMacro{res-codet5-noft-CSNm-eval-runnable-assert-stmt/test-bs10-last-rouge-r}{0.01}
\DefMacro{res-codet5-noft-CSNm-eval-runnable-assert-stmt/test-bs10-last-rouge-r-max}{2.49}
\DefMacro{res-codet5-noft-CSNm-eval-runnable-assert-stmt/test-bs10-last-rouge-r-min}{0.00}
\DefMacro{res-codet5-noft-CSNm-eval-runnable-assert-stmt/test-bs10-last-rouge-r-avg}{0.53}
\DefMacro{res-codet5-noft-CSNm-eval-runnable-assert-stmt/test-bs10-last-rouge-f}{0.01}
\DefMacro{res-codet5-noft-CSNm-eval-runnable-assert-stmt/test-bs10-last-rouge-f-max}{3.52}
\DefMacro{res-codet5-noft-CSNm-eval-runnable-assert-stmt/test-bs10-last-rouge-f-min}{0.01}
\DefMacro{res-codet5-noft-CSNm-eval-runnable-assert-stmt/test-bs10-last-rouge-f-avg}{0.74}
\DefMacro{res-codet5-noft-CSNm-eval-runnable-assert-stmt/test-bs10-last-num-seq}{10.00}
\DefMacro{res-codet5-noft-CSNm-eval-runnable-assert-stmt/test-bs10-last-compilable}{0.00}
\DefMacro{res-codet5-noft-CSNm-eval-runnable-assert-stmt/test-bs10-last-runnable}{0.00}
\DefMacro{res-codex-CSNm-eval-any-stmt/test-bs10-last-xmatch}{12.69}
\DefMacro{res-codex-CSNm-eval-any-stmt/test-bs10-last-xmatch-top3}{12.69}
\DefMacro{res-codex-CSNm-eval-any-stmt/test-bs10-last-xmatch-top5}{12.69}
\DefMacro{res-codex-CSNm-eval-any-stmt/test-bs10-last-xmatch-top10}{12.69}
\DefMacro{res-codex-CSNm-eval-any-stmt/test-bs10-last-xmatch-top100}{12.69}
\DefMacro{res-codex-CSNm-eval-any-stmt/test-bs10-last-bleu}{34.53}
\DefMacro{res-codex-CSNm-eval-any-stmt/test-bs10-last-bleu-max}{34.53}
\DefMacro{res-codex-CSNm-eval-any-stmt/test-bs10-last-bleu-min}{34.53}
\DefMacro{res-codex-CSNm-eval-any-stmt/test-bs10-last-bleu-avg}{34.53}
\DefMacro{res-codex-CSNm-eval-any-stmt/test-bs10-last-code-bleu}{29.91}
\DefMacro{res-codex-CSNm-eval-any-stmt/test-bs10-last-code-bleu-max}{29.91}
\DefMacro{res-codex-CSNm-eval-any-stmt/test-bs10-last-code-bleu-min}{29.91}
\DefMacro{res-codex-CSNm-eval-any-stmt/test-bs10-last-code-bleu-avg}{29.91}
\DefMacro{res-codex-CSNm-eval-any-stmt/test-bs10-last-edit-sim}{58.08}
\DefMacro{res-codex-CSNm-eval-any-stmt/test-bs10-last-edit-sim-max}{58.08}
\DefMacro{res-codex-CSNm-eval-any-stmt/test-bs10-last-edit-sim-min}{58.08}
\DefMacro{res-codex-CSNm-eval-any-stmt/test-bs10-last-edit-sim-avg}{58.08}
\DefMacro{res-codex-CSNm-eval-any-stmt/test-bs10-last-rouge-p}{55.83}
\DefMacro{res-codex-CSNm-eval-any-stmt/test-bs10-last-rouge-p-max}{55.83}
\DefMacro{res-codex-CSNm-eval-any-stmt/test-bs10-last-rouge-p-min}{55.83}
\DefMacro{res-codex-CSNm-eval-any-stmt/test-bs10-last-rouge-p-avg}{55.83}
\DefMacro{res-codex-CSNm-eval-any-stmt/test-bs10-last-rouge-r}{59.92}
\DefMacro{res-codex-CSNm-eval-any-stmt/test-bs10-last-rouge-r-max}{59.92}
\DefMacro{res-codex-CSNm-eval-any-stmt/test-bs10-last-rouge-r-min}{59.92}
\DefMacro{res-codex-CSNm-eval-any-stmt/test-bs10-last-rouge-r-avg}{59.92}
\DefMacro{res-codex-CSNm-eval-any-stmt/test-bs10-last-rouge-f}{56.04}
\DefMacro{res-codex-CSNm-eval-any-stmt/test-bs10-last-rouge-f-max}{56.04}
\DefMacro{res-codex-CSNm-eval-any-stmt/test-bs10-last-rouge-f-min}{56.04}
\DefMacro{res-codex-CSNm-eval-any-stmt/test-bs10-last-rouge-f-avg}{56.04}
\DefMacro{res-codex-CSNm-eval-any-stmt/test-bs10-last-compilable}{-1.00}
\DefMacro{res-codex-CSNm-eval-any-stmt/test-bs10-last-runnable}{-1.00}
\DefMacro{res-codex-CSNm-eval-assert-stmt/test-bs10-last-xmatch}{12.30}
\DefMacro{res-codex-CSNm-eval-assert-stmt/test-bs10-last-xmatch-top3}{12.30}
\DefMacro{res-codex-CSNm-eval-assert-stmt/test-bs10-last-xmatch-top5}{12.30}
\DefMacro{res-codex-CSNm-eval-assert-stmt/test-bs10-last-xmatch-top10}{12.30}
\DefMacro{res-codex-CSNm-eval-assert-stmt/test-bs10-last-xmatch-top100}{12.30}
\DefMacro{res-codex-CSNm-eval-assert-stmt/test-bs10-last-bleu}{35.09}
\DefMacro{res-codex-CSNm-eval-assert-stmt/test-bs10-last-bleu-max}{35.09}
\DefMacro{res-codex-CSNm-eval-assert-stmt/test-bs10-last-bleu-min}{35.09}
\DefMacro{res-codex-CSNm-eval-assert-stmt/test-bs10-last-bleu-avg}{35.09}
\DefMacro{res-codex-CSNm-eval-assert-stmt/test-bs10-last-code-bleu}{30.24}
\DefMacro{res-codex-CSNm-eval-assert-stmt/test-bs10-last-code-bleu-max}{30.24}
\DefMacro{res-codex-CSNm-eval-assert-stmt/test-bs10-last-code-bleu-min}{30.24}
\DefMacro{res-codex-CSNm-eval-assert-stmt/test-bs10-last-code-bleu-avg}{30.24}
\DefMacro{res-codex-CSNm-eval-assert-stmt/test-bs10-last-edit-sim}{59.59}
\DefMacro{res-codex-CSNm-eval-assert-stmt/test-bs10-last-edit-sim-max}{59.59}
\DefMacro{res-codex-CSNm-eval-assert-stmt/test-bs10-last-edit-sim-min}{59.59}
\DefMacro{res-codex-CSNm-eval-assert-stmt/test-bs10-last-edit-sim-avg}{59.59}
\DefMacro{res-codex-CSNm-eval-assert-stmt/test-bs10-last-rouge-p}{57.57}
\DefMacro{res-codex-CSNm-eval-assert-stmt/test-bs10-last-rouge-p-max}{57.57}
\DefMacro{res-codex-CSNm-eval-assert-stmt/test-bs10-last-rouge-p-min}{57.57}
\DefMacro{res-codex-CSNm-eval-assert-stmt/test-bs10-last-rouge-p-avg}{57.57}
\DefMacro{res-codex-CSNm-eval-assert-stmt/test-bs10-last-rouge-r}{61.50}
\DefMacro{res-codex-CSNm-eval-assert-stmt/test-bs10-last-rouge-r-max}{61.50}
\DefMacro{res-codex-CSNm-eval-assert-stmt/test-bs10-last-rouge-r-min}{61.50}
\DefMacro{res-codex-CSNm-eval-assert-stmt/test-bs10-last-rouge-r-avg}{61.50}
\DefMacro{res-codex-CSNm-eval-assert-stmt/test-bs10-last-rouge-f}{57.67}
\DefMacro{res-codex-CSNm-eval-assert-stmt/test-bs10-last-rouge-f-max}{57.67}
\DefMacro{res-codex-CSNm-eval-assert-stmt/test-bs10-last-rouge-f-min}{57.67}
\DefMacro{res-codex-CSNm-eval-assert-stmt/test-bs10-last-rouge-f-avg}{57.67}
\DefMacro{res-codex-CSNm-eval-assert-stmt/test-bs10-last-compilable}{-1.00}
\DefMacro{res-codex-CSNm-eval-assert-stmt/test-bs10-last-runnable}{-1.00}
\DefMacro{res-codex-CSNm-eval-runnable-any-stmt/test-bs10-last-xmatch}{12.88}
\DefMacro{res-codex-CSNm-eval-runnable-any-stmt/test-bs10-last-xmatch-top3}{12.88}
\DefMacro{res-codex-CSNm-eval-runnable-any-stmt/test-bs10-last-xmatch-top5}{12.88}
\DefMacro{res-codex-CSNm-eval-runnable-any-stmt/test-bs10-last-xmatch-top10}{12.88}
\DefMacro{res-codex-CSNm-eval-runnable-any-stmt/test-bs10-last-xmatch-top100}{12.88}
\DefMacro{res-codex-CSNm-eval-runnable-any-stmt/test-bs10-last-bleu}{34.89}
\DefMacro{res-codex-CSNm-eval-runnable-any-stmt/test-bs10-last-bleu-max}{34.89}
\DefMacro{res-codex-CSNm-eval-runnable-any-stmt/test-bs10-last-bleu-min}{34.89}
\DefMacro{res-codex-CSNm-eval-runnable-any-stmt/test-bs10-last-bleu-avg}{34.89}
\DefMacro{res-codex-CSNm-eval-runnable-any-stmt/test-bs10-last-code-bleu}{30.12}
\DefMacro{res-codex-CSNm-eval-runnable-any-stmt/test-bs10-last-code-bleu-max}{30.12}
\DefMacro{res-codex-CSNm-eval-runnable-any-stmt/test-bs10-last-code-bleu-min}{30.12}
\DefMacro{res-codex-CSNm-eval-runnable-any-stmt/test-bs10-last-code-bleu-avg}{30.12}
\DefMacro{res-codex-CSNm-eval-runnable-any-stmt/test-bs10-last-edit-sim}{58.44}
\DefMacro{res-codex-CSNm-eval-runnable-any-stmt/test-bs10-last-edit-sim-max}{58.44}
\DefMacro{res-codex-CSNm-eval-runnable-any-stmt/test-bs10-last-edit-sim-min}{58.44}
\DefMacro{res-codex-CSNm-eval-runnable-any-stmt/test-bs10-last-edit-sim-avg}{58.44}
\DefMacro{res-codex-CSNm-eval-runnable-any-stmt/test-bs10-last-rouge-p}{56.35}
\DefMacro{res-codex-CSNm-eval-runnable-any-stmt/test-bs10-last-rouge-p-max}{56.35}
\DefMacro{res-codex-CSNm-eval-runnable-any-stmt/test-bs10-last-rouge-p-min}{56.35}
\DefMacro{res-codex-CSNm-eval-runnable-any-stmt/test-bs10-last-rouge-p-avg}{56.35}
\DefMacro{res-codex-CSNm-eval-runnable-any-stmt/test-bs10-last-rouge-r}{60.38}
\DefMacro{res-codex-CSNm-eval-runnable-any-stmt/test-bs10-last-rouge-r-max}{60.38}
\DefMacro{res-codex-CSNm-eval-runnable-any-stmt/test-bs10-last-rouge-r-min}{60.38}
\DefMacro{res-codex-CSNm-eval-runnable-any-stmt/test-bs10-last-rouge-r-avg}{60.38}
\DefMacro{res-codex-CSNm-eval-runnable-any-stmt/test-bs10-last-rouge-f}{56.58}
\DefMacro{res-codex-CSNm-eval-runnable-any-stmt/test-bs10-last-rouge-f-max}{56.58}
\DefMacro{res-codex-CSNm-eval-runnable-any-stmt/test-bs10-last-rouge-f-min}{56.58}
\DefMacro{res-codex-CSNm-eval-runnable-any-stmt/test-bs10-last-rouge-f-avg}{56.58}
\DefMacro{res-codex-CSNm-eval-runnable-any-stmt/test-bs10-last-compilable}{38.80}
\DefMacro{res-codex-CSNm-eval-runnable-any-stmt/test-bs10-last-runnable}{19.12}
\DefMacro{res-codex-CSNm-eval-runnable-assert-stmt/test-bs10-last-xmatch}{12.15}
\DefMacro{res-codex-CSNm-eval-runnable-assert-stmt/test-bs10-last-xmatch-top3}{12.15}
\DefMacro{res-codex-CSNm-eval-runnable-assert-stmt/test-bs10-last-xmatch-top5}{12.15}
\DefMacro{res-codex-CSNm-eval-runnable-assert-stmt/test-bs10-last-xmatch-top10}{12.15}
\DefMacro{res-codex-CSNm-eval-runnable-assert-stmt/test-bs10-last-xmatch-top100}{12.15}
\DefMacro{res-codex-CSNm-eval-runnable-assert-stmt/test-bs10-last-bleu}{35.05}
\DefMacro{res-codex-CSNm-eval-runnable-assert-stmt/test-bs10-last-bleu-max}{35.05}
\DefMacro{res-codex-CSNm-eval-runnable-assert-stmt/test-bs10-last-bleu-min}{35.05}
\DefMacro{res-codex-CSNm-eval-runnable-assert-stmt/test-bs10-last-bleu-avg}{35.05}
\DefMacro{res-codex-CSNm-eval-runnable-assert-stmt/test-bs10-last-code-bleu}{30.12}
\DefMacro{res-codex-CSNm-eval-runnable-assert-stmt/test-bs10-last-code-bleu-max}{30.12}
\DefMacro{res-codex-CSNm-eval-runnable-assert-stmt/test-bs10-last-code-bleu-min}{30.12}
\DefMacro{res-codex-CSNm-eval-runnable-assert-stmt/test-bs10-last-code-bleu-avg}{30.12}
\DefMacro{res-codex-CSNm-eval-runnable-assert-stmt/test-bs10-last-edit-sim}{59.79}
\DefMacro{res-codex-CSNm-eval-runnable-assert-stmt/test-bs10-last-edit-sim-max}{59.79}
\DefMacro{res-codex-CSNm-eval-runnable-assert-stmt/test-bs10-last-edit-sim-min}{59.79}
\DefMacro{res-codex-CSNm-eval-runnable-assert-stmt/test-bs10-last-edit-sim-avg}{59.79}
\DefMacro{res-codex-CSNm-eval-runnable-assert-stmt/test-bs10-last-rouge-p}{57.69}
\DefMacro{res-codex-CSNm-eval-runnable-assert-stmt/test-bs10-last-rouge-p-max}{57.69}
\DefMacro{res-codex-CSNm-eval-runnable-assert-stmt/test-bs10-last-rouge-p-min}{57.69}
\DefMacro{res-codex-CSNm-eval-runnable-assert-stmt/test-bs10-last-rouge-p-avg}{57.69}
\DefMacro{res-codex-CSNm-eval-runnable-assert-stmt/test-bs10-last-rouge-r}{61.71}
\DefMacro{res-codex-CSNm-eval-runnable-assert-stmt/test-bs10-last-rouge-r-max}{61.71}
\DefMacro{res-codex-CSNm-eval-runnable-assert-stmt/test-bs10-last-rouge-r-min}{61.71}
\DefMacro{res-codex-CSNm-eval-runnable-assert-stmt/test-bs10-last-rouge-r-avg}{61.71}
\DefMacro{res-codex-CSNm-eval-runnable-assert-stmt/test-bs10-last-rouge-f}{57.86}
\DefMacro{res-codex-CSNm-eval-runnable-assert-stmt/test-bs10-last-rouge-f-max}{57.86}
\DefMacro{res-codex-CSNm-eval-runnable-assert-stmt/test-bs10-last-rouge-f-min}{57.86}
\DefMacro{res-codex-CSNm-eval-runnable-assert-stmt/test-bs10-last-rouge-f-avg}{57.86}
\DefMacro{res-codex-CSNm-eval-runnable-assert-stmt/test-bs10-last-compilable}{39.46}
\DefMacro{res-codex-CSNm-eval-runnable-assert-stmt/test-bs10-last-runnable}{20.73}
\DefMacro{res-teco-icse23-CSNm-eval-any-stmt/test-bs10-last-xmatch}{17.61}
\DefMacro{res-teco-icse23-CSNm-eval-any-stmt/test-bs10-last-xmatch-top3}{22.59}
\DefMacro{res-teco-icse23-CSNm-eval-any-stmt/test-bs10-last-xmatch-top5}{24.47}
\DefMacro{res-teco-icse23-CSNm-eval-any-stmt/test-bs10-last-xmatch-top10}{27.20}
\DefMacro{res-teco-icse23-CSNm-eval-any-stmt/test-bs10-last-xmatch-top100}{27.20}
\DefMacro{res-teco-icse23-CSNm-eval-any-stmt/test-bs10-last-bleu}{42.01}
\DefMacro{res-teco-icse23-CSNm-eval-any-stmt/test-bs10-last-bleu-max}{55.53}
\DefMacro{res-teco-icse23-CSNm-eval-any-stmt/test-bs10-last-bleu-min}{18.20}
\DefMacro{res-teco-icse23-CSNm-eval-any-stmt/test-bs10-last-bleu-avg}{31.90}
\DefMacro{res-teco-icse23-CSNm-eval-any-stmt/test-bs10-last-code-bleu}{37.61}
\DefMacro{res-teco-icse23-CSNm-eval-any-stmt/test-bs10-last-code-bleu-max}{53.45}
\DefMacro{res-teco-icse23-CSNm-eval-any-stmt/test-bs10-last-code-bleu-min}{12.14}
\DefMacro{res-teco-icse23-CSNm-eval-any-stmt/test-bs10-last-code-bleu-avg}{27.53}
\DefMacro{res-teco-icse23-CSNm-eval-any-stmt/test-bs10-last-edit-sim}{63.49}
\DefMacro{res-teco-icse23-CSNm-eval-any-stmt/test-bs10-last-edit-sim-max}{74.54}
\DefMacro{res-teco-icse23-CSNm-eval-any-stmt/test-bs10-last-edit-sim-min}{44.79}
\DefMacro{res-teco-icse23-CSNm-eval-any-stmt/test-bs10-last-edit-sim-avg}{57.82}
\DefMacro{res-teco-icse23-CSNm-eval-any-stmt/test-bs10-last-rouge-p}{66.70}
\DefMacro{res-teco-icse23-CSNm-eval-any-stmt/test-bs10-last-rouge-p-max}{78.15}
\DefMacro{res-teco-icse23-CSNm-eval-any-stmt/test-bs10-last-rouge-p-min}{44.70}
\DefMacro{res-teco-icse23-CSNm-eval-any-stmt/test-bs10-last-rouge-p-avg}{58.53}
\DefMacro{res-teco-icse23-CSNm-eval-any-stmt/test-bs10-last-rouge-r}{66.39}
\DefMacro{res-teco-icse23-CSNm-eval-any-stmt/test-bs10-last-rouge-r-max}{77.08}
\DefMacro{res-teco-icse23-CSNm-eval-any-stmt/test-bs10-last-rouge-r-min}{50.62}
\DefMacro{res-teco-icse23-CSNm-eval-any-stmt/test-bs10-last-rouge-r-avg}{63.32}
\DefMacro{res-teco-icse23-CSNm-eval-any-stmt/test-bs10-last-rouge-f}{65.23}
\DefMacro{res-teco-icse23-CSNm-eval-any-stmt/test-bs10-last-rouge-f-max}{76.06}
\DefMacro{res-teco-icse23-CSNm-eval-any-stmt/test-bs10-last-rouge-f-min}{46.45}
\DefMacro{res-teco-icse23-CSNm-eval-any-stmt/test-bs10-last-rouge-f-avg}{59.31}
\DefMacro{res-teco-icse23-CSNm-eval-any-stmt/test-bs10-last-num-seq}{10.00}
\DefMacro{res-teco-icse23-CSNm-eval-any-stmt/test-bs10-last-compilable}{-1.00}
\DefMacro{res-teco-icse23-CSNm-eval-any-stmt/test-bs10-last-runnable}{-1.00}
\DefMacro{res-teco-icse23-CSNm-eval-assert-stmt/test-bs10-last-xmatch}{16.44}
\DefMacro{res-teco-icse23-CSNm-eval-assert-stmt/test-bs10-last-xmatch-top3}{21.01}
\DefMacro{res-teco-icse23-CSNm-eval-assert-stmt/test-bs10-last-xmatch-top5}{23.24}
\DefMacro{res-teco-icse23-CSNm-eval-assert-stmt/test-bs10-last-xmatch-top10}{27.41}
\DefMacro{res-teco-icse23-CSNm-eval-assert-stmt/test-bs10-last-xmatch-top100}{27.41}
\DefMacro{res-teco-icse23-CSNm-eval-assert-stmt/test-bs10-last-bleu}{43.09}
\DefMacro{res-teco-icse23-CSNm-eval-assert-stmt/test-bs10-last-bleu-max}{59.87}
\DefMacro{res-teco-icse23-CSNm-eval-assert-stmt/test-bs10-last-bleu-min}{18.86}
\DefMacro{res-teco-icse23-CSNm-eval-assert-stmt/test-bs10-last-bleu-avg}{34.35}
\DefMacro{res-teco-icse23-CSNm-eval-assert-stmt/test-bs10-last-code-bleu}{35.88}
\DefMacro{res-teco-icse23-CSNm-eval-assert-stmt/test-bs10-last-code-bleu-max}{54.23}
\DefMacro{res-teco-icse23-CSNm-eval-assert-stmt/test-bs10-last-code-bleu-min}{11.69}
\DefMacro{res-teco-icse23-CSNm-eval-assert-stmt/test-bs10-last-code-bleu-avg}{27.13}
\DefMacro{res-teco-icse23-CSNm-eval-assert-stmt/test-bs10-last-edit-sim}{68.05}
\DefMacro{res-teco-icse23-CSNm-eval-assert-stmt/test-bs10-last-edit-sim-max}{80.64}
\DefMacro{res-teco-icse23-CSNm-eval-assert-stmt/test-bs10-last-edit-sim-min}{49.42}
\DefMacro{res-teco-icse23-CSNm-eval-assert-stmt/test-bs10-last-edit-sim-avg}{63.70}
\DefMacro{res-teco-icse23-CSNm-eval-assert-stmt/test-bs10-last-rouge-p}{69.84}
\DefMacro{res-teco-icse23-CSNm-eval-assert-stmt/test-bs10-last-rouge-p-max}{82.82}
\DefMacro{res-teco-icse23-CSNm-eval-assert-stmt/test-bs10-last-rouge-p-min}{47.08}
\DefMacro{res-teco-icse23-CSNm-eval-assert-stmt/test-bs10-last-rouge-p-avg}{62.21}
\DefMacro{res-teco-icse23-CSNm-eval-assert-stmt/test-bs10-last-rouge-r}{69.64}
\DefMacro{res-teco-icse23-CSNm-eval-assert-stmt/test-bs10-last-rouge-r-max}{81.53}
\DefMacro{res-teco-icse23-CSNm-eval-assert-stmt/test-bs10-last-rouge-r-min}{53.95}
\DefMacro{res-teco-icse23-CSNm-eval-assert-stmt/test-bs10-last-rouge-r-avg}{66.95}
\DefMacro{res-teco-icse23-CSNm-eval-assert-stmt/test-bs10-last-rouge-f}{68.71}
\DefMacro{res-teco-icse23-CSNm-eval-assert-stmt/test-bs10-last-rouge-f-max}{80.93}
\DefMacro{res-teco-icse23-CSNm-eval-assert-stmt/test-bs10-last-rouge-f-min}{49.51}
\DefMacro{res-teco-icse23-CSNm-eval-assert-stmt/test-bs10-last-rouge-f-avg}{63.36}
\DefMacro{res-teco-icse23-CSNm-eval-assert-stmt/test-bs10-last-num-seq}{10.00}
\DefMacro{res-teco-icse23-CSNm-eval-assert-stmt/test-bs10-last-compilable}{-1.00}
\DefMacro{res-teco-icse23-CSNm-eval-assert-stmt/test-bs10-last-runnable}{-1.00}
\DefMacro{res-teco-icse23-CSNm-eval-runnable-any-stmt/test-bs10-last-xmatch}{18.96}
\DefMacro{res-teco-icse23-CSNm-eval-runnable-any-stmt/test-bs10-last-xmatch-top3}{24.01}
\DefMacro{res-teco-icse23-CSNm-eval-runnable-any-stmt/test-bs10-last-xmatch-top5}{25.79}
\DefMacro{res-teco-icse23-CSNm-eval-runnable-any-stmt/test-bs10-last-xmatch-top10}{28.40}
\DefMacro{res-teco-icse23-CSNm-eval-runnable-any-stmt/test-bs10-last-xmatch-top100}{28.40}
\DefMacro{res-teco-icse23-CSNm-eval-runnable-any-stmt/test-bs10-last-bleu}{43.15}
\DefMacro{res-teco-icse23-CSNm-eval-runnable-any-stmt/test-bs10-last-bleu-max}{56.70}
\DefMacro{res-teco-icse23-CSNm-eval-runnable-any-stmt/test-bs10-last-bleu-min}{18.31}
\DefMacro{res-teco-icse23-CSNm-eval-runnable-any-stmt/test-bs10-last-bleu-avg}{32.38}
\DefMacro{res-teco-icse23-CSNm-eval-runnable-any-stmt/test-bs10-last-code-bleu}{38.45}
\DefMacro{res-teco-icse23-CSNm-eval-runnable-any-stmt/test-bs10-last-code-bleu-max}{54.36}
\DefMacro{res-teco-icse23-CSNm-eval-runnable-any-stmt/test-bs10-last-code-bleu-min}{11.90}
\DefMacro{res-teco-icse23-CSNm-eval-runnable-any-stmt/test-bs10-last-code-bleu-avg}{27.72}
\DefMacro{res-teco-icse23-CSNm-eval-runnable-any-stmt/test-bs10-last-edit-sim}{64.12}
\DefMacro{res-teco-icse23-CSNm-eval-runnable-any-stmt/test-bs10-last-edit-sim-max}{75.32}
\DefMacro{res-teco-icse23-CSNm-eval-runnable-any-stmt/test-bs10-last-edit-sim-min}{44.75}
\DefMacro{res-teco-icse23-CSNm-eval-runnable-any-stmt/test-bs10-last-edit-sim-avg}{58.13}
\DefMacro{res-teco-icse23-CSNm-eval-runnable-any-stmt/test-bs10-last-rouge-p}{67.64}
\DefMacro{res-teco-icse23-CSNm-eval-runnable-any-stmt/test-bs10-last-rouge-p-max}{78.97}
\DefMacro{res-teco-icse23-CSNm-eval-runnable-any-stmt/test-bs10-last-rouge-p-min}{44.86}
\DefMacro{res-teco-icse23-CSNm-eval-runnable-any-stmt/test-bs10-last-rouge-p-avg}{59.03}
\DefMacro{res-teco-icse23-CSNm-eval-runnable-any-stmt/test-bs10-last-rouge-r}{67.06}
\DefMacro{res-teco-icse23-CSNm-eval-runnable-any-stmt/test-bs10-last-rouge-r-max}{77.88}
\DefMacro{res-teco-icse23-CSNm-eval-runnable-any-stmt/test-bs10-last-rouge-r-min}{50.75}
\DefMacro{res-teco-icse23-CSNm-eval-runnable-any-stmt/test-bs10-last-rouge-r-avg}{63.86}
\DefMacro{res-teco-icse23-CSNm-eval-runnable-any-stmt/test-bs10-last-rouge-f}{66.09}
\DefMacro{res-teco-icse23-CSNm-eval-runnable-any-stmt/test-bs10-last-rouge-f-max}{76.96}
\DefMacro{res-teco-icse23-CSNm-eval-runnable-any-stmt/test-bs10-last-rouge-f-min}{46.65}
\DefMacro{res-teco-icse23-CSNm-eval-runnable-any-stmt/test-bs10-last-rouge-f-avg}{59.86}
\DefMacro{res-teco-icse23-CSNm-eval-runnable-any-stmt/test-bs10-last-num-seq}{10.00}
\DefMacro{res-teco-icse23-CSNm-eval-runnable-any-stmt/test-bs10-last-compilable}{76.22}
\DefMacro{res-teco-icse23-CSNm-eval-runnable-any-stmt/test-bs10-last-runnable}{28.63}
\DefMacro{res-teco-icse23-CSNm-eval-runnable-assert-stmt/test-bs10-last-xmatch}{17.37}
\DefMacro{res-teco-icse23-CSNm-eval-runnable-assert-stmt/test-bs10-last-xmatch-top3}{21.63}
\DefMacro{res-teco-icse23-CSNm-eval-runnable-assert-stmt/test-bs10-last-xmatch-top5}{23.43}
\DefMacro{res-teco-icse23-CSNm-eval-runnable-assert-stmt/test-bs10-last-xmatch-top10}{27.39}
\DefMacro{res-teco-icse23-CSNm-eval-runnable-assert-stmt/test-bs10-last-xmatch-top100}{27.39}
\DefMacro{res-teco-icse23-CSNm-eval-runnable-assert-stmt/test-bs10-last-bleu}{44.27}
\DefMacro{res-teco-icse23-CSNm-eval-runnable-assert-stmt/test-bs10-last-bleu-max}{60.21}
\DefMacro{res-teco-icse23-CSNm-eval-runnable-assert-stmt/test-bs10-last-bleu-min}{19.30}
\DefMacro{res-teco-icse23-CSNm-eval-runnable-assert-stmt/test-bs10-last-bleu-avg}{34.85}
\DefMacro{res-teco-icse23-CSNm-eval-runnable-assert-stmt/test-bs10-last-code-bleu}{36.98}
\DefMacro{res-teco-icse23-CSNm-eval-runnable-assert-stmt/test-bs10-last-code-bleu-max}{54.37}
\DefMacro{res-teco-icse23-CSNm-eval-runnable-assert-stmt/test-bs10-last-code-bleu-min}{12.15}
\DefMacro{res-teco-icse23-CSNm-eval-runnable-assert-stmt/test-bs10-last-code-bleu-avg}{27.62}
\DefMacro{res-teco-icse23-CSNm-eval-runnable-assert-stmt/test-bs10-last-edit-sim}{68.43}
\DefMacro{res-teco-icse23-CSNm-eval-runnable-assert-stmt/test-bs10-last-edit-sim-max}{80.69}
\DefMacro{res-teco-icse23-CSNm-eval-runnable-assert-stmt/test-bs10-last-edit-sim-min}{49.49}
\DefMacro{res-teco-icse23-CSNm-eval-runnable-assert-stmt/test-bs10-last-edit-sim-avg}{63.73}
\DefMacro{res-teco-icse23-CSNm-eval-runnable-assert-stmt/test-bs10-last-rouge-p}{70.36}
\DefMacro{res-teco-icse23-CSNm-eval-runnable-assert-stmt/test-bs10-last-rouge-p-max}{82.86}
\DefMacro{res-teco-icse23-CSNm-eval-runnable-assert-stmt/test-bs10-last-rouge-p-min}{47.12}
\DefMacro{res-teco-icse23-CSNm-eval-runnable-assert-stmt/test-bs10-last-rouge-p-avg}{62.40}
\DefMacro{res-teco-icse23-CSNm-eval-runnable-assert-stmt/test-bs10-last-rouge-r}{70.32}
\DefMacro{res-teco-icse23-CSNm-eval-runnable-assert-stmt/test-bs10-last-rouge-r-max}{82.08}
\DefMacro{res-teco-icse23-CSNm-eval-runnable-assert-stmt/test-bs10-last-rouge-r-min}{54.34}
\DefMacro{res-teco-icse23-CSNm-eval-runnable-assert-stmt/test-bs10-last-rouge-r-avg}{67.50}
\DefMacro{res-teco-icse23-CSNm-eval-runnable-assert-stmt/test-bs10-last-rouge-f}{69.35}
\DefMacro{res-teco-icse23-CSNm-eval-runnable-assert-stmt/test-bs10-last-rouge-f-max}{81.29}
\DefMacro{res-teco-icse23-CSNm-eval-runnable-assert-stmt/test-bs10-last-rouge-f-min}{49.73}
\DefMacro{res-teco-icse23-CSNm-eval-runnable-assert-stmt/test-bs10-last-rouge-f-avg}{63.73}
\DefMacro{res-teco-icse23-CSNm-eval-runnable-assert-stmt/test-bs10-last-num-seq}{10.00}
\DefMacro{res-teco-icse23-CSNm-eval-runnable-assert-stmt/test-bs10-last-compilable}{67.93}
\DefMacro{res-teco-icse23-CSNm-eval-runnable-assert-stmt/test-bs10-last-runnable}{30.29}
\DefMacro{res-teco-icse23-norr-CSNm-eval-any-stmt/test-bs10-last-xmatch}{15.25}
\DefMacro{res-teco-icse23-norr-CSNm-eval-any-stmt/test-bs10-last-xmatch-top3}{21.46}
\DefMacro{res-teco-icse23-norr-CSNm-eval-any-stmt/test-bs10-last-xmatch-top5}{23.90}
\DefMacro{res-teco-icse23-norr-CSNm-eval-any-stmt/test-bs10-last-xmatch-top10}{27.20}
\DefMacro{res-teco-icse23-norr-CSNm-eval-any-stmt/test-bs10-last-xmatch-top100}{27.20}
\DefMacro{res-teco-icse23-norr-CSNm-eval-any-stmt/test-bs10-last-bleu}{40.84}
\DefMacro{res-teco-icse23-norr-CSNm-eval-any-stmt/test-bs10-last-bleu-max}{55.53}
\DefMacro{res-teco-icse23-norr-CSNm-eval-any-stmt/test-bs10-last-bleu-min}{18.20}
\DefMacro{res-teco-icse23-norr-CSNm-eval-any-stmt/test-bs10-last-bleu-avg}{31.90}
\DefMacro{res-teco-icse23-norr-CSNm-eval-any-stmt/test-bs10-last-code-bleu}{36.34}
\DefMacro{res-teco-icse23-norr-CSNm-eval-any-stmt/test-bs10-last-code-bleu-max}{53.45}
\DefMacro{res-teco-icse23-norr-CSNm-eval-any-stmt/test-bs10-last-code-bleu-min}{12.14}
\DefMacro{res-teco-icse23-norr-CSNm-eval-any-stmt/test-bs10-last-code-bleu-avg}{27.53}
\DefMacro{res-teco-icse23-norr-CSNm-eval-any-stmt/test-bs10-last-edit-sim}{62.92}
\DefMacro{res-teco-icse23-norr-CSNm-eval-any-stmt/test-bs10-last-edit-sim-max}{74.54}
\DefMacro{res-teco-icse23-norr-CSNm-eval-any-stmt/test-bs10-last-edit-sim-min}{44.79}
\DefMacro{res-teco-icse23-norr-CSNm-eval-any-stmt/test-bs10-last-edit-sim-avg}{57.82}
\DefMacro{res-teco-icse23-norr-CSNm-eval-any-stmt/test-bs10-last-rouge-p}{65.90}
\DefMacro{res-teco-icse23-norr-CSNm-eval-any-stmt/test-bs10-last-rouge-p-max}{78.15}
\DefMacro{res-teco-icse23-norr-CSNm-eval-any-stmt/test-bs10-last-rouge-p-min}{44.70}
\DefMacro{res-teco-icse23-norr-CSNm-eval-any-stmt/test-bs10-last-rouge-p-avg}{58.53}
\DefMacro{res-teco-icse23-norr-CSNm-eval-any-stmt/test-bs10-last-rouge-r}{66.16}
\DefMacro{res-teco-icse23-norr-CSNm-eval-any-stmt/test-bs10-last-rouge-r-max}{77.08}
\DefMacro{res-teco-icse23-norr-CSNm-eval-any-stmt/test-bs10-last-rouge-r-min}{50.62}
\DefMacro{res-teco-icse23-norr-CSNm-eval-any-stmt/test-bs10-last-rouge-r-avg}{63.32}
\DefMacro{res-teco-icse23-norr-CSNm-eval-any-stmt/test-bs10-last-rouge-f}{64.71}
\DefMacro{res-teco-icse23-norr-CSNm-eval-any-stmt/test-bs10-last-rouge-f-max}{76.06}
\DefMacro{res-teco-icse23-norr-CSNm-eval-any-stmt/test-bs10-last-rouge-f-min}{46.45}
\DefMacro{res-teco-icse23-norr-CSNm-eval-any-stmt/test-bs10-last-rouge-f-avg}{59.31}
\DefMacro{res-teco-icse23-norr-CSNm-eval-any-stmt/test-bs10-last-num-seq}{10.00}
\DefMacro{res-teco-icse23-norr-CSNm-eval-any-stmt/test-bs10-last-compilable}{-1.00}
\DefMacro{res-teco-icse23-norr-CSNm-eval-any-stmt/test-bs10-last-runnable}{-1.00}
\DefMacro{res-teco-icse23-norr-CSNm-eval-assert-stmt/test-bs10-last-xmatch}{9.92}
\DefMacro{res-teco-icse23-norr-CSNm-eval-assert-stmt/test-bs10-last-xmatch-top3}{17.05}
\DefMacro{res-teco-icse23-norr-CSNm-eval-assert-stmt/test-bs10-last-xmatch-top5}{21.59}
\DefMacro{res-teco-icse23-norr-CSNm-eval-assert-stmt/test-bs10-last-xmatch-top10}{27.41}
\DefMacro{res-teco-icse23-norr-CSNm-eval-assert-stmt/test-bs10-last-xmatch-top100}{27.41}
\DefMacro{res-teco-icse23-norr-CSNm-eval-assert-stmt/test-bs10-last-bleu}{40.81}
\DefMacro{res-teco-icse23-norr-CSNm-eval-assert-stmt/test-bs10-last-bleu-max}{59.87}
\DefMacro{res-teco-icse23-norr-CSNm-eval-assert-stmt/test-bs10-last-bleu-min}{18.86}
\DefMacro{res-teco-icse23-norr-CSNm-eval-assert-stmt/test-bs10-last-bleu-avg}{34.35}
\DefMacro{res-teco-icse23-norr-CSNm-eval-assert-stmt/test-bs10-last-code-bleu}{32.90}
\DefMacro{res-teco-icse23-norr-CSNm-eval-assert-stmt/test-bs10-last-code-bleu-max}{54.23}
\DefMacro{res-teco-icse23-norr-CSNm-eval-assert-stmt/test-bs10-last-code-bleu-min}{11.69}
\DefMacro{res-teco-icse23-norr-CSNm-eval-assert-stmt/test-bs10-last-code-bleu-avg}{27.13}
\DefMacro{res-teco-icse23-norr-CSNm-eval-assert-stmt/test-bs10-last-edit-sim}{67.32}
\DefMacro{res-teco-icse23-norr-CSNm-eval-assert-stmt/test-bs10-last-edit-sim-max}{80.64}
\DefMacro{res-teco-icse23-norr-CSNm-eval-assert-stmt/test-bs10-last-edit-sim-min}{49.42}
\DefMacro{res-teco-icse23-norr-CSNm-eval-assert-stmt/test-bs10-last-edit-sim-avg}{63.70}
\DefMacro{res-teco-icse23-norr-CSNm-eval-assert-stmt/test-bs10-last-rouge-p}{68.82}
\DefMacro{res-teco-icse23-norr-CSNm-eval-assert-stmt/test-bs10-last-rouge-p-max}{82.82}
\DefMacro{res-teco-icse23-norr-CSNm-eval-assert-stmt/test-bs10-last-rouge-p-min}{47.08}
\DefMacro{res-teco-icse23-norr-CSNm-eval-assert-stmt/test-bs10-last-rouge-p-avg}{62.21}
\DefMacro{res-teco-icse23-norr-CSNm-eval-assert-stmt/test-bs10-last-rouge-r}{69.15}
\DefMacro{res-teco-icse23-norr-CSNm-eval-assert-stmt/test-bs10-last-rouge-r-max}{81.53}
\DefMacro{res-teco-icse23-norr-CSNm-eval-assert-stmt/test-bs10-last-rouge-r-min}{53.95}
\DefMacro{res-teco-icse23-norr-CSNm-eval-assert-stmt/test-bs10-last-rouge-r-avg}{66.95}
\DefMacro{res-teco-icse23-norr-CSNm-eval-assert-stmt/test-bs10-last-rouge-f}{67.92}
\DefMacro{res-teco-icse23-norr-CSNm-eval-assert-stmt/test-bs10-last-rouge-f-max}{80.93}
\DefMacro{res-teco-icse23-norr-CSNm-eval-assert-stmt/test-bs10-last-rouge-f-min}{49.51}
\DefMacro{res-teco-icse23-norr-CSNm-eval-assert-stmt/test-bs10-last-rouge-f-avg}{63.36}
\DefMacro{res-teco-icse23-norr-CSNm-eval-assert-stmt/test-bs10-last-num-seq}{10.00}
\DefMacro{res-teco-icse23-norr-CSNm-eval-assert-stmt/test-bs10-last-compilable}{-1.00}
\DefMacro{res-teco-icse23-norr-CSNm-eval-assert-stmt/test-bs10-last-runnable}{-1.00}
\DefMacro{res-teco-icse23-norr-CSNm-eval-runnable-any-stmt/test-bs10-last-xmatch}{15.99}
\DefMacro{res-teco-icse23-norr-CSNm-eval-runnable-any-stmt/test-bs10-last-xmatch-top3}{22.50}
\DefMacro{res-teco-icse23-norr-CSNm-eval-runnable-any-stmt/test-bs10-last-xmatch-top5}{24.96}
\DefMacro{res-teco-icse23-norr-CSNm-eval-runnable-any-stmt/test-bs10-last-xmatch-top10}{28.30}
\DefMacro{res-teco-icse23-norr-CSNm-eval-runnable-any-stmt/test-bs10-last-xmatch-top100}{28.30}
\DefMacro{res-teco-icse23-norr-CSNm-eval-runnable-any-stmt/test-bs10-last-bleu}{41.64}
\DefMacro{res-teco-icse23-norr-CSNm-eval-runnable-any-stmt/test-bs10-last-bleu-max}{56.63}
\DefMacro{res-teco-icse23-norr-CSNm-eval-runnable-any-stmt/test-bs10-last-bleu-min}{18.24}
\DefMacro{res-teco-icse23-norr-CSNm-eval-runnable-any-stmt/test-bs10-last-bleu-avg}{32.29}
\DefMacro{res-teco-icse23-norr-CSNm-eval-runnable-any-stmt/test-bs10-last-code-bleu}{36.82}
\DefMacro{res-teco-icse23-norr-CSNm-eval-runnable-any-stmt/test-bs10-last-code-bleu-max}{54.29}
\DefMacro{res-teco-icse23-norr-CSNm-eval-runnable-any-stmt/test-bs10-last-code-bleu-min}{11.83}
\DefMacro{res-teco-icse23-norr-CSNm-eval-runnable-any-stmt/test-bs10-last-code-bleu-avg}{27.62}
\DefMacro{res-teco-icse23-norr-CSNm-eval-runnable-any-stmt/test-bs10-last-edit-sim}{63.38}
\DefMacro{res-teco-icse23-norr-CSNm-eval-runnable-any-stmt/test-bs10-last-edit-sim-max}{75.27}
\DefMacro{res-teco-icse23-norr-CSNm-eval-runnable-any-stmt/test-bs10-last-edit-sim-min}{44.70}
\DefMacro{res-teco-icse23-norr-CSNm-eval-runnable-any-stmt/test-bs10-last-edit-sim-avg}{58.08}
\DefMacro{res-teco-icse23-norr-CSNm-eval-runnable-any-stmt/test-bs10-last-rouge-p}{66.60}
\DefMacro{res-teco-icse23-norr-CSNm-eval-runnable-any-stmt/test-bs10-last-rouge-p-max}{78.92}
\DefMacro{res-teco-icse23-norr-CSNm-eval-runnable-any-stmt/test-bs10-last-rouge-p-min}{44.83}
\DefMacro{res-teco-icse23-norr-CSNm-eval-runnable-any-stmt/test-bs10-last-rouge-p-avg}{58.99}
\DefMacro{res-teco-icse23-norr-CSNm-eval-runnable-any-stmt/test-bs10-last-rouge-r}{66.76}
\DefMacro{res-teco-icse23-norr-CSNm-eval-runnable-any-stmt/test-bs10-last-rouge-r-max}{77.88}
\DefMacro{res-teco-icse23-norr-CSNm-eval-runnable-any-stmt/test-bs10-last-rouge-r-min}{50.73}
\DefMacro{res-teco-icse23-norr-CSNm-eval-runnable-any-stmt/test-bs10-last-rouge-r-avg}{63.83}
\DefMacro{res-teco-icse23-norr-CSNm-eval-runnable-any-stmt/test-bs10-last-rouge-f}{65.42}
\DefMacro{res-teco-icse23-norr-CSNm-eval-runnable-any-stmt/test-bs10-last-rouge-f-max}{76.93}
\DefMacro{res-teco-icse23-norr-CSNm-eval-runnable-any-stmt/test-bs10-last-rouge-f-min}{46.62}
\DefMacro{res-teco-icse23-norr-CSNm-eval-runnable-any-stmt/test-bs10-last-rouge-f-avg}{59.82}
\DefMacro{res-teco-icse23-norr-CSNm-eval-runnable-any-stmt/test-bs10-last-num-seq}{10.00}
\DefMacro{res-teco-icse23-norr-CSNm-eval-runnable-any-stmt/test-bs10-last-compilable}{60.80}
\DefMacro{res-teco-icse23-norr-CSNm-eval-runnable-any-stmt/test-bs10-last-runnable}{19.49}
\DefMacro{res-teco-icse23-norr-CSNm-eval-runnable-assert-stmt/test-bs10-last-xmatch}{9.62}
\DefMacro{res-teco-icse23-norr-CSNm-eval-runnable-assert-stmt/test-bs10-last-xmatch-top3}{16.91}
\DefMacro{res-teco-icse23-norr-CSNm-eval-runnable-assert-stmt/test-bs10-last-xmatch-top5}{21.47}
\DefMacro{res-teco-icse23-norr-CSNm-eval-runnable-assert-stmt/test-bs10-last-xmatch-top10}{27.39}
\DefMacro{res-teco-icse23-norr-CSNm-eval-runnable-assert-stmt/test-bs10-last-xmatch-top100}{27.39}
\DefMacro{res-teco-icse23-norr-CSNm-eval-runnable-assert-stmt/test-bs10-last-bleu}{41.55}
\DefMacro{res-teco-icse23-norr-CSNm-eval-runnable-assert-stmt/test-bs10-last-bleu-max}{60.21}
\DefMacro{res-teco-icse23-norr-CSNm-eval-runnable-assert-stmt/test-bs10-last-bleu-min}{19.30}
\DefMacro{res-teco-icse23-norr-CSNm-eval-runnable-assert-stmt/test-bs10-last-bleu-avg}{34.85}
\DefMacro{res-teco-icse23-norr-CSNm-eval-runnable-assert-stmt/test-bs10-last-code-bleu}{33.44}
\DefMacro{res-teco-icse23-norr-CSNm-eval-runnable-assert-stmt/test-bs10-last-code-bleu-max}{54.36}
\DefMacro{res-teco-icse23-norr-CSNm-eval-runnable-assert-stmt/test-bs10-last-code-bleu-min}{12.15}
\DefMacro{res-teco-icse23-norr-CSNm-eval-runnable-assert-stmt/test-bs10-last-code-bleu-avg}{27.62}
\DefMacro{res-teco-icse23-norr-CSNm-eval-runnable-assert-stmt/test-bs10-last-edit-sim}{67.57}
\DefMacro{res-teco-icse23-norr-CSNm-eval-runnable-assert-stmt/test-bs10-last-edit-sim-max}{80.69}
\DefMacro{res-teco-icse23-norr-CSNm-eval-runnable-assert-stmt/test-bs10-last-edit-sim-min}{49.49}
\DefMacro{res-teco-icse23-norr-CSNm-eval-runnable-assert-stmt/test-bs10-last-edit-sim-avg}{63.73}
\DefMacro{res-teco-icse23-norr-CSNm-eval-runnable-assert-stmt/test-bs10-last-rouge-p}{69.15}
\DefMacro{res-teco-icse23-norr-CSNm-eval-runnable-assert-stmt/test-bs10-last-rouge-p-max}{82.86}
\DefMacro{res-teco-icse23-norr-CSNm-eval-runnable-assert-stmt/test-bs10-last-rouge-p-min}{47.12}
\DefMacro{res-teco-icse23-norr-CSNm-eval-runnable-assert-stmt/test-bs10-last-rouge-p-avg}{62.40}
\DefMacro{res-teco-icse23-norr-CSNm-eval-runnable-assert-stmt/test-bs10-last-rouge-r}{69.74}
\DefMacro{res-teco-icse23-norr-CSNm-eval-runnable-assert-stmt/test-bs10-last-rouge-r-max}{82.08}
\DefMacro{res-teco-icse23-norr-CSNm-eval-runnable-assert-stmt/test-bs10-last-rouge-r-min}{54.34}
\DefMacro{res-teco-icse23-norr-CSNm-eval-runnable-assert-stmt/test-bs10-last-rouge-r-avg}{67.50}
\DefMacro{res-teco-icse23-norr-CSNm-eval-runnable-assert-stmt/test-bs10-last-rouge-f}{68.41}
\DefMacro{res-teco-icse23-norr-CSNm-eval-runnable-assert-stmt/test-bs10-last-rouge-f-max}{81.29}
\DefMacro{res-teco-icse23-norr-CSNm-eval-runnable-assert-stmt/test-bs10-last-rouge-f-min}{49.73}
\DefMacro{res-teco-icse23-norr-CSNm-eval-runnable-assert-stmt/test-bs10-last-rouge-f-avg}{63.73}
\DefMacro{res-teco-icse23-norr-CSNm-eval-runnable-assert-stmt/test-bs10-last-num-seq}{10.00}
\DefMacro{res-teco-icse23-norr-CSNm-eval-runnable-assert-stmt/test-bs10-last-compilable}{48.13}
\DefMacro{res-teco-icse23-norr-CSNm-eval-runnable-assert-stmt/test-bs10-last-runnable}{18.45}
\DefMacro{res-teco-s1-CSNm-eval-any-stmt/test-bs10-last-xmatch}{13.88}
\DefMacro{res-teco-s1-CSNm-eval-any-stmt/test-bs10-last-xmatch-top3}{19.54}
\DefMacro{res-teco-s1-CSNm-eval-any-stmt/test-bs10-last-xmatch-top5}{21.87}
\DefMacro{res-teco-s1-CSNm-eval-any-stmt/test-bs10-last-xmatch-top10}{24.93}
\DefMacro{res-teco-s1-CSNm-eval-any-stmt/test-bs10-last-xmatch-top100}{24.93}
\DefMacro{res-teco-s1-CSNm-eval-any-stmt/test-bs10-last-bleu}{39.12}
\DefMacro{res-teco-s1-CSNm-eval-any-stmt/test-bs10-last-bleu-max}{53.57}
\DefMacro{res-teco-s1-CSNm-eval-any-stmt/test-bs10-last-bleu-min}{17.32}
\DefMacro{res-teco-s1-CSNm-eval-any-stmt/test-bs10-last-bleu-avg}{30.86}
\DefMacro{res-teco-s1-CSNm-eval-any-stmt/test-bs10-last-code-bleu}{34.66}
\DefMacro{res-teco-s1-CSNm-eval-any-stmt/test-bs10-last-code-bleu-max}{51.72}
\DefMacro{res-teco-s1-CSNm-eval-any-stmt/test-bs10-last-code-bleu-min}{11.05}
\DefMacro{res-teco-s1-CSNm-eval-any-stmt/test-bs10-last-code-bleu-avg}{26.41}
\DefMacro{res-teco-s1-CSNm-eval-any-stmt/test-bs10-last-edit-sim}{61.58}
\DefMacro{res-teco-s1-CSNm-eval-any-stmt/test-bs10-last-edit-sim-max}{73.06}
\DefMacro{res-teco-s1-CSNm-eval-any-stmt/test-bs10-last-edit-sim-min}{43.50}
\DefMacro{res-teco-s1-CSNm-eval-any-stmt/test-bs10-last-edit-sim-avg}{56.65}
\DefMacro{res-teco-s1-CSNm-eval-any-stmt/test-bs10-last-rouge-p}{64.89}
\DefMacro{res-teco-s1-CSNm-eval-any-stmt/test-bs10-last-rouge-p-max}{77.10}
\DefMacro{res-teco-s1-CSNm-eval-any-stmt/test-bs10-last-rouge-p-min}{43.75}
\DefMacro{res-teco-s1-CSNm-eval-any-stmt/test-bs10-last-rouge-p-avg}{57.82}
\DefMacro{res-teco-s1-CSNm-eval-any-stmt/test-bs10-last-rouge-r}{64.81}
\DefMacro{res-teco-s1-CSNm-eval-any-stmt/test-bs10-last-rouge-r-max}{75.67}
\DefMacro{res-teco-s1-CSNm-eval-any-stmt/test-bs10-last-rouge-r-min}{49.39}
\DefMacro{res-teco-s1-CSNm-eval-any-stmt/test-bs10-last-rouge-r-avg}{62.26}
\DefMacro{res-teco-s1-CSNm-eval-any-stmt/test-bs10-last-rouge-f}{63.51}
\DefMacro{res-teco-s1-CSNm-eval-any-stmt/test-bs10-last-rouge-f-max}{74.78}
\DefMacro{res-teco-s1-CSNm-eval-any-stmt/test-bs10-last-rouge-f-min}{45.36}
\DefMacro{res-teco-s1-CSNm-eval-any-stmt/test-bs10-last-rouge-f-avg}{58.38}
\DefMacro{res-teco-s1-CSNm-eval-any-stmt/test-bs10-last-num-seq}{10.00}
\DefMacro{res-teco-s1-CSNm-eval-any-stmt/test-bs10-last-compilable}{-1.00}
\DefMacro{res-teco-s1-CSNm-eval-any-stmt/test-bs10-last-runnable}{-1.00}
\DefMacro{res-teco-s1-CSNm-eval-assert-stmt/test-bs10-last-xmatch}{8.86}
\DefMacro{res-teco-s1-CSNm-eval-assert-stmt/test-bs10-last-xmatch-top3}{14.23}
\DefMacro{res-teco-s1-CSNm-eval-assert-stmt/test-bs10-last-xmatch-top5}{18.36}
\DefMacro{res-teco-s1-CSNm-eval-assert-stmt/test-bs10-last-xmatch-top10}{24.37}
\DefMacro{res-teco-s1-CSNm-eval-assert-stmt/test-bs10-last-xmatch-top100}{24.37}
\DefMacro{res-teco-s1-CSNm-eval-assert-stmt/test-bs10-last-bleu}{38.03}
\DefMacro{res-teco-s1-CSNm-eval-assert-stmt/test-bs10-last-bleu-max}{57.61}
\DefMacro{res-teco-s1-CSNm-eval-assert-stmt/test-bs10-last-bleu-min}{17.26}
\DefMacro{res-teco-s1-CSNm-eval-assert-stmt/test-bs10-last-bleu-avg}{32.56}
\DefMacro{res-teco-s1-CSNm-eval-assert-stmt/test-bs10-last-code-bleu}{29.93}
\DefMacro{res-teco-s1-CSNm-eval-assert-stmt/test-bs10-last-code-bleu-max}{51.93}
\DefMacro{res-teco-s1-CSNm-eval-assert-stmt/test-bs10-last-code-bleu-min}{9.57}
\DefMacro{res-teco-s1-CSNm-eval-assert-stmt/test-bs10-last-code-bleu-avg}{25.04}
\DefMacro{res-teco-s1-CSNm-eval-assert-stmt/test-bs10-last-edit-sim}{65.59}
\DefMacro{res-teco-s1-CSNm-eval-assert-stmt/test-bs10-last-edit-sim-max}{79.38}
\DefMacro{res-teco-s1-CSNm-eval-assert-stmt/test-bs10-last-edit-sim-min}{47.66}
\DefMacro{res-teco-s1-CSNm-eval-assert-stmt/test-bs10-last-edit-sim-avg}{62.35}
\DefMacro{res-teco-s1-CSNm-eval-assert-stmt/test-bs10-last-rouge-p}{66.93}
\DefMacro{res-teco-s1-CSNm-eval-assert-stmt/test-bs10-last-rouge-p-max}{81.70}
\DefMacro{res-teco-s1-CSNm-eval-assert-stmt/test-bs10-last-rouge-p-min}{45.69}
\DefMacro{res-teco-s1-CSNm-eval-assert-stmt/test-bs10-last-rouge-p-avg}{61.09}
\DefMacro{res-teco-s1-CSNm-eval-assert-stmt/test-bs10-last-rouge-r}{67.22}
\DefMacro{res-teco-s1-CSNm-eval-assert-stmt/test-bs10-last-rouge-r-max}{80.16}
\DefMacro{res-teco-s1-CSNm-eval-assert-stmt/test-bs10-last-rouge-r-min}{52.05}
\DefMacro{res-teco-s1-CSNm-eval-assert-stmt/test-bs10-last-rouge-r-avg}{65.45}
\DefMacro{res-teco-s1-CSNm-eval-assert-stmt/test-bs10-last-rouge-f}{65.95}
\DefMacro{res-teco-s1-CSNm-eval-assert-stmt/test-bs10-last-rouge-f-max}{79.60}
\DefMacro{res-teco-s1-CSNm-eval-assert-stmt/test-bs10-last-rouge-f-min}{47.86}
\DefMacro{res-teco-s1-CSNm-eval-assert-stmt/test-bs10-last-rouge-f-avg}{62.01}
\DefMacro{res-teco-s1-CSNm-eval-assert-stmt/test-bs10-last-num-seq}{10.00}
\DefMacro{res-teco-s1-CSNm-eval-assert-stmt/test-bs10-last-compilable}{-1.00}
\DefMacro{res-teco-s1-CSNm-eval-assert-stmt/test-bs10-last-runnable}{-1.00}
\DefMacro{res-teco-s1-CSNm-eval-runnable-any-stmt/test-bs10-last-xmatch}{14.71}
\DefMacro{res-teco-s1-CSNm-eval-runnable-any-stmt/test-bs10-last-xmatch-top3}{20.76}
\DefMacro{res-teco-s1-CSNm-eval-runnable-any-stmt/test-bs10-last-xmatch-top5}{23.15}
\DefMacro{res-teco-s1-CSNm-eval-runnable-any-stmt/test-bs10-last-xmatch-top10}{26.33}
\DefMacro{res-teco-s1-CSNm-eval-runnable-any-stmt/test-bs10-last-xmatch-top100}{26.33}
\DefMacro{res-teco-s1-CSNm-eval-runnable-any-stmt/test-bs10-last-bleu}{39.99}
\DefMacro{res-teco-s1-CSNm-eval-runnable-any-stmt/test-bs10-last-bleu-max}{54.81}
\DefMacro{res-teco-s1-CSNm-eval-runnable-any-stmt/test-bs10-last-bleu-min}{17.34}
\DefMacro{res-teco-s1-CSNm-eval-runnable-any-stmt/test-bs10-last-bleu-avg}{31.30}
\DefMacro{res-teco-s1-CSNm-eval-runnable-any-stmt/test-bs10-last-code-bleu}{35.23}
\DefMacro{res-teco-s1-CSNm-eval-runnable-any-stmt/test-bs10-last-code-bleu-max}{52.75}
\DefMacro{res-teco-s1-CSNm-eval-runnable-any-stmt/test-bs10-last-code-bleu-min}{10.70}
\DefMacro{res-teco-s1-CSNm-eval-runnable-any-stmt/test-bs10-last-code-bleu-avg}{26.53}
\DefMacro{res-teco-s1-CSNm-eval-runnable-any-stmt/test-bs10-last-edit-sim}{62.12}
\DefMacro{res-teco-s1-CSNm-eval-runnable-any-stmt/test-bs10-last-edit-sim-max}{73.87}
\DefMacro{res-teco-s1-CSNm-eval-runnable-any-stmt/test-bs10-last-edit-sim-min}{43.36}
\DefMacro{res-teco-s1-CSNm-eval-runnable-any-stmt/test-bs10-last-edit-sim-avg}{56.92}
\DefMacro{res-teco-s1-CSNm-eval-runnable-any-stmt/test-bs10-last-rouge-p}{65.65}
\DefMacro{res-teco-s1-CSNm-eval-runnable-any-stmt/test-bs10-last-rouge-p-max}{77.98}
\DefMacro{res-teco-s1-CSNm-eval-runnable-any-stmt/test-bs10-last-rouge-p-min}{43.92}
\DefMacro{res-teco-s1-CSNm-eval-runnable-any-stmt/test-bs10-last-rouge-p-avg}{58.30}
\DefMacro{res-teco-s1-CSNm-eval-runnable-any-stmt/test-bs10-last-rouge-r}{65.46}
\DefMacro{res-teco-s1-CSNm-eval-runnable-any-stmt/test-bs10-last-rouge-r-max}{76.56}
\DefMacro{res-teco-s1-CSNm-eval-runnable-any-stmt/test-bs10-last-rouge-r-min}{49.51}
\DefMacro{res-teco-s1-CSNm-eval-runnable-any-stmt/test-bs10-last-rouge-r-avg}{62.77}
\DefMacro{res-teco-s1-CSNm-eval-runnable-any-stmt/test-bs10-last-rouge-f}{64.28}
\DefMacro{res-teco-s1-CSNm-eval-runnable-any-stmt/test-bs10-last-rouge-f-max}{75.74}
\DefMacro{res-teco-s1-CSNm-eval-runnable-any-stmt/test-bs10-last-rouge-f-min}{45.55}
\DefMacro{res-teco-s1-CSNm-eval-runnable-any-stmt/test-bs10-last-rouge-f-avg}{58.92}
\DefMacro{res-teco-s1-CSNm-eval-runnable-any-stmt/test-bs10-last-num-seq}{10.00}
\DefMacro{res-teco-s1-CSNm-eval-runnable-any-stmt/test-bs10-last-compilable}{-1.00}
\DefMacro{res-teco-s1-CSNm-eval-runnable-any-stmt/test-bs10-last-runnable}{-1.00}
\DefMacro{res-teco-s1-CSNm-eval-runnable-assert-stmt/test-bs10-last-xmatch}{9.07}
\DefMacro{res-teco-s1-CSNm-eval-runnable-assert-stmt/test-bs10-last-xmatch-top3}{14.63}
\DefMacro{res-teco-s1-CSNm-eval-runnable-assert-stmt/test-bs10-last-xmatch-top5}{18.82}
\DefMacro{res-teco-s1-CSNm-eval-runnable-assert-stmt/test-bs10-last-xmatch-top10}{24.93}
\DefMacro{res-teco-s1-CSNm-eval-runnable-assert-stmt/test-bs10-last-xmatch-top100}{24.93}
\DefMacro{res-teco-s1-CSNm-eval-runnable-assert-stmt/test-bs10-last-bleu}{39.01}
\DefMacro{res-teco-s1-CSNm-eval-runnable-assert-stmt/test-bs10-last-bleu-max}{58.11}
\DefMacro{res-teco-s1-CSNm-eval-runnable-assert-stmt/test-bs10-last-bleu-min}{17.59}
\DefMacro{res-teco-s1-CSNm-eval-runnable-assert-stmt/test-bs10-last-bleu-avg}{33.10}
\DefMacro{res-teco-s1-CSNm-eval-runnable-assert-stmt/test-bs10-last-code-bleu}{30.85}
\DefMacro{res-teco-s1-CSNm-eval-runnable-assert-stmt/test-bs10-last-code-bleu-max}{52.32}
\DefMacro{res-teco-s1-CSNm-eval-runnable-assert-stmt/test-bs10-last-code-bleu-min}{9.84}
\DefMacro{res-teco-s1-CSNm-eval-runnable-assert-stmt/test-bs10-last-code-bleu-avg}{25.56}
\DefMacro{res-teco-s1-CSNm-eval-runnable-assert-stmt/test-bs10-last-edit-sim}{66.11}
\DefMacro{res-teco-s1-CSNm-eval-runnable-assert-stmt/test-bs10-last-edit-sim-max}{79.61}
\DefMacro{res-teco-s1-CSNm-eval-runnable-assert-stmt/test-bs10-last-edit-sim-min}{47.63}
\DefMacro{res-teco-s1-CSNm-eval-runnable-assert-stmt/test-bs10-last-edit-sim-avg}{62.56}
\DefMacro{res-teco-s1-CSNm-eval-runnable-assert-stmt/test-bs10-last-rouge-p}{67.36}
\DefMacro{res-teco-s1-CSNm-eval-runnable-assert-stmt/test-bs10-last-rouge-p-max}{81.72}
\DefMacro{res-teco-s1-CSNm-eval-runnable-assert-stmt/test-bs10-last-rouge-p-min}{45.73}
\DefMacro{res-teco-s1-CSNm-eval-runnable-assert-stmt/test-bs10-last-rouge-p-avg}{61.25}
\DefMacro{res-teco-s1-CSNm-eval-runnable-assert-stmt/test-bs10-last-rouge-r}{68.03}
\DefMacro{res-teco-s1-CSNm-eval-runnable-assert-stmt/test-bs10-last-rouge-r-max}{80.79}
\DefMacro{res-teco-s1-CSNm-eval-runnable-assert-stmt/test-bs10-last-rouge-r-min}{52.34}
\DefMacro{res-teco-s1-CSNm-eval-runnable-assert-stmt/test-bs10-last-rouge-r-avg}{66.06}
\DefMacro{res-teco-s1-CSNm-eval-runnable-assert-stmt/test-bs10-last-rouge-f}{66.60}
\DefMacro{res-teco-s1-CSNm-eval-runnable-assert-stmt/test-bs10-last-rouge-f-max}{80.00}
\DefMacro{res-teco-s1-CSNm-eval-runnable-assert-stmt/test-bs10-last-rouge-f-min}{48.04}
\DefMacro{res-teco-s1-CSNm-eval-runnable-assert-stmt/test-bs10-last-rouge-f-avg}{62.41}
\DefMacro{res-teco-s1-CSNm-eval-runnable-assert-stmt/test-bs10-last-num-seq}{10.00}
\DefMacro{res-teco-s1-CSNm-eval-runnable-assert-stmt/test-bs10-last-compilable}{-1.00}
\DefMacro{res-teco-s1-CSNm-eval-runnable-assert-stmt/test-bs10-last-runnable}{-1.00}
\DefMacro{res-teco-s2-CSNm-eval-any-stmt/test-bs10-last-xmatch}{14.06}
\DefMacro{res-teco-s2-CSNm-eval-any-stmt/test-bs10-last-xmatch-top3}{20.00}
\DefMacro{res-teco-s2-CSNm-eval-any-stmt/test-bs10-last-xmatch-top5}{22.14}
\DefMacro{res-teco-s2-CSNm-eval-any-stmt/test-bs10-last-xmatch-top10}{25.11}
\DefMacro{res-teco-s2-CSNm-eval-any-stmt/test-bs10-last-xmatch-top100}{25.11}
\DefMacro{res-teco-s2-CSNm-eval-any-stmt/test-bs10-last-bleu}{39.56}
\DefMacro{res-teco-s2-CSNm-eval-any-stmt/test-bs10-last-bleu-max}{53.98}
\DefMacro{res-teco-s2-CSNm-eval-any-stmt/test-bs10-last-bleu-min}{17.73}
\DefMacro{res-teco-s2-CSNm-eval-any-stmt/test-bs10-last-bleu-avg}{31.44}
\DefMacro{res-teco-s2-CSNm-eval-any-stmt/test-bs10-last-code-bleu}{35.17}
\DefMacro{res-teco-s2-CSNm-eval-any-stmt/test-bs10-last-code-bleu-max}{52.00}
\DefMacro{res-teco-s2-CSNm-eval-any-stmt/test-bs10-last-code-bleu-min}{11.65}
\DefMacro{res-teco-s2-CSNm-eval-any-stmt/test-bs10-last-code-bleu-avg}{27.09}
\DefMacro{res-teco-s2-CSNm-eval-any-stmt/test-bs10-last-edit-sim}{62.20}
\DefMacro{res-teco-s2-CSNm-eval-any-stmt/test-bs10-last-edit-sim-max}{73.68}
\DefMacro{res-teco-s2-CSNm-eval-any-stmt/test-bs10-last-edit-sim-min}{44.42}
\DefMacro{res-teco-s2-CSNm-eval-any-stmt/test-bs10-last-edit-sim-avg}{57.62}
\DefMacro{res-teco-s2-CSNm-eval-any-stmt/test-bs10-last-rouge-p}{65.34}
\DefMacro{res-teco-s2-CSNm-eval-any-stmt/test-bs10-last-rouge-p-max}{77.40}
\DefMacro{res-teco-s2-CSNm-eval-any-stmt/test-bs10-last-rouge-p-min}{44.35}
\DefMacro{res-teco-s2-CSNm-eval-any-stmt/test-bs10-last-rouge-p-avg}{58.38}
\DefMacro{res-teco-s2-CSNm-eval-any-stmt/test-bs10-last-rouge-r}{65.20}
\DefMacro{res-teco-s2-CSNm-eval-any-stmt/test-bs10-last-rouge-r-max}{75.77}
\DefMacro{res-teco-s2-CSNm-eval-any-stmt/test-bs10-last-rouge-r-min}{50.11}
\DefMacro{res-teco-s2-CSNm-eval-any-stmt/test-bs10-last-rouge-r-avg}{62.76}
\DefMacro{res-teco-s2-CSNm-eval-any-stmt/test-bs10-last-rouge-f}{63.92}
\DefMacro{res-teco-s2-CSNm-eval-any-stmt/test-bs10-last-rouge-f-max}{74.97}
\DefMacro{res-teco-s2-CSNm-eval-any-stmt/test-bs10-last-rouge-f-min}{45.97}
\DefMacro{res-teco-s2-CSNm-eval-any-stmt/test-bs10-last-rouge-f-avg}{58.91}
\DefMacro{res-teco-s2-CSNm-eval-any-stmt/test-bs10-last-num-seq}{10.00}
\DefMacro{res-teco-s2-CSNm-eval-any-stmt/test-bs10-last-compilable}{-1.00}
\DefMacro{res-teco-s2-CSNm-eval-any-stmt/test-bs10-last-runnable}{-1.00}
\DefMacro{res-teco-s2-CSNm-eval-assert-stmt/test-bs10-last-xmatch}{8.23}
\DefMacro{res-teco-s2-CSNm-eval-assert-stmt/test-bs10-last-xmatch-top3}{14.59}
\DefMacro{res-teco-s2-CSNm-eval-assert-stmt/test-bs10-last-xmatch-top5}{18.13}
\DefMacro{res-teco-s2-CSNm-eval-assert-stmt/test-bs10-last-xmatch-top10}{23.69}
\DefMacro{res-teco-s2-CSNm-eval-assert-stmt/test-bs10-last-xmatch-top100}{23.69}
\DefMacro{res-teco-s2-CSNm-eval-assert-stmt/test-bs10-last-bleu}{38.13}
\DefMacro{res-teco-s2-CSNm-eval-assert-stmt/test-bs10-last-bleu-max}{57.55}
\DefMacro{res-teco-s2-CSNm-eval-assert-stmt/test-bs10-last-bleu-min}{18.41}
\DefMacro{res-teco-s2-CSNm-eval-assert-stmt/test-bs10-last-bleu-avg}{33.35}
\DefMacro{res-teco-s2-CSNm-eval-assert-stmt/test-bs10-last-code-bleu}{30.27}
\DefMacro{res-teco-s2-CSNm-eval-assert-stmt/test-bs10-last-code-bleu-max}{51.99}
\DefMacro{res-teco-s2-CSNm-eval-assert-stmt/test-bs10-last-code-bleu-min}{10.98}
\DefMacro{res-teco-s2-CSNm-eval-assert-stmt/test-bs10-last-code-bleu-avg}{26.05}
\DefMacro{res-teco-s2-CSNm-eval-assert-stmt/test-bs10-last-edit-sim}{65.52}
\DefMacro{res-teco-s2-CSNm-eval-assert-stmt/test-bs10-last-edit-sim-max}{79.31}
\DefMacro{res-teco-s2-CSNm-eval-assert-stmt/test-bs10-last-edit-sim-min}{49.14}
\DefMacro{res-teco-s2-CSNm-eval-assert-stmt/test-bs10-last-edit-sim-avg}{63.13}
\DefMacro{res-teco-s2-CSNm-eval-assert-stmt/test-bs10-last-rouge-p}{67.10}
\DefMacro{res-teco-s2-CSNm-eval-assert-stmt/test-bs10-last-rouge-p-max}{81.45}
\DefMacro{res-teco-s2-CSNm-eval-assert-stmt/test-bs10-last-rouge-p-min}{46.44}
\DefMacro{res-teco-s2-CSNm-eval-assert-stmt/test-bs10-last-rouge-p-avg}{61.68}
\DefMacro{res-teco-s2-CSNm-eval-assert-stmt/test-bs10-last-rouge-r}{67.12}
\DefMacro{res-teco-s2-CSNm-eval-assert-stmt/test-bs10-last-rouge-r-max}{80.00}
\DefMacro{res-teco-s2-CSNm-eval-assert-stmt/test-bs10-last-rouge-r-min}{52.87}
\DefMacro{res-teco-s2-CSNm-eval-assert-stmt/test-bs10-last-rouge-r-avg}{65.91}
\DefMacro{res-teco-s2-CSNm-eval-assert-stmt/test-bs10-last-rouge-f}{65.98}
\DefMacro{res-teco-s2-CSNm-eval-assert-stmt/test-bs10-last-rouge-f-max}{79.45}
\DefMacro{res-teco-s2-CSNm-eval-assert-stmt/test-bs10-last-rouge-f-min}{48.63}
\DefMacro{res-teco-s2-CSNm-eval-assert-stmt/test-bs10-last-rouge-f-avg}{62.57}
\DefMacro{res-teco-s2-CSNm-eval-assert-stmt/test-bs10-last-num-seq}{10.00}
\DefMacro{res-teco-s2-CSNm-eval-assert-stmt/test-bs10-last-compilable}{-1.00}
\DefMacro{res-teco-s2-CSNm-eval-assert-stmt/test-bs10-last-runnable}{-1.00}
\DefMacro{res-teco-s2-CSNm-eval-runnable-any-stmt/test-bs10-last-xmatch}{14.91}
\DefMacro{res-teco-s2-CSNm-eval-runnable-any-stmt/test-bs10-last-xmatch-top3}{21.17}
\DefMacro{res-teco-s2-CSNm-eval-runnable-any-stmt/test-bs10-last-xmatch-top5}{23.40}
\DefMacro{res-teco-s2-CSNm-eval-runnable-any-stmt/test-bs10-last-xmatch-top10}{26.48}
\DefMacro{res-teco-s2-CSNm-eval-runnable-any-stmt/test-bs10-last-xmatch-top100}{26.48}
\DefMacro{res-teco-s2-CSNm-eval-runnable-any-stmt/test-bs10-last-bleu}{40.48}
\DefMacro{res-teco-s2-CSNm-eval-runnable-any-stmt/test-bs10-last-bleu-max}{55.27}
\DefMacro{res-teco-s2-CSNm-eval-runnable-any-stmt/test-bs10-last-bleu-min}{17.84}
\DefMacro{res-teco-s2-CSNm-eval-runnable-any-stmt/test-bs10-last-bleu-avg}{31.94}
\DefMacro{res-teco-s2-CSNm-eval-runnable-any-stmt/test-bs10-last-code-bleu}{35.84}
\DefMacro{res-teco-s2-CSNm-eval-runnable-any-stmt/test-bs10-last-code-bleu-max}{53.09}
\DefMacro{res-teco-s2-CSNm-eval-runnable-any-stmt/test-bs10-last-code-bleu-min}{11.38}
\DefMacro{res-teco-s2-CSNm-eval-runnable-any-stmt/test-bs10-last-code-bleu-avg}{27.29}
\DefMacro{res-teco-s2-CSNm-eval-runnable-any-stmt/test-bs10-last-edit-sim}{62.76}
\DefMacro{res-teco-s2-CSNm-eval-runnable-any-stmt/test-bs10-last-edit-sim-max}{74.51}
\DefMacro{res-teco-s2-CSNm-eval-runnable-any-stmt/test-bs10-last-edit-sim-min}{44.41}
\DefMacro{res-teco-s2-CSNm-eval-runnable-any-stmt/test-bs10-last-edit-sim-avg}{57.98}
\DefMacro{res-teco-s2-CSNm-eval-runnable-any-stmt/test-bs10-last-rouge-p}{66.00}
\DefMacro{res-teco-s2-CSNm-eval-runnable-any-stmt/test-bs10-last-rouge-p-max}{78.18}
\DefMacro{res-teco-s2-CSNm-eval-runnable-any-stmt/test-bs10-last-rouge-p-min}{44.48}
\DefMacro{res-teco-s2-CSNm-eval-runnable-any-stmt/test-bs10-last-rouge-p-avg}{58.80}
\DefMacro{res-teco-s2-CSNm-eval-runnable-any-stmt/test-bs10-last-rouge-r}{65.86}
\DefMacro{res-teco-s2-CSNm-eval-runnable-any-stmt/test-bs10-last-rouge-r-max}{76.67}
\DefMacro{res-teco-s2-CSNm-eval-runnable-any-stmt/test-bs10-last-rouge-r-min}{50.27}
\DefMacro{res-teco-s2-CSNm-eval-runnable-any-stmt/test-bs10-last-rouge-r-avg}{63.31}
\DefMacro{res-teco-s2-CSNm-eval-runnable-any-stmt/test-bs10-last-rouge-f}{64.63}
\DefMacro{res-teco-s2-CSNm-eval-runnable-any-stmt/test-bs10-last-rouge-f-max}{75.88}
\DefMacro{res-teco-s2-CSNm-eval-runnable-any-stmt/test-bs10-last-rouge-f-min}{46.16}
\DefMacro{res-teco-s2-CSNm-eval-runnable-any-stmt/test-bs10-last-rouge-f-avg}{59.43}
\DefMacro{res-teco-s2-CSNm-eval-runnable-any-stmt/test-bs10-last-num-seq}{10.00}
\DefMacro{res-teco-s2-CSNm-eval-runnable-any-stmt/test-bs10-last-compilable}{-1.00}
\DefMacro{res-teco-s2-CSNm-eval-runnable-any-stmt/test-bs10-last-runnable}{-1.00}
\DefMacro{res-teco-s2-CSNm-eval-runnable-assert-stmt/test-bs10-last-xmatch}{8.45}
\DefMacro{res-teco-s2-CSNm-eval-runnable-assert-stmt/test-bs10-last-xmatch-top3}{14.83}
\DefMacro{res-teco-s2-CSNm-eval-runnable-assert-stmt/test-bs10-last-xmatch-top5}{18.37}
\DefMacro{res-teco-s2-CSNm-eval-runnable-assert-stmt/test-bs10-last-xmatch-top10}{23.93}
\DefMacro{res-teco-s2-CSNm-eval-runnable-assert-stmt/test-bs10-last-xmatch-top100}{23.93}
\DefMacro{res-teco-s2-CSNm-eval-runnable-assert-stmt/test-bs10-last-bleu}{39.20}
\DefMacro{res-teco-s2-CSNm-eval-runnable-assert-stmt/test-bs10-last-bleu-max}{57.75}
\DefMacro{res-teco-s2-CSNm-eval-runnable-assert-stmt/test-bs10-last-bleu-min}{18.86}
\DefMacro{res-teco-s2-CSNm-eval-runnable-assert-stmt/test-bs10-last-bleu-avg}{33.82}
\DefMacro{res-teco-s2-CSNm-eval-runnable-assert-stmt/test-bs10-last-code-bleu}{31.32}
\DefMacro{res-teco-s2-CSNm-eval-runnable-assert-stmt/test-bs10-last-code-bleu-max}{52.06}
\DefMacro{res-teco-s2-CSNm-eval-runnable-assert-stmt/test-bs10-last-code-bleu-min}{11.46}
\DefMacro{res-teco-s2-CSNm-eval-runnable-assert-stmt/test-bs10-last-code-bleu-avg}{26.54}
\DefMacro{res-teco-s2-CSNm-eval-runnable-assert-stmt/test-bs10-last-edit-sim}{66.04}
\DefMacro{res-teco-s2-CSNm-eval-runnable-assert-stmt/test-bs10-last-edit-sim-max}{79.37}
\DefMacro{res-teco-s2-CSNm-eval-runnable-assert-stmt/test-bs10-last-edit-sim-min}{49.29}
\DefMacro{res-teco-s2-CSNm-eval-runnable-assert-stmt/test-bs10-last-edit-sim-avg}{63.27}
\DefMacro{res-teco-s2-CSNm-eval-runnable-assert-stmt/test-bs10-last-rouge-p}{67.43}
\DefMacro{res-teco-s2-CSNm-eval-runnable-assert-stmt/test-bs10-last-rouge-p-max}{81.23}
\DefMacro{res-teco-s2-CSNm-eval-runnable-assert-stmt/test-bs10-last-rouge-p-min}{46.41}
\DefMacro{res-teco-s2-CSNm-eval-runnable-assert-stmt/test-bs10-last-rouge-p-avg}{61.65}
\DefMacro{res-teco-s2-CSNm-eval-runnable-assert-stmt/test-bs10-last-rouge-r}{67.82}
\DefMacro{res-teco-s2-CSNm-eval-runnable-assert-stmt/test-bs10-last-rouge-r-max}{80.26}
\DefMacro{res-teco-s2-CSNm-eval-runnable-assert-stmt/test-bs10-last-rouge-r-min}{53.23}
\DefMacro{res-teco-s2-CSNm-eval-runnable-assert-stmt/test-bs10-last-rouge-r-avg}{66.33}
\DefMacro{res-teco-s2-CSNm-eval-runnable-assert-stmt/test-bs10-last-rouge-f}{66.54}
\DefMacro{res-teco-s2-CSNm-eval-runnable-assert-stmt/test-bs10-last-rouge-f-max}{79.48}
\DefMacro{res-teco-s2-CSNm-eval-runnable-assert-stmt/test-bs10-last-rouge-f-min}{48.78}
\DefMacro{res-teco-s2-CSNm-eval-runnable-assert-stmt/test-bs10-last-rouge-f-avg}{62.78}
\DefMacro{res-teco-s2-CSNm-eval-runnable-assert-stmt/test-bs10-last-num-seq}{10.00}
\DefMacro{res-teco-s2-CSNm-eval-runnable-assert-stmt/test-bs10-last-compilable}{-1.00}
\DefMacro{res-teco-s2-CSNm-eval-runnable-assert-stmt/test-bs10-last-runnable}{-1.00}
\DefMacro{res-teco-s3-CSNm-eval-any-stmt/test-bs10-last-xmatch}{14.04}
\DefMacro{res-teco-s3-CSNm-eval-any-stmt/test-bs10-last-xmatch-top3}{19.47}
\DefMacro{res-teco-s3-CSNm-eval-any-stmt/test-bs10-last-xmatch-top5}{21.62}
\DefMacro{res-teco-s3-CSNm-eval-any-stmt/test-bs10-last-xmatch-top10}{24.40}
\DefMacro{res-teco-s3-CSNm-eval-any-stmt/test-bs10-last-xmatch-top100}{24.40}
\DefMacro{res-teco-s3-CSNm-eval-any-stmt/test-bs10-last-bleu}{38.81}
\DefMacro{res-teco-s3-CSNm-eval-any-stmt/test-bs10-last-bleu-max}{52.39}
\DefMacro{res-teco-s3-CSNm-eval-any-stmt/test-bs10-last-bleu-min}{17.62}
\DefMacro{res-teco-s3-CSNm-eval-any-stmt/test-bs10-last-bleu-avg}{30.67}
\DefMacro{res-teco-s3-CSNm-eval-any-stmt/test-bs10-last-code-bleu}{34.24}
\DefMacro{res-teco-s3-CSNm-eval-any-stmt/test-bs10-last-code-bleu-max}{50.19}
\DefMacro{res-teco-s3-CSNm-eval-any-stmt/test-bs10-last-code-bleu-min}{11.73}
\DefMacro{res-teco-s3-CSNm-eval-any-stmt/test-bs10-last-code-bleu-avg}{26.31}
\DefMacro{res-teco-s3-CSNm-eval-any-stmt/test-bs10-last-edit-sim}{61.21}
\DefMacro{res-teco-s3-CSNm-eval-any-stmt/test-bs10-last-edit-sim-max}{72.10}
\DefMacro{res-teco-s3-CSNm-eval-any-stmt/test-bs10-last-edit-sim-min}{43.92}
\DefMacro{res-teco-s3-CSNm-eval-any-stmt/test-bs10-last-edit-sim-avg}{56.46}
\DefMacro{res-teco-s3-CSNm-eval-any-stmt/test-bs10-last-rouge-p}{64.05}
\DefMacro{res-teco-s3-CSNm-eval-any-stmt/test-bs10-last-rouge-p-max}{75.92}
\DefMacro{res-teco-s3-CSNm-eval-any-stmt/test-bs10-last-rouge-p-min}{43.79}
\DefMacro{res-teco-s3-CSNm-eval-any-stmt/test-bs10-last-rouge-p-avg}{57.40}
\DefMacro{res-teco-s3-CSNm-eval-any-stmt/test-bs10-last-rouge-r}{64.42}
\DefMacro{res-teco-s3-CSNm-eval-any-stmt/test-bs10-last-rouge-r-max}{74.67}
\DefMacro{res-teco-s3-CSNm-eval-any-stmt/test-bs10-last-rouge-r-min}{49.77}
\DefMacro{res-teco-s3-CSNm-eval-any-stmt/test-bs10-last-rouge-r-avg}{61.98}
\DefMacro{res-teco-s3-CSNm-eval-any-stmt/test-bs10-last-rouge-f}{62.87}
\DefMacro{res-teco-s3-CSNm-eval-any-stmt/test-bs10-last-rouge-f-max}{73.62}
\DefMacro{res-teco-s3-CSNm-eval-any-stmt/test-bs10-last-rouge-f-min}{45.56}
\DefMacro{res-teco-s3-CSNm-eval-any-stmt/test-bs10-last-rouge-f-avg}{58.02}
\DefMacro{res-teco-s3-CSNm-eval-any-stmt/test-bs10-last-num-seq}{10.00}
\DefMacro{res-teco-s3-CSNm-eval-any-stmt/test-bs10-last-compilable}{-1.00}
\DefMacro{res-teco-s3-CSNm-eval-any-stmt/test-bs10-last-runnable}{-1.00}
\DefMacro{res-teco-s3-CSNm-eval-assert-stmt/test-bs10-last-xmatch}{8.72}
\DefMacro{res-teco-s3-CSNm-eval-assert-stmt/test-bs10-last-xmatch-top3}{14.55}
\DefMacro{res-teco-s3-CSNm-eval-assert-stmt/test-bs10-last-xmatch-top5}{18.20}
\DefMacro{res-teco-s3-CSNm-eval-assert-stmt/test-bs10-last-xmatch-top10}{23.57}
\DefMacro{res-teco-s3-CSNm-eval-assert-stmt/test-bs10-last-xmatch-top100}{23.57}
\DefMacro{res-teco-s3-CSNm-eval-assert-stmt/test-bs10-last-bleu}{39.90}
\DefMacro{res-teco-s3-CSNm-eval-assert-stmt/test-bs10-last-bleu-max}{58.09}
\DefMacro{res-teco-s3-CSNm-eval-assert-stmt/test-bs10-last-bleu-min}{18.99}
\DefMacro{res-teco-s3-CSNm-eval-assert-stmt/test-bs10-last-bleu-avg}{34.11}
\DefMacro{res-teco-s3-CSNm-eval-assert-stmt/test-bs10-last-code-bleu}{31.96}
\DefMacro{res-teco-s3-CSNm-eval-assert-stmt/test-bs10-last-code-bleu-max}{52.26}
\DefMacro{res-teco-s3-CSNm-eval-assert-stmt/test-bs10-last-code-bleu-min}{12.07}
\DefMacro{res-teco-s3-CSNm-eval-assert-stmt/test-bs10-last-code-bleu-avg}{27.00}
\DefMacro{res-teco-s3-CSNm-eval-assert-stmt/test-bs10-last-edit-sim}{67.20}
\DefMacro{res-teco-s3-CSNm-eval-assert-stmt/test-bs10-last-edit-sim-max}{80.08}
\DefMacro{res-teco-s3-CSNm-eval-assert-stmt/test-bs10-last-edit-sim-min}{49.64}
\DefMacro{res-teco-s3-CSNm-eval-assert-stmt/test-bs10-last-edit-sim-avg}{63.76}
\DefMacro{res-teco-s3-CSNm-eval-assert-stmt/test-bs10-last-rouge-p}{68.17}
\DefMacro{res-teco-s3-CSNm-eval-assert-stmt/test-bs10-last-rouge-p-max}{81.90}
\DefMacro{res-teco-s3-CSNm-eval-assert-stmt/test-bs10-last-rouge-p-min}{46.87}
\DefMacro{res-teco-s3-CSNm-eval-assert-stmt/test-bs10-last-rouge-p-avg}{62.11}
\DefMacro{res-teco-s3-CSNm-eval-assert-stmt/test-bs10-last-rouge-r}{68.81}
\DefMacro{res-teco-s3-CSNm-eval-assert-stmt/test-bs10-last-rouge-r-max}{80.64}
\DefMacro{res-teco-s3-CSNm-eval-assert-stmt/test-bs10-last-rouge-r-min}{53.75}
\DefMacro{res-teco-s3-CSNm-eval-assert-stmt/test-bs10-last-rouge-r-avg}{67.00}
\DefMacro{res-teco-s3-CSNm-eval-assert-stmt/test-bs10-last-rouge-f}{67.44}
\DefMacro{res-teco-s3-CSNm-eval-assert-stmt/test-bs10-last-rouge-f-max}{80.02}
\DefMacro{res-teco-s3-CSNm-eval-assert-stmt/test-bs10-last-rouge-f-min}{49.27}
\DefMacro{res-teco-s3-CSNm-eval-assert-stmt/test-bs10-last-rouge-f-avg}{63.33}
\DefMacro{res-teco-s3-CSNm-eval-assert-stmt/test-bs10-last-num-seq}{10.00}
\DefMacro{res-teco-s3-CSNm-eval-assert-stmt/test-bs10-last-compilable}{-1.00}
\DefMacro{res-teco-s3-CSNm-eval-assert-stmt/test-bs10-last-runnable}{-1.00}
\DefMacro{res-teco-s3-CSNm-eval-runnable-any-stmt/test-bs10-last-xmatch}{14.84}
\DefMacro{res-teco-s3-CSNm-eval-runnable-any-stmt/test-bs10-last-xmatch-top3}{20.60}
\DefMacro{res-teco-s3-CSNm-eval-runnable-any-stmt/test-bs10-last-xmatch-top5}{22.87}
\DefMacro{res-teco-s3-CSNm-eval-runnable-any-stmt/test-bs10-last-xmatch-top10}{25.76}
\DefMacro{res-teco-s3-CSNm-eval-runnable-any-stmt/test-bs10-last-xmatch-top100}{25.76}
\DefMacro{res-teco-s3-CSNm-eval-runnable-any-stmt/test-bs10-last-bleu}{39.77}
\DefMacro{res-teco-s3-CSNm-eval-runnable-any-stmt/test-bs10-last-bleu-max}{53.66}
\DefMacro{res-teco-s3-CSNm-eval-runnable-any-stmt/test-bs10-last-bleu-min}{17.71}
\DefMacro{res-teco-s3-CSNm-eval-runnable-any-stmt/test-bs10-last-bleu-avg}{31.18}
\DefMacro{res-teco-s3-CSNm-eval-runnable-any-stmt/test-bs10-last-code-bleu}{34.89}
\DefMacro{res-teco-s3-CSNm-eval-runnable-any-stmt/test-bs10-last-code-bleu-max}{51.25}
\DefMacro{res-teco-s3-CSNm-eval-runnable-any-stmt/test-bs10-last-code-bleu-min}{11.47}
\DefMacro{res-teco-s3-CSNm-eval-runnable-any-stmt/test-bs10-last-code-bleu-avg}{26.52}
\DefMacro{res-teco-s3-CSNm-eval-runnable-any-stmt/test-bs10-last-edit-sim}{61.83}
\DefMacro{res-teco-s3-CSNm-eval-runnable-any-stmt/test-bs10-last-edit-sim-max}{72.92}
\DefMacro{res-teco-s3-CSNm-eval-runnable-any-stmt/test-bs10-last-edit-sim-min}{43.96}
\DefMacro{res-teco-s3-CSNm-eval-runnable-any-stmt/test-bs10-last-edit-sim-avg}{56.84}
\DefMacro{res-teco-s3-CSNm-eval-runnable-any-stmt/test-bs10-last-rouge-p}{64.79}
\DefMacro{res-teco-s3-CSNm-eval-runnable-any-stmt/test-bs10-last-rouge-p-max}{76.78}
\DefMacro{res-teco-s3-CSNm-eval-runnable-any-stmt/test-bs10-last-rouge-p-min}{44.00}
\DefMacro{res-teco-s3-CSNm-eval-runnable-any-stmt/test-bs10-last-rouge-p-avg}{57.93}
\DefMacro{res-teco-s3-CSNm-eval-runnable-any-stmt/test-bs10-last-rouge-r}{65.18}
\DefMacro{res-teco-s3-CSNm-eval-runnable-any-stmt/test-bs10-last-rouge-r-max}{75.60}
\DefMacro{res-teco-s3-CSNm-eval-runnable-any-stmt/test-bs10-last-rouge-r-min}{49.96}
\DefMacro{res-teco-s3-CSNm-eval-runnable-any-stmt/test-bs10-last-rouge-r-avg}{62.59}
\DefMacro{res-teco-s3-CSNm-eval-runnable-any-stmt/test-bs10-last-rouge-f}{63.68}
\DefMacro{res-teco-s3-CSNm-eval-runnable-any-stmt/test-bs10-last-rouge-f-max}{74.59}
\DefMacro{res-teco-s3-CSNm-eval-runnable-any-stmt/test-bs10-last-rouge-f-min}{45.82}
\DefMacro{res-teco-s3-CSNm-eval-runnable-any-stmt/test-bs10-last-rouge-f-avg}{58.62}
\DefMacro{res-teco-s3-CSNm-eval-runnable-any-stmt/test-bs10-last-num-seq}{10.00}
\DefMacro{res-teco-s3-CSNm-eval-runnable-any-stmt/test-bs10-last-compilable}{-1.00}
\DefMacro{res-teco-s3-CSNm-eval-runnable-any-stmt/test-bs10-last-runnable}{-1.00}
\DefMacro{res-teco-s3-CSNm-eval-runnable-assert-stmt/test-bs10-last-xmatch}{9.09}
\DefMacro{res-teco-s3-CSNm-eval-runnable-assert-stmt/test-bs10-last-xmatch-top3}{14.73}
\DefMacro{res-teco-s3-CSNm-eval-runnable-assert-stmt/test-bs10-last-xmatch-top5}{18.39}
\DefMacro{res-teco-s3-CSNm-eval-runnable-assert-stmt/test-bs10-last-xmatch-top10}{23.89}
\DefMacro{res-teco-s3-CSNm-eval-runnable-assert-stmt/test-bs10-last-xmatch-top100}{23.89}
\DefMacro{res-teco-s3-CSNm-eval-runnable-assert-stmt/test-bs10-last-bleu}{41.13}
\DefMacro{res-teco-s3-CSNm-eval-runnable-assert-stmt/test-bs10-last-bleu-max}{58.36}
\DefMacro{res-teco-s3-CSNm-eval-runnable-assert-stmt/test-bs10-last-bleu-min}{19.49}
\DefMacro{res-teco-s3-CSNm-eval-runnable-assert-stmt/test-bs10-last-bleu-avg}{34.69}
\DefMacro{res-teco-s3-CSNm-eval-runnable-assert-stmt/test-bs10-last-code-bleu}{33.18}
\DefMacro{res-teco-s3-CSNm-eval-runnable-assert-stmt/test-bs10-last-code-bleu-max}{52.38}
\DefMacro{res-teco-s3-CSNm-eval-runnable-assert-stmt/test-bs10-last-code-bleu-min}{12.58}
\DefMacro{res-teco-s3-CSNm-eval-runnable-assert-stmt/test-bs10-last-code-bleu-avg}{27.62}
\DefMacro{res-teco-s3-CSNm-eval-runnable-assert-stmt/test-bs10-last-edit-sim}{67.82}
\DefMacro{res-teco-s3-CSNm-eval-runnable-assert-stmt/test-bs10-last-edit-sim-max}{80.12}
\DefMacro{res-teco-s3-CSNm-eval-runnable-assert-stmt/test-bs10-last-edit-sim-min}{49.87}
\DefMacro{res-teco-s3-CSNm-eval-runnable-assert-stmt/test-bs10-last-edit-sim-avg}{63.97}
\DefMacro{res-teco-s3-CSNm-eval-runnable-assert-stmt/test-bs10-last-rouge-p}{68.71}
\DefMacro{res-teco-s3-CSNm-eval-runnable-assert-stmt/test-bs10-last-rouge-p-max}{81.89}
\DefMacro{res-teco-s3-CSNm-eval-runnable-assert-stmt/test-bs10-last-rouge-p-min}{47.13}
\DefMacro{res-teco-s3-CSNm-eval-runnable-assert-stmt/test-bs10-last-rouge-p-avg}{62.38}
\DefMacro{res-teco-s3-CSNm-eval-runnable-assert-stmt/test-bs10-last-rouge-r}{69.73}
\DefMacro{res-teco-s3-CSNm-eval-runnable-assert-stmt/test-bs10-last-rouge-r-max}{81.01}
\DefMacro{res-teco-s3-CSNm-eval-runnable-assert-stmt/test-bs10-last-rouge-r-min}{54.21}
\DefMacro{res-teco-s3-CSNm-eval-runnable-assert-stmt/test-bs10-last-rouge-r-avg}{67.55}
\DefMacro{res-teco-s3-CSNm-eval-runnable-assert-stmt/test-bs10-last-rouge-f}{68.21}
\DefMacro{res-teco-s3-CSNm-eval-runnable-assert-stmt/test-bs10-last-rouge-f-max}{80.27}
\DefMacro{res-teco-s3-CSNm-eval-runnable-assert-stmt/test-bs10-last-rouge-f-min}{49.64}
\DefMacro{res-teco-s3-CSNm-eval-runnable-assert-stmt/test-bs10-last-rouge-f-avg}{63.75}
\DefMacro{res-teco-s3-CSNm-eval-runnable-assert-stmt/test-bs10-last-num-seq}{10.00}
\DefMacro{res-teco-s3-CSNm-eval-runnable-assert-stmt/test-bs10-last-compilable}{-1.00}
\DefMacro{res-teco-s3-CSNm-eval-runnable-assert-stmt/test-bs10-last-runnable}{-1.00}
\DefMacro{res-teco-s4-CSNm-eval-any-stmt/test-bs10-last-xmatch}{14.44}
\DefMacro{res-teco-s4-CSNm-eval-any-stmt/test-bs10-last-xmatch-top3}{20.32}
\DefMacro{res-teco-s4-CSNm-eval-any-stmt/test-bs10-last-xmatch-top5}{22.57}
\DefMacro{res-teco-s4-CSNm-eval-any-stmt/test-bs10-last-xmatch-top10}{25.55}
\DefMacro{res-teco-s4-CSNm-eval-any-stmt/test-bs10-last-xmatch-top100}{25.55}
\DefMacro{res-teco-s4-CSNm-eval-any-stmt/test-bs10-last-bleu}{39.39}
\DefMacro{res-teco-s4-CSNm-eval-any-stmt/test-bs10-last-bleu-max}{53.45}
\DefMacro{res-teco-s4-CSNm-eval-any-stmt/test-bs10-last-bleu-min}{17.55}
\DefMacro{res-teco-s4-CSNm-eval-any-stmt/test-bs10-last-bleu-avg}{30.95}
\DefMacro{res-teco-s4-CSNm-eval-any-stmt/test-bs10-last-code-bleu}{35.00}
\DefMacro{res-teco-s4-CSNm-eval-any-stmt/test-bs10-last-code-bleu-max}{51.41}
\DefMacro{res-teco-s4-CSNm-eval-any-stmt/test-bs10-last-code-bleu-min}{11.73}
\DefMacro{res-teco-s4-CSNm-eval-any-stmt/test-bs10-last-code-bleu-avg}{26.77}
\DefMacro{res-teco-s4-CSNm-eval-any-stmt/test-bs10-last-edit-sim}{61.63}
\DefMacro{res-teco-s4-CSNm-eval-any-stmt/test-bs10-last-edit-sim-max}{72.77}
\DefMacro{res-teco-s4-CSNm-eval-any-stmt/test-bs10-last-edit-sim-min}{44.02}
\DefMacro{res-teco-s4-CSNm-eval-any-stmt/test-bs10-last-edit-sim-avg}{56.69}
\DefMacro{res-teco-s4-CSNm-eval-any-stmt/test-bs10-last-rouge-p}{64.47}
\DefMacro{res-teco-s4-CSNm-eval-any-stmt/test-bs10-last-rouge-p-max}{76.61}
\DefMacro{res-teco-s4-CSNm-eval-any-stmt/test-bs10-last-rouge-p-min}{44.00}
\DefMacro{res-teco-s4-CSNm-eval-any-stmt/test-bs10-last-rouge-p-avg}{57.72}
\DefMacro{res-teco-s4-CSNm-eval-any-stmt/test-bs10-last-rouge-r}{65.00}
\DefMacro{res-teco-s4-CSNm-eval-any-stmt/test-bs10-last-rouge-r-max}{75.57}
\DefMacro{res-teco-s4-CSNm-eval-any-stmt/test-bs10-last-rouge-r-min}{50.06}
\DefMacro{res-teco-s4-CSNm-eval-any-stmt/test-bs10-last-rouge-r-avg}{62.55}
\DefMacro{res-teco-s4-CSNm-eval-any-stmt/test-bs10-last-rouge-f}{63.40}
\DefMacro{res-teco-s4-CSNm-eval-any-stmt/test-bs10-last-rouge-f-max}{74.47}
\DefMacro{res-teco-s4-CSNm-eval-any-stmt/test-bs10-last-rouge-f-min}{45.79}
\DefMacro{res-teco-s4-CSNm-eval-any-stmt/test-bs10-last-rouge-f-avg}{58.50}
\DefMacro{res-teco-s4-CSNm-eval-any-stmt/test-bs10-last-num-seq}{10.00}
\DefMacro{res-teco-s4-CSNm-eval-any-stmt/test-bs10-last-compilable}{-1.00}
\DefMacro{res-teco-s4-CSNm-eval-any-stmt/test-bs10-last-runnable}{-1.00}
\DefMacro{res-teco-s4-CSNm-eval-assert-stmt/test-bs10-last-xmatch}{8.14}
\DefMacro{res-teco-s4-CSNm-eval-assert-stmt/test-bs10-last-xmatch-top3}{13.94}
\DefMacro{res-teco-s4-CSNm-eval-assert-stmt/test-bs10-last-xmatch-top5}{17.93}
\DefMacro{res-teco-s4-CSNm-eval-assert-stmt/test-bs10-last-xmatch-top10}{23.84}
\DefMacro{res-teco-s4-CSNm-eval-assert-stmt/test-bs10-last-xmatch-top100}{23.84}
\DefMacro{res-teco-s4-CSNm-eval-assert-stmt/test-bs10-last-bleu}{38.54}
\DefMacro{res-teco-s4-CSNm-eval-assert-stmt/test-bs10-last-bleu-max}{57.35}
\DefMacro{res-teco-s4-CSNm-eval-assert-stmt/test-bs10-last-bleu-min}{18.04}
\DefMacro{res-teco-s4-CSNm-eval-assert-stmt/test-bs10-last-bleu-avg}{33.13}
\DefMacro{res-teco-s4-CSNm-eval-assert-stmt/test-bs10-last-code-bleu}{30.77}
\DefMacro{res-teco-s4-CSNm-eval-assert-stmt/test-bs10-last-code-bleu-max}{51.72}
\DefMacro{res-teco-s4-CSNm-eval-assert-stmt/test-bs10-last-code-bleu-min}{11.10}
\DefMacro{res-teco-s4-CSNm-eval-assert-stmt/test-bs10-last-code-bleu-avg}{26.20}
\DefMacro{res-teco-s4-CSNm-eval-assert-stmt/test-bs10-last-edit-sim}{65.59}
\DefMacro{res-teco-s4-CSNm-eval-assert-stmt/test-bs10-last-edit-sim-max}{79.11}
\DefMacro{res-teco-s4-CSNm-eval-assert-stmt/test-bs10-last-edit-sim-min}{48.47}
\DefMacro{res-teco-s4-CSNm-eval-assert-stmt/test-bs10-last-edit-sim-avg}{62.60}
\DefMacro{res-teco-s4-CSNm-eval-assert-stmt/test-bs10-last-rouge-p}{66.35}
\DefMacro{res-teco-s4-CSNm-eval-assert-stmt/test-bs10-last-rouge-p-max}{81.08}
\DefMacro{res-teco-s4-CSNm-eval-assert-stmt/test-bs10-last-rouge-p-min}{45.80}
\DefMacro{res-teco-s4-CSNm-eval-assert-stmt/test-bs10-last-rouge-p-avg}{61.22}
\DefMacro{res-teco-s4-CSNm-eval-assert-stmt/test-bs10-last-rouge-r}{67.61}
\DefMacro{res-teco-s4-CSNm-eval-assert-stmt/test-bs10-last-rouge-r-max}{80.32}
\DefMacro{res-teco-s4-CSNm-eval-assert-stmt/test-bs10-last-rouge-r-min}{52.67}
\DefMacro{res-teco-s4-CSNm-eval-assert-stmt/test-bs10-last-rouge-r-avg}{66.07}
\DefMacro{res-teco-s4-CSNm-eval-assert-stmt/test-bs10-last-rouge-f}{65.92}
\DefMacro{res-teco-s4-CSNm-eval-assert-stmt/test-bs10-last-rouge-f-max}{79.45}
\DefMacro{res-teco-s4-CSNm-eval-assert-stmt/test-bs10-last-rouge-f-min}{48.19}
\DefMacro{res-teco-s4-CSNm-eval-assert-stmt/test-bs10-last-rouge-f-avg}{62.43}
\DefMacro{res-teco-s4-CSNm-eval-assert-stmt/test-bs10-last-num-seq}{10.00}
\DefMacro{res-teco-s4-CSNm-eval-assert-stmt/test-bs10-last-compilable}{-1.00}
\DefMacro{res-teco-s4-CSNm-eval-assert-stmt/test-bs10-last-runnable}{-1.00}
\DefMacro{res-teco-s4-CSNm-eval-runnable-any-stmt/test-bs10-last-xmatch}{15.33}
\DefMacro{res-teco-s4-CSNm-eval-runnable-any-stmt/test-bs10-last-xmatch-top3}{21.54}
\DefMacro{res-teco-s4-CSNm-eval-runnable-any-stmt/test-bs10-last-xmatch-top5}{23.90}
\DefMacro{res-teco-s4-CSNm-eval-runnable-any-stmt/test-bs10-last-xmatch-top10}{26.96}
\DefMacro{res-teco-s4-CSNm-eval-runnable-any-stmt/test-bs10-last-xmatch-top100}{26.96}
\DefMacro{res-teco-s4-CSNm-eval-runnable-any-stmt/test-bs10-last-bleu}{40.43}
\DefMacro{res-teco-s4-CSNm-eval-runnable-any-stmt/test-bs10-last-bleu-max}{54.76}
\DefMacro{res-teco-s4-CSNm-eval-runnable-any-stmt/test-bs10-last-bleu-min}{17.63}
\DefMacro{res-teco-s4-CSNm-eval-runnable-any-stmt/test-bs10-last-bleu-avg}{31.47}
\DefMacro{res-teco-s4-CSNm-eval-runnable-any-stmt/test-bs10-last-code-bleu}{35.73}
\DefMacro{res-teco-s4-CSNm-eval-runnable-any-stmt/test-bs10-last-code-bleu-max}{52.51}
\DefMacro{res-teco-s4-CSNm-eval-runnable-any-stmt/test-bs10-last-code-bleu-min}{11.46}
\DefMacro{res-teco-s4-CSNm-eval-runnable-any-stmt/test-bs10-last-code-bleu-avg}{27.02}
\DefMacro{res-teco-s4-CSNm-eval-runnable-any-stmt/test-bs10-last-edit-sim}{62.31}
\DefMacro{res-teco-s4-CSNm-eval-runnable-any-stmt/test-bs10-last-edit-sim-max}{73.63}
\DefMacro{res-teco-s4-CSNm-eval-runnable-any-stmt/test-bs10-last-edit-sim-min}{44.02}
\DefMacro{res-teco-s4-CSNm-eval-runnable-any-stmt/test-bs10-last-edit-sim-avg}{57.07}
\DefMacro{res-teco-s4-CSNm-eval-runnable-any-stmt/test-bs10-last-rouge-p}{65.29}
\DefMacro{res-teco-s4-CSNm-eval-runnable-any-stmt/test-bs10-last-rouge-p-max}{77.50}
\DefMacro{res-teco-s4-CSNm-eval-runnable-any-stmt/test-bs10-last-rouge-p-min}{44.19}
\DefMacro{res-teco-s4-CSNm-eval-runnable-any-stmt/test-bs10-last-rouge-p-avg}{58.25}
\DefMacro{res-teco-s4-CSNm-eval-runnable-any-stmt/test-bs10-last-rouge-r}{65.74}
\DefMacro{res-teco-s4-CSNm-eval-runnable-any-stmt/test-bs10-last-rouge-r-max}{76.50}
\DefMacro{res-teco-s4-CSNm-eval-runnable-any-stmt/test-bs10-last-rouge-r-min}{50.23}
\DefMacro{res-teco-s4-CSNm-eval-runnable-any-stmt/test-bs10-last-rouge-r-avg}{63.15}
\DefMacro{res-teco-s4-CSNm-eval-runnable-any-stmt/test-bs10-last-rouge-f}{64.23}
\DefMacro{res-teco-s4-CSNm-eval-runnable-any-stmt/test-bs10-last-rouge-f-max}{75.45}
\DefMacro{res-teco-s4-CSNm-eval-runnable-any-stmt/test-bs10-last-rouge-f-min}{46.02}
\DefMacro{res-teco-s4-CSNm-eval-runnable-any-stmt/test-bs10-last-rouge-f-avg}{59.11}
\DefMacro{res-teco-s4-CSNm-eval-runnable-any-stmt/test-bs10-last-num-seq}{10.00}
\DefMacro{res-teco-s4-CSNm-eval-runnable-any-stmt/test-bs10-last-compilable}{-1.00}
\DefMacro{res-teco-s4-CSNm-eval-runnable-any-stmt/test-bs10-last-runnable}{-1.00}
\DefMacro{res-teco-s4-CSNm-eval-runnable-assert-stmt/test-bs10-last-xmatch}{8.37}
\DefMacro{res-teco-s4-CSNm-eval-runnable-assert-stmt/test-bs10-last-xmatch-top3}{14.51}
\DefMacro{res-teco-s4-CSNm-eval-runnable-assert-stmt/test-bs10-last-xmatch-top5}{18.58}
\DefMacro{res-teco-s4-CSNm-eval-runnable-assert-stmt/test-bs10-last-xmatch-top10}{24.31}
\DefMacro{res-teco-s4-CSNm-eval-runnable-assert-stmt/test-bs10-last-xmatch-top100}{24.31}
\DefMacro{res-teco-s4-CSNm-eval-runnable-assert-stmt/test-bs10-last-bleu}{39.68}
\DefMacro{res-teco-s4-CSNm-eval-runnable-assert-stmt/test-bs10-last-bleu-max}{57.65}
\DefMacro{res-teco-s4-CSNm-eval-runnable-assert-stmt/test-bs10-last-bleu-min}{18.44}
\DefMacro{res-teco-s4-CSNm-eval-runnable-assert-stmt/test-bs10-last-bleu-avg}{33.69}
\DefMacro{res-teco-s4-CSNm-eval-runnable-assert-stmt/test-bs10-last-code-bleu}{31.83}
\DefMacro{res-teco-s4-CSNm-eval-runnable-assert-stmt/test-bs10-last-code-bleu-max}{51.87}
\DefMacro{res-teco-s4-CSNm-eval-runnable-assert-stmt/test-bs10-last-code-bleu-min}{11.55}
\DefMacro{res-teco-s4-CSNm-eval-runnable-assert-stmt/test-bs10-last-code-bleu-avg}{26.80}
\DefMacro{res-teco-s4-CSNm-eval-runnable-assert-stmt/test-bs10-last-edit-sim}{66.21}
\DefMacro{res-teco-s4-CSNm-eval-runnable-assert-stmt/test-bs10-last-edit-sim-max}{79.23}
\DefMacro{res-teco-s4-CSNm-eval-runnable-assert-stmt/test-bs10-last-edit-sim-min}{48.56}
\DefMacro{res-teco-s4-CSNm-eval-runnable-assert-stmt/test-bs10-last-edit-sim-avg}{62.78}
\DefMacro{res-teco-s4-CSNm-eval-runnable-assert-stmt/test-bs10-last-rouge-p}{67.08}
\DefMacro{res-teco-s4-CSNm-eval-runnable-assert-stmt/test-bs10-last-rouge-p-max}{80.94}
\DefMacro{res-teco-s4-CSNm-eval-runnable-assert-stmt/test-bs10-last-rouge-p-min}{45.94}
\DefMacro{res-teco-s4-CSNm-eval-runnable-assert-stmt/test-bs10-last-rouge-p-avg}{61.41}
\DefMacro{res-teco-s4-CSNm-eval-runnable-assert-stmt/test-bs10-last-rouge-r}{68.49}
\DefMacro{res-teco-s4-CSNm-eval-runnable-assert-stmt/test-bs10-last-rouge-r-max}{80.70}
\DefMacro{res-teco-s4-CSNm-eval-runnable-assert-stmt/test-bs10-last-rouge-r-min}{53.03}
\DefMacro{res-teco-s4-CSNm-eval-runnable-assert-stmt/test-bs10-last-rouge-r-avg}{66.66}
\DefMacro{res-teco-s4-CSNm-eval-runnable-assert-stmt/test-bs10-last-rouge-f}{66.75}
\DefMacro{res-teco-s4-CSNm-eval-runnable-assert-stmt/test-bs10-last-rouge-f-max}{79.59}
\DefMacro{res-teco-s4-CSNm-eval-runnable-assert-stmt/test-bs10-last-rouge-f-min}{48.43}
\DefMacro{res-teco-s4-CSNm-eval-runnable-assert-stmt/test-bs10-last-rouge-f-avg}{62.82}
\DefMacro{res-teco-s4-CSNm-eval-runnable-assert-stmt/test-bs10-last-num-seq}{10.00}
\DefMacro{res-teco-s4-CSNm-eval-runnable-assert-stmt/test-bs10-last-compilable}{-1.00}
\DefMacro{res-teco-s4-CSNm-eval-runnable-assert-stmt/test-bs10-last-runnable}{-1.00}
\DefMacro{res-teco-s5-CSNm-eval-any-stmt/test-bs10-last-xmatch}{14.05}
\DefMacro{res-teco-s5-CSNm-eval-any-stmt/test-bs10-last-xmatch-top3}{19.45}
\DefMacro{res-teco-s5-CSNm-eval-any-stmt/test-bs10-last-xmatch-top5}{21.62}
\DefMacro{res-teco-s5-CSNm-eval-any-stmt/test-bs10-last-xmatch-top10}{24.78}
\DefMacro{res-teco-s5-CSNm-eval-any-stmt/test-bs10-last-xmatch-top100}{24.78}
\DefMacro{res-teco-s5-CSNm-eval-any-stmt/test-bs10-last-bleu}{38.74}
\DefMacro{res-teco-s5-CSNm-eval-any-stmt/test-bs10-last-bleu-max}{52.98}
\DefMacro{res-teco-s5-CSNm-eval-any-stmt/test-bs10-last-bleu-min}{17.43}
\DefMacro{res-teco-s5-CSNm-eval-any-stmt/test-bs10-last-bleu-avg}{30.75}
\DefMacro{res-teco-s5-CSNm-eval-any-stmt/test-bs10-last-code-bleu}{34.34}
\DefMacro{res-teco-s5-CSNm-eval-any-stmt/test-bs10-last-code-bleu-max}{50.94}
\DefMacro{res-teco-s5-CSNm-eval-any-stmt/test-bs10-last-code-bleu-min}{11.58}
\DefMacro{res-teco-s5-CSNm-eval-any-stmt/test-bs10-last-code-bleu-avg}{26.46}
\DefMacro{res-teco-s5-CSNm-eval-any-stmt/test-bs10-last-edit-sim}{61.26}
\DefMacro{res-teco-s5-CSNm-eval-any-stmt/test-bs10-last-edit-sim-max}{72.49}
\DefMacro{res-teco-s5-CSNm-eval-any-stmt/test-bs10-last-edit-sim-min}{43.77}
\DefMacro{res-teco-s5-CSNm-eval-any-stmt/test-bs10-last-edit-sim-avg}{56.59}
\DefMacro{res-teco-s5-CSNm-eval-any-stmt/test-bs10-last-rouge-p}{64.07}
\DefMacro{res-teco-s5-CSNm-eval-any-stmt/test-bs10-last-rouge-p-max}{76.18}
\DefMacro{res-teco-s5-CSNm-eval-any-stmt/test-bs10-last-rouge-p-min}{43.45}
\DefMacro{res-teco-s5-CSNm-eval-any-stmt/test-bs10-last-rouge-p-avg}{57.28}
\DefMacro{res-teco-s5-CSNm-eval-any-stmt/test-bs10-last-rouge-r}{64.64}
\DefMacro{res-teco-s5-CSNm-eval-any-stmt/test-bs10-last-rouge-r-max}{75.06}
\DefMacro{res-teco-s5-CSNm-eval-any-stmt/test-bs10-last-rouge-r-min}{49.86}
\DefMacro{res-teco-s5-CSNm-eval-any-stmt/test-bs10-last-rouge-r-avg}{62.23}
\DefMacro{res-teco-s5-CSNm-eval-any-stmt/test-bs10-last-rouge-f}{63.00}
\DefMacro{res-teco-s5-CSNm-eval-any-stmt/test-bs10-last-rouge-f-max}{73.99}
\DefMacro{res-teco-s5-CSNm-eval-any-stmt/test-bs10-last-rouge-f-min}{45.40}
\DefMacro{res-teco-s5-CSNm-eval-any-stmt/test-bs10-last-rouge-f-avg}{58.09}
\DefMacro{res-teco-s5-CSNm-eval-any-stmt/test-bs10-last-num-seq}{10.00}
\DefMacro{res-teco-s5-CSNm-eval-any-stmt/test-bs10-last-compilable}{-1.00}
\DefMacro{res-teco-s5-CSNm-eval-any-stmt/test-bs10-last-runnable}{-1.00}
\DefMacro{res-teco-s5-CSNm-eval-assert-stmt/test-bs10-last-xmatch}{8.43}
\DefMacro{res-teco-s5-CSNm-eval-assert-stmt/test-bs10-last-xmatch-top3}{13.96}
\DefMacro{res-teco-s5-CSNm-eval-assert-stmt/test-bs10-last-xmatch-top5}{17.63}
\DefMacro{res-teco-s5-CSNm-eval-assert-stmt/test-bs10-last-xmatch-top10}{24.10}
\DefMacro{res-teco-s5-CSNm-eval-assert-stmt/test-bs10-last-xmatch-top100}{24.10}
\DefMacro{res-teco-s5-CSNm-eval-assert-stmt/test-bs10-last-bleu}{38.67}
\DefMacro{res-teco-s5-CSNm-eval-assert-stmt/test-bs10-last-bleu-max}{58.31}
\DefMacro{res-teco-s5-CSNm-eval-assert-stmt/test-bs10-last-bleu-min}{18.25}
\DefMacro{res-teco-s5-CSNm-eval-assert-stmt/test-bs10-last-bleu-avg}{33.78}
\DefMacro{res-teco-s5-CSNm-eval-assert-stmt/test-bs10-last-code-bleu}{30.85}
\DefMacro{res-teco-s5-CSNm-eval-assert-stmt/test-bs10-last-code-bleu-max}{52.75}
\DefMacro{res-teco-s5-CSNm-eval-assert-stmt/test-bs10-last-code-bleu-min}{11.21}
\DefMacro{res-teco-s5-CSNm-eval-assert-stmt/test-bs10-last-code-bleu-avg}{26.73}
\DefMacro{res-teco-s5-CSNm-eval-assert-stmt/test-bs10-last-edit-sim}{66.17}
\DefMacro{res-teco-s5-CSNm-eval-assert-stmt/test-bs10-last-edit-sim-max}{79.81}
\DefMacro{res-teco-s5-CSNm-eval-assert-stmt/test-bs10-last-edit-sim-min}{48.84}
\DefMacro{res-teco-s5-CSNm-eval-assert-stmt/test-bs10-last-edit-sim-avg}{63.27}
\DefMacro{res-teco-s5-CSNm-eval-assert-stmt/test-bs10-last-rouge-p}{67.08}
\DefMacro{res-teco-s5-CSNm-eval-assert-stmt/test-bs10-last-rouge-p-max}{81.83}
\DefMacro{res-teco-s5-CSNm-eval-assert-stmt/test-bs10-last-rouge-p-min}{46.11}
\DefMacro{res-teco-s5-CSNm-eval-assert-stmt/test-bs10-last-rouge-p-avg}{61.74}
\DefMacro{res-teco-s5-CSNm-eval-assert-stmt/test-bs10-last-rouge-r}{67.59}
\DefMacro{res-teco-s5-CSNm-eval-assert-stmt/test-bs10-last-rouge-r-max}{80.35}
\DefMacro{res-teco-s5-CSNm-eval-assert-stmt/test-bs10-last-rouge-r-min}{53.03}
\DefMacro{res-teco-s5-CSNm-eval-assert-stmt/test-bs10-last-rouge-r-avg}{66.38}
\DefMacro{res-teco-s5-CSNm-eval-assert-stmt/test-bs10-last-rouge-f}{66.23}
\DefMacro{res-teco-s5-CSNm-eval-assert-stmt/test-bs10-last-rouge-f-max}{79.82}
\DefMacro{res-teco-s5-CSNm-eval-assert-stmt/test-bs10-last-rouge-f-min}{48.59}
\DefMacro{res-teco-s5-CSNm-eval-assert-stmt/test-bs10-last-rouge-f-avg}{62.82}
\DefMacro{res-teco-s5-CSNm-eval-assert-stmt/test-bs10-last-num-seq}{10.00}
\DefMacro{res-teco-s5-CSNm-eval-assert-stmt/test-bs10-last-compilable}{-1.00}
\DefMacro{res-teco-s5-CSNm-eval-assert-stmt/test-bs10-last-runnable}{-1.00}
\DefMacro{res-teco-s5-CSNm-eval-runnable-any-stmt/test-bs10-last-xmatch}{14.87}
\DefMacro{res-teco-s5-CSNm-eval-runnable-any-stmt/test-bs10-last-xmatch-top3}{20.47}
\DefMacro{res-teco-s5-CSNm-eval-runnable-any-stmt/test-bs10-last-xmatch-top5}{22.70}
\DefMacro{res-teco-s5-CSNm-eval-runnable-any-stmt/test-bs10-last-xmatch-top10}{26.13}
\DefMacro{res-teco-s5-CSNm-eval-runnable-any-stmt/test-bs10-last-xmatch-top100}{26.13}
\DefMacro{res-teco-s5-CSNm-eval-runnable-any-stmt/test-bs10-last-bleu}{39.67}
\DefMacro{res-teco-s5-CSNm-eval-runnable-any-stmt/test-bs10-last-bleu-max}{54.22}
\DefMacro{res-teco-s5-CSNm-eval-runnable-any-stmt/test-bs10-last-bleu-min}{17.54}
\DefMacro{res-teco-s5-CSNm-eval-runnable-any-stmt/test-bs10-last-bleu-avg}{31.27}
\DefMacro{res-teco-s5-CSNm-eval-runnable-any-stmt/test-bs10-last-code-bleu}{35.01}
\DefMacro{res-teco-s5-CSNm-eval-runnable-any-stmt/test-bs10-last-code-bleu-max}{51.97}
\DefMacro{res-teco-s5-CSNm-eval-runnable-any-stmt/test-bs10-last-code-bleu-min}{11.33}
\DefMacro{res-teco-s5-CSNm-eval-runnable-any-stmt/test-bs10-last-code-bleu-avg}{26.69}
\DefMacro{res-teco-s5-CSNm-eval-runnable-any-stmt/test-bs10-last-edit-sim}{61.88}
\DefMacro{res-teco-s5-CSNm-eval-runnable-any-stmt/test-bs10-last-edit-sim-max}{73.31}
\DefMacro{res-teco-s5-CSNm-eval-runnable-any-stmt/test-bs10-last-edit-sim-min}{43.85}
\DefMacro{res-teco-s5-CSNm-eval-runnable-any-stmt/test-bs10-last-edit-sim-avg}{57.00}
\DefMacro{res-teco-s5-CSNm-eval-runnable-any-stmt/test-bs10-last-rouge-p}{64.82}
\DefMacro{res-teco-s5-CSNm-eval-runnable-any-stmt/test-bs10-last-rouge-p-max}{77.03}
\DefMacro{res-teco-s5-CSNm-eval-runnable-any-stmt/test-bs10-last-rouge-p-min}{43.67}
\DefMacro{res-teco-s5-CSNm-eval-runnable-any-stmt/test-bs10-last-rouge-p-avg}{57.79}
\DefMacro{res-teco-s5-CSNm-eval-runnable-any-stmt/test-bs10-last-rouge-r}{65.35}
\DefMacro{res-teco-s5-CSNm-eval-runnable-any-stmt/test-bs10-last-rouge-r-max}{75.93}
\DefMacro{res-teco-s5-CSNm-eval-runnable-any-stmt/test-bs10-last-rouge-r-min}{50.11}
\DefMacro{res-teco-s5-CSNm-eval-runnable-any-stmt/test-bs10-last-rouge-r-avg}{62.84}
\DefMacro{res-teco-s5-CSNm-eval-runnable-any-stmt/test-bs10-last-rouge-f}{63.77}
\DefMacro{res-teco-s5-CSNm-eval-runnable-any-stmt/test-bs10-last-rouge-f-max}{74.90}
\DefMacro{res-teco-s5-CSNm-eval-runnable-any-stmt/test-bs10-last-rouge-f-min}{45.68}
\DefMacro{res-teco-s5-CSNm-eval-runnable-any-stmt/test-bs10-last-rouge-f-avg}{58.68}
\DefMacro{res-teco-s5-CSNm-eval-runnable-any-stmt/test-bs10-last-num-seq}{10.00}
\DefMacro{res-teco-s5-CSNm-eval-runnable-any-stmt/test-bs10-last-compilable}{-1.00}
\DefMacro{res-teco-s5-CSNm-eval-runnable-any-stmt/test-bs10-last-runnable}{-1.00}
\DefMacro{res-teco-s5-CSNm-eval-runnable-assert-stmt/test-bs10-last-xmatch}{8.63}
\DefMacro{res-teco-s5-CSNm-eval-runnable-assert-stmt/test-bs10-last-xmatch-top3}{14.04}
\DefMacro{res-teco-s5-CSNm-eval-runnable-assert-stmt/test-bs10-last-xmatch-top5}{17.63}
\DefMacro{res-teco-s5-CSNm-eval-runnable-assert-stmt/test-bs10-last-xmatch-top10}{24.56}
\DefMacro{res-teco-s5-CSNm-eval-runnable-assert-stmt/test-bs10-last-xmatch-top100}{24.56}
\DefMacro{res-teco-s5-CSNm-eval-runnable-assert-stmt/test-bs10-last-bleu}{39.66}
\DefMacro{res-teco-s5-CSNm-eval-runnable-assert-stmt/test-bs10-last-bleu-max}{58.55}
\DefMacro{res-teco-s5-CSNm-eval-runnable-assert-stmt/test-bs10-last-bleu-min}{18.71}
\DefMacro{res-teco-s5-CSNm-eval-runnable-assert-stmt/test-bs10-last-bleu-avg}{34.31}
\DefMacro{res-teco-s5-CSNm-eval-runnable-assert-stmt/test-bs10-last-code-bleu}{31.76}
\DefMacro{res-teco-s5-CSNm-eval-runnable-assert-stmt/test-bs10-last-code-bleu-max}{52.86}
\DefMacro{res-teco-s5-CSNm-eval-runnable-assert-stmt/test-bs10-last-code-bleu-min}{11.63}
\DefMacro{res-teco-s5-CSNm-eval-runnable-assert-stmt/test-bs10-last-code-bleu-avg}{27.26}
\DefMacro{res-teco-s5-CSNm-eval-runnable-assert-stmt/test-bs10-last-edit-sim}{66.64}
\DefMacro{res-teco-s5-CSNm-eval-runnable-assert-stmt/test-bs10-last-edit-sim-max}{79.90}
\DefMacro{res-teco-s5-CSNm-eval-runnable-assert-stmt/test-bs10-last-edit-sim-min}{49.02}
\DefMacro{res-teco-s5-CSNm-eval-runnable-assert-stmt/test-bs10-last-edit-sim-avg}{63.46}
\DefMacro{res-teco-s5-CSNm-eval-runnable-assert-stmt/test-bs10-last-rouge-p}{67.41}
\DefMacro{res-teco-s5-CSNm-eval-runnable-assert-stmt/test-bs10-last-rouge-p-max}{81.77}
\DefMacro{res-teco-s5-CSNm-eval-runnable-assert-stmt/test-bs10-last-rouge-p-min}{46.25}
\DefMacro{res-teco-s5-CSNm-eval-runnable-assert-stmt/test-bs10-last-rouge-p-avg}{61.90}
\DefMacro{res-teco-s5-CSNm-eval-runnable-assert-stmt/test-bs10-last-rouge-r}{68.34}
\DefMacro{res-teco-s5-CSNm-eval-runnable-assert-stmt/test-bs10-last-rouge-r-max}{80.71}
\DefMacro{res-teco-s5-CSNm-eval-runnable-assert-stmt/test-bs10-last-rouge-r-min}{53.42}
\DefMacro{res-teco-s5-CSNm-eval-runnable-assert-stmt/test-bs10-last-rouge-r-avg}{66.88}
\DefMacro{res-teco-s5-CSNm-eval-runnable-assert-stmt/test-bs10-last-rouge-f}{66.80}
\DefMacro{res-teco-s5-CSNm-eval-runnable-assert-stmt/test-bs10-last-rouge-f-max}{80.00}
\DefMacro{res-teco-s5-CSNm-eval-runnable-assert-stmt/test-bs10-last-rouge-f-min}{48.83}
\DefMacro{res-teco-s5-CSNm-eval-runnable-assert-stmt/test-bs10-last-rouge-f-avg}{63.17}
\DefMacro{res-teco-s5-CSNm-eval-runnable-assert-stmt/test-bs10-last-num-seq}{10.00}
\DefMacro{res-teco-s5-CSNm-eval-runnable-assert-stmt/test-bs10-last-compilable}{-1.00}
\DefMacro{res-teco-s5-CSNm-eval-runnable-assert-stmt/test-bs10-last-runnable}{-1.00}
\DefMacro{res-teco-s6-CSNm-eval-any-stmt/test-bs10-last-xmatch}{14.13}
\DefMacro{res-teco-s6-CSNm-eval-any-stmt/test-bs10-last-xmatch-top3}{19.77}
\DefMacro{res-teco-s6-CSNm-eval-any-stmt/test-bs10-last-xmatch-top5}{21.97}
\DefMacro{res-teco-s6-CSNm-eval-any-stmt/test-bs10-last-xmatch-top10}{24.74}
\DefMacro{res-teco-s6-CSNm-eval-any-stmt/test-bs10-last-xmatch-top100}{24.74}
\DefMacro{res-teco-s6-CSNm-eval-any-stmt/test-bs10-last-bleu}{38.70}
\DefMacro{res-teco-s6-CSNm-eval-any-stmt/test-bs10-last-bleu-max}{52.39}
\DefMacro{res-teco-s6-CSNm-eval-any-stmt/test-bs10-last-bleu-min}{17.36}
\DefMacro{res-teco-s6-CSNm-eval-any-stmt/test-bs10-last-bleu-avg}{30.44}
\DefMacro{res-teco-s6-CSNm-eval-any-stmt/test-bs10-last-code-bleu}{34.36}
\DefMacro{res-teco-s6-CSNm-eval-any-stmt/test-bs10-last-code-bleu-max}{50.43}
\DefMacro{res-teco-s6-CSNm-eval-any-stmt/test-bs10-last-code-bleu-min}{11.62}
\DefMacro{res-teco-s6-CSNm-eval-any-stmt/test-bs10-last-code-bleu-avg}{26.28}
\DefMacro{res-teco-s6-CSNm-eval-any-stmt/test-bs10-last-edit-sim}{60.86}
\DefMacro{res-teco-s6-CSNm-eval-any-stmt/test-bs10-last-edit-sim-max}{71.89}
\DefMacro{res-teco-s6-CSNm-eval-any-stmt/test-bs10-last-edit-sim-min}{43.48}
\DefMacro{res-teco-s6-CSNm-eval-any-stmt/test-bs10-last-edit-sim-avg}{56.06}
\DefMacro{res-teco-s6-CSNm-eval-any-stmt/test-bs10-last-rouge-p}{63.57}
\DefMacro{res-teco-s6-CSNm-eval-any-stmt/test-bs10-last-rouge-p-max}{75.57}
\DefMacro{res-teco-s6-CSNm-eval-any-stmt/test-bs10-last-rouge-p-min}{43.29}
\DefMacro{res-teco-s6-CSNm-eval-any-stmt/test-bs10-last-rouge-p-avg}{56.86}
\DefMacro{res-teco-s6-CSNm-eval-any-stmt/test-bs10-last-rouge-r}{64.09}
\DefMacro{res-teco-s6-CSNm-eval-any-stmt/test-bs10-last-rouge-r-max}{74.36}
\DefMacro{res-teco-s6-CSNm-eval-any-stmt/test-bs10-last-rouge-r-min}{49.31}
\DefMacro{res-teco-s6-CSNm-eval-any-stmt/test-bs10-last-rouge-r-avg}{61.48}
\DefMacro{res-teco-s6-CSNm-eval-any-stmt/test-bs10-last-rouge-f}{62.52}
\DefMacro{res-teco-s6-CSNm-eval-any-stmt/test-bs10-last-rouge-f-max}{73.35}
\DefMacro{res-teco-s6-CSNm-eval-any-stmt/test-bs10-last-rouge-f-min}{45.09}
\DefMacro{res-teco-s6-CSNm-eval-any-stmt/test-bs10-last-rouge-f-avg}{57.53}
\DefMacro{res-teco-s6-CSNm-eval-any-stmt/test-bs10-last-num-seq}{10.00}
\DefMacro{res-teco-s6-CSNm-eval-any-stmt/test-bs10-last-compilable}{-1.00}
\DefMacro{res-teco-s6-CSNm-eval-any-stmt/test-bs10-last-runnable}{-1.00}
\DefMacro{res-teco-s6-CSNm-eval-assert-stmt/test-bs10-last-xmatch}{9.81}
\DefMacro{res-teco-s6-CSNm-eval-assert-stmt/test-bs10-last-xmatch-top3}{16.09}
\DefMacro{res-teco-s6-CSNm-eval-assert-stmt/test-bs10-last-xmatch-top5}{20.23}
\DefMacro{res-teco-s6-CSNm-eval-assert-stmt/test-bs10-last-xmatch-top10}{25.47}
\DefMacro{res-teco-s6-CSNm-eval-assert-stmt/test-bs10-last-xmatch-top100}{25.47}
\DefMacro{res-teco-s6-CSNm-eval-assert-stmt/test-bs10-last-bleu}{39.88}
\DefMacro{res-teco-s6-CSNm-eval-assert-stmt/test-bs10-last-bleu-max}{58.61}
\DefMacro{res-teco-s6-CSNm-eval-assert-stmt/test-bs10-last-bleu-min}{18.46}
\DefMacro{res-teco-s6-CSNm-eval-assert-stmt/test-bs10-last-bleu-avg}{33.84}
\DefMacro{res-teco-s6-CSNm-eval-assert-stmt/test-bs10-last-code-bleu}{32.20}
\DefMacro{res-teco-s6-CSNm-eval-assert-stmt/test-bs10-last-code-bleu-max}{53.28}
\DefMacro{res-teco-s6-CSNm-eval-assert-stmt/test-bs10-last-code-bleu-min}{11.43}
\DefMacro{res-teco-s6-CSNm-eval-assert-stmt/test-bs10-last-code-bleu-avg}{26.83}
\DefMacro{res-teco-s6-CSNm-eval-assert-stmt/test-bs10-last-edit-sim}{66.89}
\DefMacro{res-teco-s6-CSNm-eval-assert-stmt/test-bs10-last-edit-sim-max}{79.94}
\DefMacro{res-teco-s6-CSNm-eval-assert-stmt/test-bs10-last-edit-sim-min}{48.94}
\DefMacro{res-teco-s6-CSNm-eval-assert-stmt/test-bs10-last-edit-sim-avg}{63.36}
\DefMacro{res-teco-s6-CSNm-eval-assert-stmt/test-bs10-last-rouge-p}{67.92}
\DefMacro{res-teco-s6-CSNm-eval-assert-stmt/test-bs10-last-rouge-p-max}{81.93}
\DefMacro{res-teco-s6-CSNm-eval-assert-stmt/test-bs10-last-rouge-p-min}{46.29}
\DefMacro{res-teco-s6-CSNm-eval-assert-stmt/test-bs10-last-rouge-p-avg}{61.72}
\DefMacro{res-teco-s6-CSNm-eval-assert-stmt/test-bs10-last-rouge-r}{68.15}
\DefMacro{res-teco-s6-CSNm-eval-assert-stmt/test-bs10-last-rouge-r-max}{80.44}
\DefMacro{res-teco-s6-CSNm-eval-assert-stmt/test-bs10-last-rouge-r-min}{52.90}
\DefMacro{res-teco-s6-CSNm-eval-assert-stmt/test-bs10-last-rouge-r-avg}{66.19}
\DefMacro{res-teco-s6-CSNm-eval-assert-stmt/test-bs10-last-rouge-f}{66.97}
\DefMacro{res-teco-s6-CSNm-eval-assert-stmt/test-bs10-last-rouge-f-max}{79.97}
\DefMacro{res-teco-s6-CSNm-eval-assert-stmt/test-bs10-last-rouge-f-min}{48.57}
\DefMacro{res-teco-s6-CSNm-eval-assert-stmt/test-bs10-last-rouge-f-avg}{62.75}
\DefMacro{res-teco-s6-CSNm-eval-assert-stmt/test-bs10-last-num-seq}{10.00}
\DefMacro{res-teco-s6-CSNm-eval-assert-stmt/test-bs10-last-compilable}{-1.00}
\DefMacro{res-teco-s6-CSNm-eval-assert-stmt/test-bs10-last-runnable}{-1.00}
\DefMacro{res-teco-s6-CSNm-eval-runnable-any-stmt/test-bs10-last-xmatch}{14.76}
\DefMacro{res-teco-s6-CSNm-eval-runnable-any-stmt/test-bs10-last-xmatch-top3}{20.70}
\DefMacro{res-teco-s6-CSNm-eval-runnable-any-stmt/test-bs10-last-xmatch-top5}{22.99}
\DefMacro{res-teco-s6-CSNm-eval-runnable-any-stmt/test-bs10-last-xmatch-top10}{25.89}
\DefMacro{res-teco-s6-CSNm-eval-runnable-any-stmt/test-bs10-last-xmatch-top100}{25.89}
\DefMacro{res-teco-s6-CSNm-eval-runnable-any-stmt/test-bs10-last-bleu}{39.47}
\DefMacro{res-teco-s6-CSNm-eval-runnable-any-stmt/test-bs10-last-bleu-max}{53.55}
\DefMacro{res-teco-s6-CSNm-eval-runnable-any-stmt/test-bs10-last-bleu-min}{17.40}
\DefMacro{res-teco-s6-CSNm-eval-runnable-any-stmt/test-bs10-last-bleu-avg}{30.89}
\DefMacro{res-teco-s6-CSNm-eval-runnable-any-stmt/test-bs10-last-code-bleu}{34.83}
\DefMacro{res-teco-s6-CSNm-eval-runnable-any-stmt/test-bs10-last-code-bleu-max}{51.36}
\DefMacro{res-teco-s6-CSNm-eval-runnable-any-stmt/test-bs10-last-code-bleu-min}{11.33}
\DefMacro{res-teco-s6-CSNm-eval-runnable-any-stmt/test-bs10-last-code-bleu-avg}{26.41}
\DefMacro{res-teco-s6-CSNm-eval-runnable-any-stmt/test-bs10-last-edit-sim}{61.34}
\DefMacro{res-teco-s6-CSNm-eval-runnable-any-stmt/test-bs10-last-edit-sim-max}{72.67}
\DefMacro{res-teco-s6-CSNm-eval-runnable-any-stmt/test-bs10-last-edit-sim-min}{43.44}
\DefMacro{res-teco-s6-CSNm-eval-runnable-any-stmt/test-bs10-last-edit-sim-avg}{56.37}
\DefMacro{res-teco-s6-CSNm-eval-runnable-any-stmt/test-bs10-last-rouge-p}{64.17}
\DefMacro{res-teco-s6-CSNm-eval-runnable-any-stmt/test-bs10-last-rouge-p-max}{76.39}
\DefMacro{res-teco-s6-CSNm-eval-runnable-any-stmt/test-bs10-last-rouge-p-min}{43.39}
\DefMacro{res-teco-s6-CSNm-eval-runnable-any-stmt/test-bs10-last-rouge-p-avg}{57.28}
\DefMacro{res-teco-s6-CSNm-eval-runnable-any-stmt/test-bs10-last-rouge-r}{64.64}
\DefMacro{res-teco-s6-CSNm-eval-runnable-any-stmt/test-bs10-last-rouge-r-max}{75.15}
\DefMacro{res-teco-s6-CSNm-eval-runnable-any-stmt/test-bs10-last-rouge-r-min}{49.41}
\DefMacro{res-teco-s6-CSNm-eval-runnable-any-stmt/test-bs10-last-rouge-r-avg}{61.96}
\DefMacro{res-teco-s6-CSNm-eval-runnable-any-stmt/test-bs10-last-rouge-f}{63.15}
\DefMacro{res-teco-s6-CSNm-eval-runnable-any-stmt/test-bs10-last-rouge-f-max}{74.22}
\DefMacro{res-teco-s6-CSNm-eval-runnable-any-stmt/test-bs10-last-rouge-f-min}{45.23}
\DefMacro{res-teco-s6-CSNm-eval-runnable-any-stmt/test-bs10-last-rouge-f-avg}{58.02}
\DefMacro{res-teco-s6-CSNm-eval-runnable-any-stmt/test-bs10-last-num-seq}{10.00}
\DefMacro{res-teco-s6-CSNm-eval-runnable-any-stmt/test-bs10-last-compilable}{-1.00}
\DefMacro{res-teco-s6-CSNm-eval-runnable-any-stmt/test-bs10-last-runnable}{-1.00}
\DefMacro{res-teco-s6-CSNm-eval-runnable-assert-stmt/test-bs10-last-xmatch}{9.26}
\DefMacro{res-teco-s6-CSNm-eval-runnable-assert-stmt/test-bs10-last-xmatch-top3}{15.48}
\DefMacro{res-teco-s6-CSNm-eval-runnable-assert-stmt/test-bs10-last-xmatch-top5}{19.74}
\DefMacro{res-teco-s6-CSNm-eval-runnable-assert-stmt/test-bs10-last-xmatch-top10}{25.19}
\DefMacro{res-teco-s6-CSNm-eval-runnable-assert-stmt/test-bs10-last-xmatch-top100}{25.19}
\DefMacro{res-teco-s6-CSNm-eval-runnable-assert-stmt/test-bs10-last-bleu}{40.28}
\DefMacro{res-teco-s6-CSNm-eval-runnable-assert-stmt/test-bs10-last-bleu-max}{58.48}
\DefMacro{res-teco-s6-CSNm-eval-runnable-assert-stmt/test-bs10-last-bleu-min}{18.79}
\DefMacro{res-teco-s6-CSNm-eval-runnable-assert-stmt/test-bs10-last-bleu-avg}{34.10}
\DefMacro{res-teco-s6-CSNm-eval-runnable-assert-stmt/test-bs10-last-code-bleu}{32.45}
\DefMacro{res-teco-s6-CSNm-eval-runnable-assert-stmt/test-bs10-last-code-bleu-max}{52.97}
\DefMacro{res-teco-s6-CSNm-eval-runnable-assert-stmt/test-bs10-last-code-bleu-min}{11.78}
\DefMacro{res-teco-s6-CSNm-eval-runnable-assert-stmt/test-bs10-last-code-bleu-avg}{27.09}
\DefMacro{res-teco-s6-CSNm-eval-runnable-assert-stmt/test-bs10-last-edit-sim}{66.98}
\DefMacro{res-teco-s6-CSNm-eval-runnable-assert-stmt/test-bs10-last-edit-sim-max}{79.86}
\DefMacro{res-teco-s6-CSNm-eval-runnable-assert-stmt/test-bs10-last-edit-sim-min}{48.98}
\DefMacro{res-teco-s6-CSNm-eval-runnable-assert-stmt/test-bs10-last-edit-sim-avg}{63.35}
\DefMacro{res-teco-s6-CSNm-eval-runnable-assert-stmt/test-bs10-last-rouge-p}{67.80}
\DefMacro{res-teco-s6-CSNm-eval-runnable-assert-stmt/test-bs10-last-rouge-p-max}{81.67}
\DefMacro{res-teco-s6-CSNm-eval-runnable-assert-stmt/test-bs10-last-rouge-p-min}{46.17}
\DefMacro{res-teco-s6-CSNm-eval-runnable-assert-stmt/test-bs10-last-rouge-p-avg}{61.59}
\DefMacro{res-teco-s6-CSNm-eval-runnable-assert-stmt/test-bs10-last-rouge-r}{68.46}
\DefMacro{res-teco-s6-CSNm-eval-runnable-assert-stmt/test-bs10-last-rouge-r-max}{80.55}
\DefMacro{res-teco-s6-CSNm-eval-runnable-assert-stmt/test-bs10-last-rouge-r-min}{53.09}
\DefMacro{res-teco-s6-CSNm-eval-runnable-assert-stmt/test-bs10-last-rouge-r-avg}{66.41}
\DefMacro{res-teco-s6-CSNm-eval-runnable-assert-stmt/test-bs10-last-rouge-f}{67.10}
\DefMacro{res-teco-s6-CSNm-eval-runnable-assert-stmt/test-bs10-last-rouge-f-max}{79.92}
\DefMacro{res-teco-s6-CSNm-eval-runnable-assert-stmt/test-bs10-last-rouge-f-min}{48.58}
\DefMacro{res-teco-s6-CSNm-eval-runnable-assert-stmt/test-bs10-last-rouge-f-avg}{62.80}
\DefMacro{res-teco-s6-CSNm-eval-runnable-assert-stmt/test-bs10-last-num-seq}{10.00}
\DefMacro{res-teco-s6-CSNm-eval-runnable-assert-stmt/test-bs10-last-compilable}{-1.00}
\DefMacro{res-teco-s6-CSNm-eval-runnable-assert-stmt/test-bs10-last-runnable}{-1.00}
\DefMacro{res-toga-CSNm-eval-any-stmt/test-bs10-last-xmatch}{-1.00}
\DefMacro{res-toga-CSNm-eval-any-stmt/test-bs10-last-xmatch-top10}{-1.00}
\DefMacro{res-toga-CSNm-eval-any-stmt/test-bs10-last-bleu}{-1.00}
\DefMacro{res-toga-CSNm-eval-any-stmt/test-bs10-last-code-bleu}{-1.00}
\DefMacro{res-toga-CSNm-eval-any-stmt/test-bs10-last-edit-sim}{-1.00}
\DefMacro{res-toga-CSNm-eval-any-stmt/test-bs10-last-rouge-f}{-1.00}
\DefMacro{res-toga-CSNm-eval-any-stmt/test-bs10-last-compilable}{-1.00}
\DefMacro{res-toga-CSNm-eval-any-stmt/test-bs10-last-runnable}{-1.00}
\DefMacro{res-toga-CSNm-eval-assert-stmt/test-bs10-last-xmatch}{9.01}
\DefMacro{res-toga-CSNm-eval-assert-stmt/test-bs10-last-xmatch-top3}{9.01}
\DefMacro{res-toga-CSNm-eval-assert-stmt/test-bs10-last-xmatch-top5}{9.01}
\DefMacro{res-toga-CSNm-eval-assert-stmt/test-bs10-last-xmatch-top10}{9.01}
\DefMacro{res-toga-CSNm-eval-assert-stmt/test-bs10-last-xmatch-top100}{9.01}
\DefMacro{res-toga-CSNm-eval-assert-stmt/test-bs10-last-bleu}{25.46}
\DefMacro{res-toga-CSNm-eval-assert-stmt/test-bs10-last-bleu-max}{25.46}
\DefMacro{res-toga-CSNm-eval-assert-stmt/test-bs10-last-bleu-min}{25.46}
\DefMacro{res-toga-CSNm-eval-assert-stmt/test-bs10-last-bleu-avg}{25.46}
\DefMacro{res-toga-CSNm-eval-assert-stmt/test-bs10-last-code-bleu}{24.73}
\DefMacro{res-toga-CSNm-eval-assert-stmt/test-bs10-last-code-bleu-max}{24.73}
\DefMacro{res-toga-CSNm-eval-assert-stmt/test-bs10-last-code-bleu-min}{24.73}
\DefMacro{res-toga-CSNm-eval-assert-stmt/test-bs10-last-code-bleu-avg}{24.73}
\DefMacro{res-toga-CSNm-eval-assert-stmt/test-bs10-last-edit-sim}{29.60}
\DefMacro{res-toga-CSNm-eval-assert-stmt/test-bs10-last-edit-sim-max}{29.60}
\DefMacro{res-toga-CSNm-eval-assert-stmt/test-bs10-last-edit-sim-min}{29.60}
\DefMacro{res-toga-CSNm-eval-assert-stmt/test-bs10-last-edit-sim-avg}{29.60}
\DefMacro{res-toga-CSNm-eval-assert-stmt/test-bs10-last-rouge-p}{28.91}
\DefMacro{res-toga-CSNm-eval-assert-stmt/test-bs10-last-rouge-p-max}{28.91}
\DefMacro{res-toga-CSNm-eval-assert-stmt/test-bs10-last-rouge-p-min}{28.91}
\DefMacro{res-toga-CSNm-eval-assert-stmt/test-bs10-last-rouge-p-avg}{28.91}
\DefMacro{res-toga-CSNm-eval-assert-stmt/test-bs10-last-rouge-r}{27.43}
\DefMacro{res-toga-CSNm-eval-assert-stmt/test-bs10-last-rouge-r-max}{27.43}
\DefMacro{res-toga-CSNm-eval-assert-stmt/test-bs10-last-rouge-r-min}{27.43}
\DefMacro{res-toga-CSNm-eval-assert-stmt/test-bs10-last-rouge-r-avg}{27.43}
\DefMacro{res-toga-CSNm-eval-assert-stmt/test-bs10-last-rouge-f}{28.06}
\DefMacro{res-toga-CSNm-eval-assert-stmt/test-bs10-last-rouge-f-max}{28.06}
\DefMacro{res-toga-CSNm-eval-assert-stmt/test-bs10-last-rouge-f-min}{28.06}
\DefMacro{res-toga-CSNm-eval-assert-stmt/test-bs10-last-rouge-f-avg}{28.06}
\DefMacro{res-toga-CSNm-eval-assert-stmt/test-bs10-last-compilable}{-1.00}
\DefMacro{res-toga-CSNm-eval-assert-stmt/test-bs10-last-runnable}{-1.00}
\DefMacro{res-toga-CSNm-eval-runnable-any-stmt/test-bs10-last-xmatch}{-1.00}
\DefMacro{res-toga-CSNm-eval-runnable-any-stmt/test-bs10-last-xmatch-top10}{-1.00}
\DefMacro{res-toga-CSNm-eval-runnable-any-stmt/test-bs10-last-bleu}{-1.00}
\DefMacro{res-toga-CSNm-eval-runnable-any-stmt/test-bs10-last-code-bleu}{-1.00}
\DefMacro{res-toga-CSNm-eval-runnable-any-stmt/test-bs10-last-edit-sim}{-1.00}
\DefMacro{res-toga-CSNm-eval-runnable-any-stmt/test-bs10-last-rouge-f}{-1.00}
\DefMacro{res-toga-CSNm-eval-runnable-any-stmt/test-bs10-last-compilable}{-1.00}
\DefMacro{res-toga-CSNm-eval-runnable-any-stmt/test-bs10-last-runnable}{-1.00}
\DefMacro{res-toga-CSNm-eval-runnable-assert-stmt/test-bs10-last-xmatch}{9.10}
\DefMacro{res-toga-CSNm-eval-runnable-assert-stmt/test-bs10-last-xmatch-top3}{9.10}
\DefMacro{res-toga-CSNm-eval-runnable-assert-stmt/test-bs10-last-xmatch-top5}{9.10}
\DefMacro{res-toga-CSNm-eval-runnable-assert-stmt/test-bs10-last-xmatch-top10}{9.10}
\DefMacro{res-toga-CSNm-eval-runnable-assert-stmt/test-bs10-last-xmatch-top100}{9.10}
\DefMacro{res-toga-CSNm-eval-runnable-assert-stmt/test-bs10-last-bleu}{26.51}
\DefMacro{res-toga-CSNm-eval-runnable-assert-stmt/test-bs10-last-bleu-max}{26.51}
\DefMacro{res-toga-CSNm-eval-runnable-assert-stmt/test-bs10-last-bleu-min}{26.51}
\DefMacro{res-toga-CSNm-eval-runnable-assert-stmt/test-bs10-last-bleu-avg}{26.51}
\DefMacro{res-toga-CSNm-eval-runnable-assert-stmt/test-bs10-last-code-bleu}{25.74}
\DefMacro{res-toga-CSNm-eval-runnable-assert-stmt/test-bs10-last-code-bleu-max}{25.74}
\DefMacro{res-toga-CSNm-eval-runnable-assert-stmt/test-bs10-last-code-bleu-min}{25.74}
\DefMacro{res-toga-CSNm-eval-runnable-assert-stmt/test-bs10-last-code-bleu-avg}{25.74}
\DefMacro{res-toga-CSNm-eval-runnable-assert-stmt/test-bs10-last-edit-sim}{31.00}
\DefMacro{res-toga-CSNm-eval-runnable-assert-stmt/test-bs10-last-edit-sim-max}{31.00}
\DefMacro{res-toga-CSNm-eval-runnable-assert-stmt/test-bs10-last-edit-sim-min}{31.00}
\DefMacro{res-toga-CSNm-eval-runnable-assert-stmt/test-bs10-last-edit-sim-avg}{31.00}
\DefMacro{res-toga-CSNm-eval-runnable-assert-stmt/test-bs10-last-rouge-p}{30.22}
\DefMacro{res-toga-CSNm-eval-runnable-assert-stmt/test-bs10-last-rouge-p-max}{30.22}
\DefMacro{res-toga-CSNm-eval-runnable-assert-stmt/test-bs10-last-rouge-p-min}{30.22}
\DefMacro{res-toga-CSNm-eval-runnable-assert-stmt/test-bs10-last-rouge-p-avg}{30.22}
\DefMacro{res-toga-CSNm-eval-runnable-assert-stmt/test-bs10-last-rouge-r}{28.74}
\DefMacro{res-toga-CSNm-eval-runnable-assert-stmt/test-bs10-last-rouge-r-max}{28.74}
\DefMacro{res-toga-CSNm-eval-runnable-assert-stmt/test-bs10-last-rouge-r-min}{28.74}
\DefMacro{res-toga-CSNm-eval-runnable-assert-stmt/test-bs10-last-rouge-r-avg}{28.74}
\DefMacro{res-toga-CSNm-eval-runnable-assert-stmt/test-bs10-last-rouge-f}{29.38}
\DefMacro{res-toga-CSNm-eval-runnable-assert-stmt/test-bs10-last-rouge-f-max}{29.38}
\DefMacro{res-toga-CSNm-eval-runnable-assert-stmt/test-bs10-last-rouge-f-min}{29.38}
\DefMacro{res-toga-CSNm-eval-runnable-assert-stmt/test-bs10-last-rouge-f-avg}{29.38}
\DefMacro{res-toga-CSNm-eval-runnable-assert-stmt/test-bs10-last-compilable}{25.61}
\DefMacro{res-toga-CSNm-eval-runnable-assert-stmt/test-bs10-last-runnable}{9.37}

\DefMacro{res-ours-CSNm-eval-any-stmt/test-bs10-last-xmatch}{17.61}
\DefMacro{res-baseline-CSNm-eval-any-stmt/test-bs10-last-xmatch}{13.57}
\DefMacro{res-improv-pct-CSNm-eval-any-stmt/test-bs10-last-xmatch}{29}
\DefMacro{res-improv-abs-CSNm-eval-any-stmt/test-bs10-last-xmatch}{4.04}
\DefMacro{res-ours-CSNm-eval-any-stmt/test-bs10-last-xmatch-top10}{27.20}
\DefMacro{res-baseline-CSNm-eval-any-stmt/test-bs10-last-xmatch-top10}{24.11}
\DefMacro{res-improv-pct-CSNm-eval-any-stmt/test-bs10-last-xmatch-top10}{12}
\DefMacro{res-improv-abs-CSNm-eval-any-stmt/test-bs10-last-xmatch-top10}{3.09}
\DefMacro{res-ours-CSNm-eval-any-stmt/test-bs10-last-bleu}{42.01}
\DefMacro{res-baseline-CSNm-eval-any-stmt/test-bs10-last-bleu}{38.33}
\DefMacro{res-improv-pct-CSNm-eval-any-stmt/test-bs10-last-bleu}{9}
\DefMacro{res-improv-abs-CSNm-eval-any-stmt/test-bs10-last-bleu}{3.68}
\DefMacro{res-ours-CSNm-eval-any-stmt/test-bs10-last-code-bleu}{37.61}
\DefMacro{res-baseline-CSNm-eval-any-stmt/test-bs10-last-code-bleu}{33.88}
\DefMacro{res-improv-pct-CSNm-eval-any-stmt/test-bs10-last-code-bleu}{11}
\DefMacro{res-improv-abs-CSNm-eval-any-stmt/test-bs10-last-code-bleu}{3.73}
\DefMacro{res-ours-CSNm-eval-any-stmt/test-bs10-last-edit-sim}{63.49}
\DefMacro{res-baseline-CSNm-eval-any-stmt/test-bs10-last-edit-sim}{60.81}
\DefMacro{res-improv-pct-CSNm-eval-any-stmt/test-bs10-last-edit-sim}{4}
\DefMacro{res-improv-abs-CSNm-eval-any-stmt/test-bs10-last-edit-sim}{2.68}
\DefMacro{res-ours-CSNm-eval-any-stmt/test-bs10-last-rouge-f}{65.23}
\DefMacro{res-baseline-CSNm-eval-any-stmt/test-bs10-last-rouge-f}{62.40}
\DefMacro{res-improv-pct-CSNm-eval-any-stmt/test-bs10-last-rouge-f}{4}
\DefMacro{res-improv-abs-CSNm-eval-any-stmt/test-bs10-last-rouge-f}{2.83}
\DefMacro{res-ours-CSNm-eval-any-stmt/test-bs10-last-compilable}{-1.00}
\DefMacro{res-baseline-CSNm-eval-any-stmt/test-bs10-last-compilable}{-1.00}
\DefMacro{res-improv-pct-CSNm-eval-any-stmt/test-bs10-last-compilable}{0}
\DefMacro{res-improv-abs-CSNm-eval-any-stmt/test-bs10-last-compilable}{0.00}
\DefMacro{res-ours-CSNm-eval-any-stmt/test-bs10-last-runnable}{-1.00}
\DefMacro{res-baseline-CSNm-eval-any-stmt/test-bs10-last-runnable}{-1.00}
\DefMacro{res-improv-pct-CSNm-eval-any-stmt/test-bs10-last-runnable}{0}
\DefMacro{res-improv-abs-CSNm-eval-any-stmt/test-bs10-last-runnable}{0.00}
\DefMacro{res-ours-CSNm-eval-assert-stmt/test-bs10-last-xmatch}{16.44}
\DefMacro{res-baseline-CSNm-eval-assert-stmt/test-bs10-last-xmatch}{9.01}
\DefMacro{res-improv-pct-CSNm-eval-assert-stmt/test-bs10-last-xmatch}{82}
\DefMacro{res-improv-abs-CSNm-eval-assert-stmt/test-bs10-last-xmatch}{7.43}
\DefMacro{res-ours-CSNm-eval-assert-stmt/test-bs10-last-xmatch-top10}{27.41}
\DefMacro{res-baseline-CSNm-eval-assert-stmt/test-bs10-last-xmatch-top10}{24.04}
\DefMacro{res-improv-pct-CSNm-eval-assert-stmt/test-bs10-last-xmatch-top10}{14}
\DefMacro{res-improv-abs-CSNm-eval-assert-stmt/test-bs10-last-xmatch-top10}{3.37}
\DefMacro{res-ours-CSNm-eval-assert-stmt/test-bs10-last-bleu}{43.09}
\DefMacro{res-baseline-CSNm-eval-assert-stmt/test-bs10-last-bleu}{39.03}
\DefMacro{res-improv-pct-CSNm-eval-assert-stmt/test-bs10-last-bleu}{10}
\DefMacro{res-improv-abs-CSNm-eval-assert-stmt/test-bs10-last-bleu}{4.06}
\DefMacro{res-ours-CSNm-eval-assert-stmt/test-bs10-last-code-bleu}{35.88}
\DefMacro{res-baseline-CSNm-eval-assert-stmt/test-bs10-last-code-bleu}{31.03}
\DefMacro{res-improv-pct-CSNm-eval-assert-stmt/test-bs10-last-code-bleu}{15}
\DefMacro{res-improv-abs-CSNm-eval-assert-stmt/test-bs10-last-code-bleu}{4.85}
\DefMacro{res-ours-CSNm-eval-assert-stmt/test-bs10-last-edit-sim}{68.05}
\DefMacro{res-baseline-CSNm-eval-assert-stmt/test-bs10-last-edit-sim}{66.50}
\DefMacro{res-improv-pct-CSNm-eval-assert-stmt/test-bs10-last-edit-sim}{2}
\DefMacro{res-improv-abs-CSNm-eval-assert-stmt/test-bs10-last-edit-sim}{1.55}
\DefMacro{res-ours-CSNm-eval-assert-stmt/test-bs10-last-rouge-f}{68.71}
\DefMacro{res-baseline-CSNm-eval-assert-stmt/test-bs10-last-rouge-f}{66.63}
\DefMacro{res-improv-pct-CSNm-eval-assert-stmt/test-bs10-last-rouge-f}{3}
\DefMacro{res-improv-abs-CSNm-eval-assert-stmt/test-bs10-last-rouge-f}{2.08}
\DefMacro{res-ours-CSNm-eval-assert-stmt/test-bs10-last-compilable}{-1.00}
\DefMacro{res-baseline-CSNm-eval-assert-stmt/test-bs10-last-compilable}{-1.00}
\DefMacro{res-improv-pct-CSNm-eval-assert-stmt/test-bs10-last-compilable}{0}
\DefMacro{res-improv-abs-CSNm-eval-assert-stmt/test-bs10-last-compilable}{0.00}
\DefMacro{res-ours-CSNm-eval-assert-stmt/test-bs10-last-runnable}{-1.00}
\DefMacro{res-baseline-CSNm-eval-assert-stmt/test-bs10-last-runnable}{-1.00}
\DefMacro{res-improv-pct-CSNm-eval-assert-stmt/test-bs10-last-runnable}{0}
\DefMacro{res-improv-abs-CSNm-eval-assert-stmt/test-bs10-last-runnable}{0.00}
\DefMacro{res-ours-CSNm-eval-runnable-any-stmt/test-bs10-last-xmatch}{18.96}
\DefMacro{res-baseline-CSNm-eval-runnable-any-stmt/test-bs10-last-xmatch}{14.38}
\DefMacro{res-improv-pct-CSNm-eval-runnable-any-stmt/test-bs10-last-xmatch}{31}
\DefMacro{res-improv-abs-CSNm-eval-runnable-any-stmt/test-bs10-last-xmatch}{4.58}
\DefMacro{res-ours-CSNm-eval-runnable-any-stmt/test-bs10-last-xmatch-top10}{28.40}
\DefMacro{res-baseline-CSNm-eval-runnable-any-stmt/test-bs10-last-xmatch-top10}{25.39}
\DefMacro{res-improv-pct-CSNm-eval-runnable-any-stmt/test-bs10-last-xmatch-top10}{11}
\DefMacro{res-improv-abs-CSNm-eval-runnable-any-stmt/test-bs10-last-xmatch-top10}{3.01}
\DefMacro{res-ours-CSNm-eval-runnable-any-stmt/test-bs10-last-bleu}{43.15}
\DefMacro{res-baseline-CSNm-eval-runnable-any-stmt/test-bs10-last-bleu}{39.26}
\DefMacro{res-improv-pct-CSNm-eval-runnable-any-stmt/test-bs10-last-bleu}{9}
\DefMacro{res-improv-abs-CSNm-eval-runnable-any-stmt/test-bs10-last-bleu}{3.89}
\DefMacro{res-ours-CSNm-eval-runnable-any-stmt/test-bs10-last-code-bleu}{38.45}
\DefMacro{res-baseline-CSNm-eval-runnable-any-stmt/test-bs10-last-code-bleu}{34.55}
\DefMacro{res-improv-pct-CSNm-eval-runnable-any-stmt/test-bs10-last-code-bleu}{11}
\DefMacro{res-improv-abs-CSNm-eval-runnable-any-stmt/test-bs10-last-code-bleu}{3.90}
\DefMacro{res-ours-CSNm-eval-runnable-any-stmt/test-bs10-last-edit-sim}{64.12}
\DefMacro{res-baseline-CSNm-eval-runnable-any-stmt/test-bs10-last-edit-sim}{61.36}
\DefMacro{res-improv-pct-CSNm-eval-runnable-any-stmt/test-bs10-last-edit-sim}{4}
\DefMacro{res-improv-abs-CSNm-eval-runnable-any-stmt/test-bs10-last-edit-sim}{2.76}
\DefMacro{res-ours-CSNm-eval-runnable-any-stmt/test-bs10-last-rouge-f}{66.09}
\DefMacro{res-baseline-CSNm-eval-runnable-any-stmt/test-bs10-last-rouge-f}{63.15}
\DefMacro{res-improv-pct-CSNm-eval-runnable-any-stmt/test-bs10-last-rouge-f}{4}
\DefMacro{res-improv-abs-CSNm-eval-runnable-any-stmt/test-bs10-last-rouge-f}{2.94}
\DefMacro{res-ours-CSNm-eval-runnable-any-stmt/test-bs10-last-compilable}{76.22}
\DefMacro{res-baseline-CSNm-eval-runnable-any-stmt/test-bs10-last-compilable}{54.84}
\DefMacro{res-improv-pct-CSNm-eval-runnable-any-stmt/test-bs10-last-compilable}{38}
\DefMacro{res-improv-abs-CSNm-eval-runnable-any-stmt/test-bs10-last-compilable}{21.38}
\DefMacro{res-ours-CSNm-eval-runnable-any-stmt/test-bs10-last-runnable}{28.63}
\DefMacro{res-baseline-CSNm-eval-runnable-any-stmt/test-bs10-last-runnable}{17.62}
\DefMacro{res-improv-pct-CSNm-eval-runnable-any-stmt/test-bs10-last-runnable}{62}
\DefMacro{res-improv-abs-CSNm-eval-runnable-any-stmt/test-bs10-last-runnable}{11.01}
\DefMacro{res-ours-CSNm-eval-runnable-assert-stmt/test-bs10-last-xmatch}{17.37}
\DefMacro{res-baseline-CSNm-eval-runnable-assert-stmt/test-bs10-last-xmatch}{9.10}
\DefMacro{res-improv-pct-CSNm-eval-runnable-assert-stmt/test-bs10-last-xmatch}{90}
\DefMacro{res-improv-abs-CSNm-eval-runnable-assert-stmt/test-bs10-last-xmatch}{8.27}
\DefMacro{res-ours-CSNm-eval-runnable-assert-stmt/test-bs10-last-xmatch-top10}{27.39}
\DefMacro{res-baseline-CSNm-eval-runnable-assert-stmt/test-bs10-last-xmatch-top10}{24.37}
\DefMacro{res-improv-pct-CSNm-eval-runnable-assert-stmt/test-bs10-last-xmatch-top10}{12}
\DefMacro{res-improv-abs-CSNm-eval-runnable-assert-stmt/test-bs10-last-xmatch-top10}{3.02}
\DefMacro{res-ours-CSNm-eval-runnable-assert-stmt/test-bs10-last-bleu}{44.27}
\DefMacro{res-baseline-CSNm-eval-runnable-assert-stmt/test-bs10-last-bleu}{40.26}
\DefMacro{res-improv-pct-CSNm-eval-runnable-assert-stmt/test-bs10-last-bleu}{9}
\DefMacro{res-improv-abs-CSNm-eval-runnable-assert-stmt/test-bs10-last-bleu}{4.01}
\DefMacro{res-ours-CSNm-eval-runnable-assert-stmt/test-bs10-last-code-bleu}{36.98}
\DefMacro{res-baseline-CSNm-eval-runnable-assert-stmt/test-bs10-last-code-bleu}{32.23}
\DefMacro{res-improv-pct-CSNm-eval-runnable-assert-stmt/test-bs10-last-code-bleu}{14}
\DefMacro{res-improv-abs-CSNm-eval-runnable-assert-stmt/test-bs10-last-code-bleu}{4.75}
\DefMacro{res-ours-CSNm-eval-runnable-assert-stmt/test-bs10-last-edit-sim}{68.43}
\DefMacro{res-baseline-CSNm-eval-runnable-assert-stmt/test-bs10-last-edit-sim}{67.08}
\DefMacro{res-improv-pct-CSNm-eval-runnable-assert-stmt/test-bs10-last-edit-sim}{2}
\DefMacro{res-improv-abs-CSNm-eval-runnable-assert-stmt/test-bs10-last-edit-sim}{1.35}
\DefMacro{res-ours-CSNm-eval-runnable-assert-stmt/test-bs10-last-rouge-f}{69.35}
\DefMacro{res-baseline-CSNm-eval-runnable-assert-stmt/test-bs10-last-rouge-f}{67.42}
\DefMacro{res-improv-pct-CSNm-eval-runnable-assert-stmt/test-bs10-last-rouge-f}{2}
\DefMacro{res-improv-abs-CSNm-eval-runnable-assert-stmt/test-bs10-last-rouge-f}{1.93}
\DefMacro{res-ours-CSNm-eval-runnable-assert-stmt/test-bs10-last-compilable}{67.93}
\DefMacro{res-baseline-CSNm-eval-runnable-assert-stmt/test-bs10-last-compilable}{44.45}
\DefMacro{res-improv-pct-CSNm-eval-runnable-assert-stmt/test-bs10-last-compilable}{52}
\DefMacro{res-improv-abs-CSNm-eval-runnable-assert-stmt/test-bs10-last-compilable}{23.48}
\DefMacro{res-ours-CSNm-eval-runnable-assert-stmt/test-bs10-last-runnable}{30.29}
\DefMacro{res-baseline-CSNm-eval-runnable-assert-stmt/test-bs10-last-runnable}{16.87}
\DefMacro{res-improv-pct-CSNm-eval-runnable-assert-stmt/test-bs10-last-runnable}{79}
\DefMacro{res-improv-abs-CSNm-eval-runnable-assert-stmt/test-bs10-last-runnable}{13.42}

\newcommand{\NumTrials}{three\xspace}
\newcommand{\NumSemData}{six\xspace}
\newcommand{\NumSemDataSA}{six\xspace}

\newcommand{\NumMaxSeqLen}{512\xspace}
\newcommand{\NumBeamSize}{10\xspace}
\newcommand{\NumCollectionPerProject}{136\xspace}  %
\newcommand{\NumAnalysisPerTestMin}{0.018\xspace}  %
\newcommand{\NumAnalysisPerTestMax}{0.247\xspace}  %
\newcommand{\NumRepoCSNOrigin}{4,767\xspace}
\newcommand{\NumRepoCSNFiltered}{1,535\xspace}
\newcommand{\NumPctHasControlFlow}{10\xspace}  %

\begin{document}

\title{\Title}

\author{
\IEEEauthorblockN{Pengyu Nie, Rahul Banerjee, Junyi Jessy Li, Raymond J. Mooney, Milos Gligoric}
\IEEEauthorblockA{\textit{UT Austin, USA}\\
\{pynie,rahulb517,jessy,mooney,gligoric\}@utexas.edu}
}

\maketitle

\begin{abstract}
Writing tests is a time-consuming yet essential task during software
development.
We propose to leverage recent advances in deep learning for text and code generation to assist
developers in writing tests.  We formalize the novel task of \testcomp
to automatically complete the next statement in a \testm based on the
context of \priorstmts and the \codeut.  We develop \TecoTool---a deep
learning model using \semdata for \testcomp. The key insight
underlying \TecoTool is that predicting the next statement in a \testm
requires reasoning about code execution,
which is hard to do with only \sufdata that existing code completion
models use.  \TecoTool extracts and uses \NumSemData kinds of \semdata
data, including the \exeres of \priorstmts and the \exectx of the
\testm.
To provide a testbed for this new task, as well as to evaluate
\TecoTool, we collect a corpus of \UseMacro{corpus-all-num_test}
\testms from \UseMacro{corpus-all-num_proj} open-source Java projects.
Our results show that \TecoTool achieves an \xmatch of
\numprint{\UseMacro{res-ours-CSNm-eval-any-stmt/test-bs10-last-xmatch}}, which is
\numprint{\UseMacro{res-improv-pct-CSNm-eval-any-stmt/test-bs10-last-xmatch}}\%
higher than the best baseline using \sufdata only.
When measuring functional correctness of generated next \stmt,
\TecoTool can generate runnable code in
\numprint{\UseMacro{res-ours-CSNm-eval-runnable-any-stmt/test-bs10-last-runnable}}\%
of the cases compared to
\numprint{\UseMacro{res-baseline-CSNm-eval-runnable-any-stmt/test-bs10-last-runnable}}\% obtained by the best baseline.
Moreover, \TecoTool is significantly better than prior work on
\oraclegen.

\end{abstract}

\begin{IEEEkeywords}
test completion, deep neural networks, programming language semantics
\end{IEEEkeywords}

\thispagestyle{plain}
\pagestyle{plain}

\section{Introduction}
\label{sec:intro}

Software testing is the most common approach in industry to check the
correctness of software.  However, manually writing tests is tiresome
and time-consuming.

One option is to automatically generate tests.  Researchers have
proposed a number of techniques in this domain,
including fuzz testing~\cite{PachecoETAL07Randoop,
ZellerETAL19Fuzzing, ZangETAL22JAttack}, property-based testing~\cite{ClaessenAndHughes00QuickCheck, BoyapatiETAL02Korat,
GligoricETAL10Test, IvanETAL15Programming, CelikETAL17intKorat,
HolmesETAL20TestGeneration, AlAwarETAL21Tempo}, search-based
testing~\cite{HarmanAndMcMinn10Theoretical, FraserAndArcuri11EvoSuite}, combinatorial
testing~\cite{CohenETAL06Testing}, etc.
Despite being effective in detecting software bugs,
these techniques generate tests with stylistic issues, as test code
generated through these techniques rarely resemble manually-written
tests~\cite{DakaETAL17Generating, RobinsonETAL11Scaling,
  ZhangETAL16Automatically} and can be hard to maintain.
As a result, these automated techniques end up being used only as
supplements to
manually-written tests.

Another option is to use machine learning (ML), namely training a model on
existing manually-written tests and applying it when writing new
tests, which is a plausible methodology supported by the
naturalness of
software~\cite{HindleETAL12Naturalness,MusfiqurETAL19Natural}.
Advances in deep learning such as recurrent neural
networks~\cite{rumelhart1986learning,hochreiter1997long} and
large-scale \pretrained transformer
models~\cite{vaswani2017attention,radford2019language,lewis2019bart,raffel2020exploring}
have led to promising new research in a variety of software
engineering tasks, such as code
completion~\cite{ProkschETAL15Intelligent, SvyatkovskiyETAL19Pythia,
LiETAL18Code, RaychevETAL14Code, SvyatkovskiyETAL20IntelliCode,
ChenETAL21Codex} and code summarization~\cite{IyerETAL16Summarizing,
AhmadETAL20Transformer-based, LeClairETAL19Neural, AhmadETAL21PLBART,
WangETAL21CodeT5}.  Code generated with modern models are intelligible
to humans, yet we cannot fully rely on them to generate large chunks
of meaningful code, or expect them to understand larger project
context.

Our goal is to design machine learning approaches to aid
developer productivity when \emph{writing tests}.
We present a novel task---\emph{\testcomp}---to help developers write
tests faster.  Specifically, once a developer starts writing a \testm,
she can leverage \testcomp to automatically obtain the next statement
in the test code (at any point she desires).

Despite being closely related to code
completion~\cite{ProkschETAL15Intelligent, SvyatkovskiyETAL19Pythia,
  LiETAL18Code, RaychevETAL14Code, SvyatkovskiyETAL20IntelliCode},
\testcomp is distinct in that
test code has several unique characteristics.
First, the \mut provides extra context that can be leveraged when
completing a \testm.
Second, test code follows a different programming style
that focuses on exercising the \mut.
Specifically, a \testm usually consists of a sequence of statements in
the following order: prepare inputs to the \mut, execute \mut, and
check the results of the execution using assert statements (i.e., test oracles).

We present the first deep learning solution---\TecoTool{}---that takes
into account these unique characteristics of tests.

\begin{tcolorbox}[notitle,boxrule=0pt,colback=gray!30,colframe=gray!20]
\textit{\TecoTool uses \semdata as inputs for novel ML models and performs reranking via test execution.}
\end{tcolorbox}

\Semdata refers to the information related to test/code execution not
available in the \sufdata (i.e., source code).
\TecoTool extracts \semdata (e.g., types of local variables) using
software engineering tools and feeds them directly to the model.
Once top-k predictions are produced, \TecoTool further ensures the
output quality by \emph{executing} the generated \stmts, and
prioritize the runnable and compilable \stmts over the others.

We design the \semdata used by \TecoTool based on our experience with
software analysis in order to best capture the unique
characteristics of the \testcomp task.  In total, we consider
\NumSemData different kinds of \semdata that can be grouped to two
categories: (1)~\exeres, including the types of the local variables
and whether fields are initialized;
(2)~\exectx, including the setup and teardown methods, the last called
method in the \testm, and statements in non-test code with similar
previous statements.

We implemented \TecoTool to support \testms written in Java.  We
evaluate \TecoTool on a newly collected corpus consisting of
\UseMacro{corpus-all-num_test} \testms with
\UseMacro{corpus-all-num_stmt} \stmts from
\UseMacro{corpus-all-num_proj} projects.  We release this corpus to
the community as a testbed for the \testcomp task.

We performed extensive evaluations of \TecoTool on this
corpus---covering lexical similarity,
functional correctness, and downstream application---to show the
importance of combining \semdata with deep learning.  We report
results comparing the generated \stmts against the gold
manually-written \stmts using a suite of automatic metrics: \xmatch,
\xmatchk, \BLEU~\cite{PapineniETAL02BLEU},
\CodeBLEU~\cite{RenETAL20CodeBLEU},
\editsim~\cite{SvyatkovskiyETAL20IntelliCode}, and
\ROUGEL~\cite{LinAndOch04Automatic}.
\TecoTool significantly outperforms baselines that use only \sufdata
on all metrics.  We also measure functional correctness by trying to
compile and run the generated \stmts. \TecoTool can produce a
runnable next \stmt
\numprint{\UseMacro{res-ours-CSNm-eval-runnable-any-stmt/test-bs10-last-runnable}}\%
of the time, while the figure for the best baseline model is only
\numprint{\UseMacro{res-baseline-CSNm-eval-runnable-any-stmt/test-bs10-last-runnable}}\%.
Moreover, we also evaluated \TecoTool on the task of
\oraclegen~\cite{WatsonETAL20ATLAS,DinellaETAL22TOGA}, which is
a downstream application of \testcomp.
\TecoTool achieves an \xmatch of
\numprint{\UseMacro{res-ours-CSNm-eval-assert-stmt/test-bs10-last-xmatch}}, which
significantly outperforms the prior state-of-the-art's \xmatch of
\numprint{\UseMacro{res-baseline-CSNm-eval-assert-stmt/test-bs10-last-xmatch}}.

\vspace{3pt}
\noindent
The main contributions of this paper include the following:

\begin{itemize}[topsep=3pt,itemsep=5pt,partopsep=0ex,parsep=0ex,leftmargin=*]
\item \textbf{Task}.  We propose a novel task, \testcomp, with the
goal to help developers write \testms faster.

\item \textbf{Idea}.  We propose using \semdata and code execution
  when designing ML models targeting code-related tasks.

\item \textbf{Model}.  We developed \TecoTool, the first transformer
  model trained on large \semdata data for \testcomp.  Furthermore,
  \TecoTool performs reranking by execution.
  The use of \semdata is vital for correctly modeling the execution
  process in the \testms.

\item \textbf{Corpus}.  We created a large corpus of
\UseMacro{corpus-all-num_test} \testms from
\UseMacro{corpus-all-num_proj} open-source projects.  We believe this
corpus will also be useful to many other tasks related to testing.

\item \textbf{Evaluation}.
  Our extensive evaluation shows that \TecoTool significantly
  outperforms strong baselines on all automatic metrics, both on
  \testcomp and its downstream application: \oraclegen.  We 
  also evaluate the functional correctness of generated code
  by compiling and running the generated \stmts.
\end{itemize}

\noindent
\TecoTool and our corpus are publicly available on GitHub:\\
\TecoURL.

\begin{figure}[t]
  \centering
  \newsavebox{\lstExampleMUT}
\begin{lrbox}{\lstExampleMUT}
\begin{lstlisting}[language=java-pretty-no-number]
public GMOperation addImage(final File file) {
  if (file == null) {
    throw new IllegalArgumentException(
      "file must be defined"); }
  getCmdArgs().add(file.getPath());
  return this; }
\end{lstlisting}
\end{lrbox}

\newsavebox{\lstExampleSign}
\begin{lrbox}{\lstExampleSign}
\begin{lstlisting}[language=java-pretty-no-number]
@Test 
public void addImage_ThrowsException_WhenFileIsNull() 
throws Exception
\end{lstlisting}
\end{lrbox}

\newsavebox{\lstExamplePrevStmts}
\begin{lrbox}{\lstExamplePrevStmts}
\begin{lstlisting}[language=java-pretty-no-number]
exception.except(IllegalArgumentException.class);
\end{lstlisting}
\end{lrbox}

\newsavebox{\lstExampleGold}
\begin{lrbox}{\lstExampleGold}
\begin{lstlisting}[language=java-pretty-no-number]
sut.addImage((File) null);
\end{lstlisting}
\end{lrbox}

\DefMacro{w-boxleftsep}{1.2em}
\DefMacro{h-boxsep}{3.5ex}
\DefMacro{h-boxannosep}{0ex}

\begin{tikzpicture}[
  font=\footnotesize, 
  box/.style={rectangle, scale=1, inner sep=1pt},
  box-bg/.style={box},
  box-label/.style={box, inner sep=2pt, draw=black, fill=white},
  color-suf/.style={draw=yellow!80!black, fill=yellow!10},
  color-sem/.style={draw=blue!80!black, fill=blue!10},
  color-output/.style={draw=green!80!black, fill=green!10},
  anno/.style={font={\footnotesize}},
]
  \node (c-north) at (0,0) [draw=none, fill=none] {};
  
  % \node (c-code) at (c-north) [coordinate] {};

  % \node (bg-mut) [below right = 2.5ex and .5em of c-code] [box-bg, color-suf, minimum height=14.5ex, minimum width=20em] {};
  % \node (l-mut) [above left = 0pt and 3pt of bg-mut.north east] [box-label] {\mut};

  % \node (bg-sign) [below right = 29ex and .5em of c-code] [box-bg, color-suf, minimum height=5.7ex, minimum width=23em] {};
  % \node (l-sign) [above left = -3pt and 3pt of bg-sign.north east] [box-label] {test signature};

  % % \node (bg-sign) [below right = 29ex and .5em of c-code] [box-bg, color-suf, minimum height=5.7ex, minimum width=23em] {};
  % % \node (l-sign) [above left = 3pt and 3pt of bg-sign.south east] [box-label] {test signature};

  % \node (b-code) [below right = 0 and 0 of c-code] [box] {\usebox\lstExampleFull};

  \node (bg-mut) [below right = 0 and 0 of c-north] [box-bg, color-suf, minimum height=13ex, minimum width=23em] {};
  \node (b-mut) [below right = 0 and .2em of bg-mut.north west] [box] {\usebox\lstExampleMUT};
  \node (anno-mut) [above right = \UseMacro{h-boxannosep} and 0 of bg-mut.north west] [anno] {\mut};

  \node (bg-sign) [below right = \UseMacro{h-boxsep} and 0 of bg-mut.south west] [box-bg, color-suf, minimum height=6.5ex, minimum width=23em] {};
  \node (b-sign) [below right = 0 and .2em of bg-sign.north west] [box] {\usebox\lstExampleSign};
  \node (anno-sign) [above right = \UseMacro{h-boxannosep} and 0 of bg-sign.north west] [anno] {\testsign};

  \node (bg-prevstmts) [below right = \UseMacro{h-boxsep} and 0 of bg-sign.south west] [box-bg, color-suf, minimum height=3ex, minimum width=23em] {};
  \node (b-prevstmts) [below right = 0 and .2em of bg-prevstmts.north west] [box] {\usebox\lstExamplePrevStmts};
  \node (anno-prevstmts) [above right = \UseMacro{h-boxannosep} and 0 of bg-prevstmts.north west] [anno] {\priorstmts};

  \node (bg-gold) [below right = \UseMacro{h-boxsep} and 0 of bg-prevstmts.south west] [box-bg, color-output, minimum height=3ex, minimum width=23em] {};
  \node (b-gold) [below right = 0 and .2em of bg-gold.north west] [box] {\usebox\lstExampleGold\hfill};
  \node (anno-gold) [above right = \UseMacro{h-boxannosep} and 0 of b-gold.north west] [anno] {next \stmt};
\end{tikzpicture}
\caption{Example of \testcomp: given the \codeut (represented by the
\mut), \testsign, and \priorstmts, the goal is to generate the next
\stmt.  Code from \CodeIn{sharneng/gm4java} in class
\CodeIn{GMOperationTest}.\label{fig:example-data}}
\end{figure}

\section{Task}
\label{sec:task}

In this section, we more formally describe the \testcomp task and
illustrate the task using an example.

Given an incomplete \testm, our goal is to automatically generate the
next \stmt in that \testm.
We assume that the following inputs are provided to a \testcomp
system: (1)~the \codeut, which includes both the \testm's associated
\mut as well as other non-test-method code in the project, 
(2)~the
\testsign, (3)~\priorstmts in the incomplete \testm (which can be zero
or more \stmts).

We illustrate our task in \figurename~\ref{fig:example-data}.
The example shows (in the yellow boxes) the \mut, the \testsign,
and the \priorstmts (only one \stmt in this example), as well as (in
the last green box) the next \stmt that should be generated by a
\testcomp system.

We seek to generate \stmts in the body of the \testm, thus the
\testsign (including the annotation and the name of \testm) are only
used as inputs and they are not the prediction target of the \testcomp
task.  We also do not consider the
context of other already available \testms from the same project when
completing a \testm, to prevent any model from cheating by copying
code from other similar \testms.  Our defined \testcomp task is
applicable to the situation when a developer already knows what to
test (thus knows the \mut and the \testsign), and wants to complete
the next \stmt at any point when writing the \testm, regardless of
whether the project has existing tests or not.
We also focus on modeling the body of \testms as a sequence of \stmts,
because \testms with control flows (e.g., if statements, loops, and
try blocks) are rare; we found less than \NumPctHasControlFlow{\%}
\testms have control flows in our experiments. Most testing frameworks
recommend sequential \testm body and provide annotations to replace
control flows, for example, \CodeIn{@ParameterizedTest} for replacing
loops in JUnit~5~\cite{JUnit5}.

\section{Extraction of \SemData}
\label{sec:semdata}

In this section, we describe the \NumSemData kinds of \semdata
extracted and used by \TecoTool.  

For each kind of \semdata, we design and implement 
a \sanalysis algorithm to extract it.  \Sanalysis is the analysis of
code without executing it, guided by the grammar and semantics of
programming languages.  The advantage of using \sanalysis is that it
does not require configuring the runtime environment which can be
cumbersome for some projects, and can be applied on partial code (for
example, without accessing the dependency libraries of a project,
which is needed when executing the code).  It is also much faster than
executing the code directly, which enables us to collect \semdata on a
large corpus of code.  However, \sanalysis can sometimes be
inaccurate;   
for example, when some values are unknown without executing code
(e.g., user inputs), static analysis has to over-estimate the analysis
results (e.g., assuming both branches of an if statement may be
executed).  This may not be a problem for \TecoTool as the deep
learning model can learn to ignore the inaccurate parts.

\begin{figure}[t]
\centering
\includegraphics[width=\linewidth]{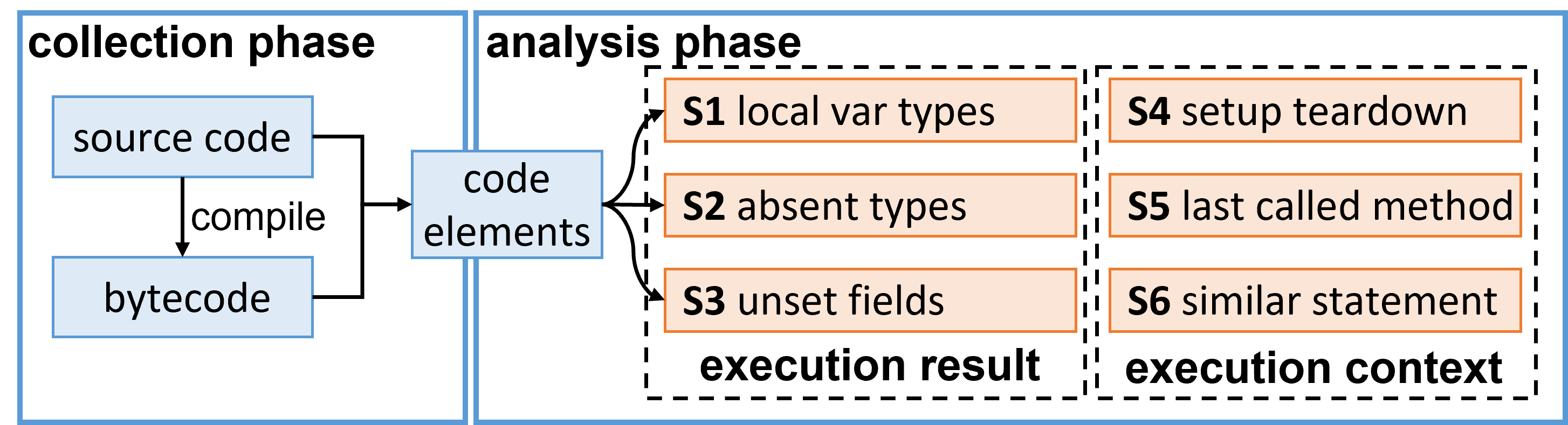}
\caption{\TecoTool's workflow of using \sanalysis to extract \semdata.\label{fig:semdata:sa}}
\vspace{-5pt}
\end{figure}

\figurename~\ref{fig:semdata:sa} illustrates the general workflow of
\TecoTool's \sanalysis which consists of two phases: given a project,
in the collection phase \TecoTool collects a shared set of all code
elements (classes, methods, fields), and in the analysis phase
\TecoTool extracts each kind of \semdata from the code elements set
using a specific algorithm.
The \semdata can be organized into two categories based on their
content: \emph{\exeres} and \emph{\exectx}.  \Exeres includes: 
(\UseMacro{no-types-local})~\UseMacro{name-types-local},
(\UseMacro{no-types-absent})~\UseMacro{name-types-absent}, and
(\UseMacro{no-fields-notset})~\UseMacro{name-fields-notset};
\exectx includes:
(\UseMacro{no-setup-teardown})~\UseMacro{name-setup-teardown},
(\UseMacro{no-last-called-method})~\UseMacro{name-last-called-method}, and
(\UseMacro{no-similar-stmt})~\UseMacro{name-similar-stmt}.
In the following paragraphs we describe the collection phase and
the analysis phase for each kind of \semdata in more detail.

\MyPara{The collection phase} The goal of this phase is to collect a
set of code elements that will be shared by all \testms and all
\NumSemDataSA kinds of \semdata.  \TecoTool collects three kinds of
code elements: classes, methods, and fields; each method and field
should have one class as its parent.  For each element, \TecoTool
collects its metadata, including name, type, access modifiers, annotations (if any), etc.  For each
class, \TecoTool records if it is a non-test class, test class, or a
class in a dependency.  \TecoTool additionally collects the
source code and bytecode for all their methods of non-test and test
classes.  We do not utilize the source code or bytecode in dependency
libraries because they are not needed for the analysis.

\MyParaOnly{(\UseMacro{no-types-local})~\UseMacro{name-types-local}}
data refers to the types of the local variables in the \testm.  The
types are extracted by partially interpreting the bytecode
of the \testm without considering
the values of variables.  
Note that this is more accurate than reading the local variable table
which contains the declared types of local variables; for example,
after interpreting the statement \CodeIn{AbstractWComponent\ comp\ =\
new\ SimpleComponent()}, the type of \CodeIn{comp} is
\CodeIn{SimpleComponent}, which is more accurate than its declared
type \CodeIn{AbstractWComponent}.  This data provides information
on what types of test inputs are available to be used in the next
statement.

\MyParaOnly{(\UseMacro{no-types-absent})~\UseMacro{name-types-absent}}
are the types of the variables that are needed by calling the \mut but
have not been prepared in the \testm.
Types needed by calling the \mut include its parameter types, plus
the \clzut (the declaring class of the \mut) if the \mut is not a
static method.  
A type is prepared if a local variable or a field of the \testc
with this type is initialized.  This data focuses on what types
of test inputs are missing, and thus may likely need to be prepared in
the next statement.

\MyParaOnly{(\UseMacro{no-fields-notset})~\UseMacro{name-fields-notset}} data refers to the fields of the \testc
and the \clzut that have not been initialized.  We deem a field as is initialized if there is any statement for setting
the value of the field in the \testm, the setup methods, or methods
transitively called from the \testm or the setup methods (up to 4
jumps, as initializations of the fields of our interest in more
in-depth calls are rare).  The fields in this data may likely need to
be initialized in the next statement.

\MyParaOnly{(\UseMacro{no-setup-teardown})~\UseMacro{name-setup-teardown}} data refers to the source code of the
setup and teardown methods in the \testc.  When a test framework
executes tests, setup methods are executed before the \testm to set up
the environment (e.g., connection to a database), and teardown methods
are executed after the \testm to clean up the environment.  By
providing this context, the \testcomp system can know what environment
is available to use in the \testm, and also can avoid duplicating the
\stmts already in setup/teardown methods.

\MyParaOnly{(\UseMacro{no-last-called-method})~\UseMacro{name-last-called-method}} data is the source code of the
last called method in the \priorstmts, which could be empty if no
method has been called yet.  This data provides more context on what
has been executed in \priorstmts.

\MyParaOnly{(\UseMacro{no-similar-stmt})~\UseMacro{name-similar-stmt}}
data is a statement in the non-test code of the project that has the
most similar \priorstmts context to the \priorstmts in the incomplete
\testm.  \TecoTool uses the BM25
algorithm~\cite{robertson2000experimentation} to search for similar
\priorstmts.  Only 2 \priorstmts are considered during the search, as
increasing the window size leads to much longer search time without
improving the quality of the returned similar \stmt.  We expect this
data to be similar to the next statement to be predicted.

\MyPara{Implementation} In the collection phase, \TecoTool uses
JavaParser~\cite{JavaParser} to collect source code and
ASM~\cite{bruneton2002asm} to collect bytecode.  The collection phase
takes \NumCollectionPerProject{s} per project on average.
All analysis algorithms are implemented in Python, with the help of
ASM for partially interpreting bytecode (for
\UseMacro{no-types-local}) and scikit-learn~\cite{scikit-learn} for
the BM25 algorithm.
The analysis phase takes
\NumAnalysisPerTestMin{s}--\NumAnalysisPerTestMax{s} per \testm on
average, depending on \semdata data.

\begin{figure}[t]
\centering

\newsavebox{\lstExampleFull}
\begin{lrbox}{\lstExampleFull}
\begin{lstlisting}[language=java-pretty]
public class GMOperation 
    extends org.im4java.core.GMOperation {
  public GMOperation addImage(final File file) {...}(*@\label{line:mut-sign}@*)
...}

public class GMOperationTest {
  GMOperation sut;(*@\label{line:field-sut}@*)
  @Before 
  public void setup() {... sut = new GMOperation(); ...}(*@\label{line:setup}@*)

  @Test 
  public void addImage_ThrowsException_WhenFileIsNull()
      throws Exception {
    exception.except(IllegalArgumentException.class);
    sut.addImage((File) null);(*@\label{line:next-stmt}@*)
  }
...}
\end{lstlisting}
\end{lrbox}

\begin{tikzpicture}[
  font=\footnotesize, 
  box/.style={rectangle, scale=1, inner sep=1pt},
  box-bg/.style={box},
  box-label/.style={box, inner sep=2pt, draw=none, fill=blue!10, rounded corners},
  color-suf/.style={draw=yellow!80!black, fill=yellow!10},
  color-sem/.style={draw=blue!80!black, fill=blue!10},
  color-output/.style={draw=green!80!black, fill=green!10},
  anno/.style={font={\footnotesize}},
]
  \node (c-nw) at (0,0) [coordinate] {};

  \node (bg-s2) [below right = 4ex and 14.85em of c-nw] [box-bg, color-sem, minimum height=2ex, minimum width=2em] {};
  \node (l-s2) [below right = .1ex and 0 of bg-s2.south west] [box-label] {(\UseMacro{no-types-absent})~\UseMacro{name-types-absent}};
  \node (bg-s4) [below right = 13ex and .65em of c-nw] [box-bg, color-sem, minimum height=4ex, minimum width=23em] {};
  \node (l-s4) [below left = .1ex and 0 of bg-s4.south east] [box-label] {(\UseMacro{no-setup-teardown})~\UseMacro{name-setup-teardown}};
  \node (bg-output) [below right = 25.5ex and 1.35em of c-nw] [box-bg, color-output, minimum height=2ex, minimum width=11em] {};
  \node (b-full) [below right = 0 and 0 of c-nw] [box] {\usebox\lstExampleFull};

\end{tikzpicture}

\caption{Some kinds of \semdata (\UseMacro{no-types-absent} and
\UseMacro{no-setup-teardown}, highlighted in the first two blue boxes)
extracted for the example in \figurename~\ref{fig:example-data}, which
help \TecoTool generate the correct next \stmt (highlighted in the
last green box).\label{fig:semdata:example}}
\end{figure}

\MyPara{Example} \figurename~\ref{fig:semdata:example} shows two kinds
of \semdata (\UseMacro{no-types-absent} and
\UseMacro{no-setup-teardown}) extracted for the example in
\figurename~\ref{fig:example-data} that help the generating the
correct next \stmt. 
The (\UseMacro{no-types-absent})~\UseMacro{name-types-absent} data is
\CodeIn{File}, because calling the \mut{} \CodeIn{addImage} requires
\CodeIn{GMOperation} (because \CodeIn{addImage} is not a static
method) and \CodeIn{File}, but \CodeIn{GMOperation} is already
available as a field \CodeIn{sut} in the test class (on line~\ref{line:field-sut}).
The (\UseMacro{no-setup-teardown})~\UseMacro{name-setup-teardown} data
is the method \CodeIn{setup} in the test class
\CodeIn{GMOperationTest} which initializes the \CodeIn{sut} field (on
line~\ref{line:setup}).
Note that the other kinds of \semdata are either empty or not useful
for this example, but are useful for some other examples.

\section{\TecoTool's Deep Learning Model}
\label{sec:model}

\begin{figure}[t]
  \centering
  \includegraphics[width=.9\linewidth]{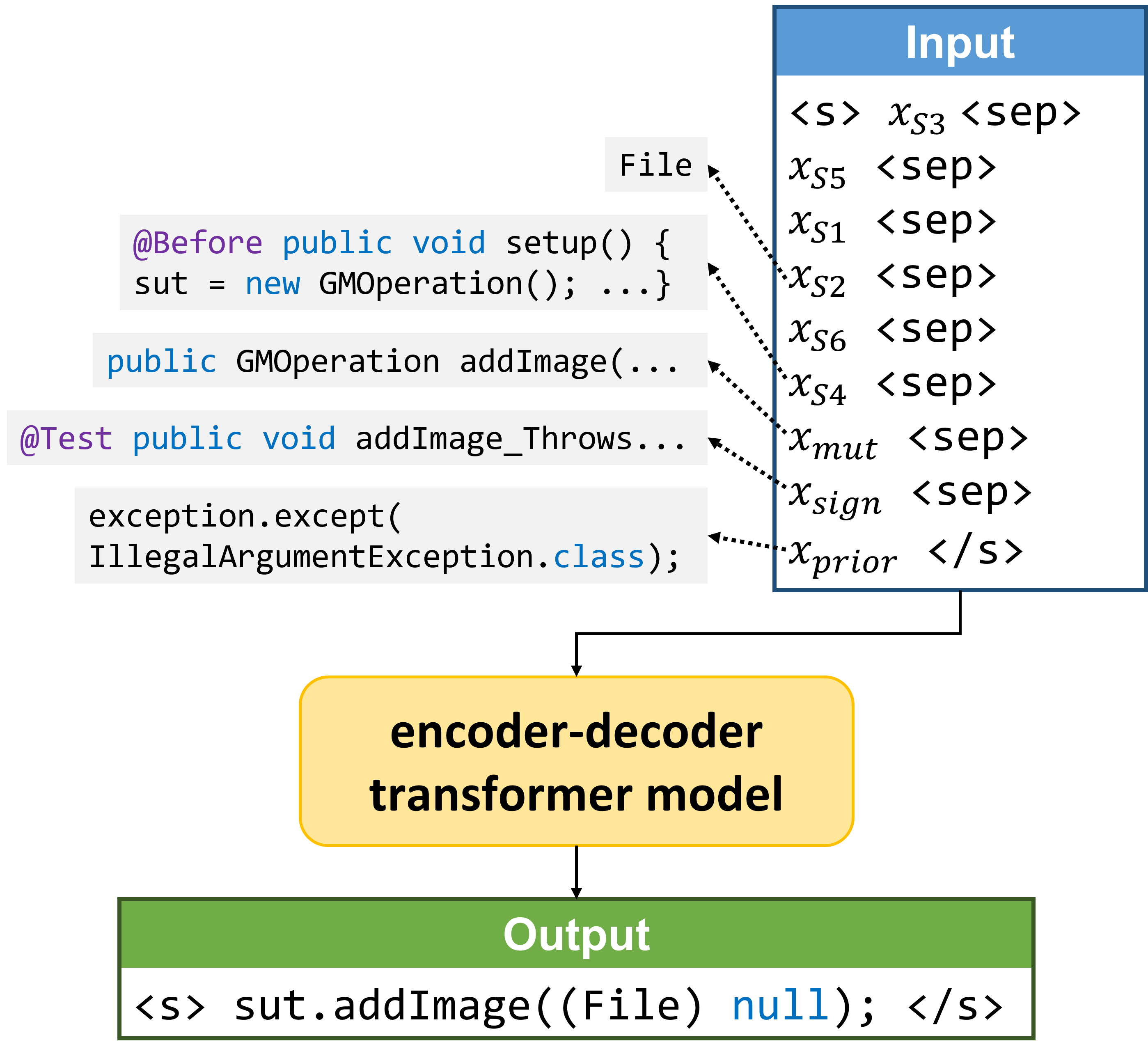}
  \caption{\TecoTool's model architecture.\label{fig:model}}
\end{figure}

This section describes the deep learning approach that \TecoTool uses
to solve the \testcomp task.  \figurename~\ref{fig:model} illustrates
the overall model architecture: an encoder-decoder transformer model
whose input includes both \semdata and \sufdata and output is the next
\stmt.

\subsection{Encoder-Decoder Transformer Model}
\label{sec:model:encdec}

Our model is based on the encoder-decoder architecture that considers
both input and output as sequences, which has been applied to many
sequence generation tasks including code
summarization~\cite{IyerETAL16Summarizing,
LeClairETAL19Neural, AhmadETAL20Transformer-based} and code
generation~\cite{LinETAL18NL2Bash, WeiETAL19CodeGen}.  In the context
of \TecoTool, the input is the \sufdata (\mut, test signature, and
\priorstmts) plus the \semdata extracted from the test to be
completed (\UseMacro{no-types-local}-\UseMacro{no-similar-stmt}),
and the output is the next statement.

More formally, \TecoTool is given the test to be
completed \aTest and the \codeut \aCodeUT as inputs, where 
\aCodeUT includes the \mut \aMUT,
and \aTest consists of two parts: the test signature \aSign and
\priorstmts \aPriorStmts.  The goal is to generate the next statement
\aOutput.  \TecoTool extracts \semdata as described in
Section~\ref{sec:semdata}: 
\begin{flalign*}
\UseMacro{a-types-local},
\UseMacro{a-types-absent}, \UseMacro{a-fields-notset},
\UseMacro{a-setup-teardown}, \UseMacro{a-last-called-method},
\UseMacro{a-similar-stmt} &= \aSanalysis(\aTest, \aCodeUT)
\end{flalign*}

Each input piece \aInputPiece and output \aOutput is a sequence of
\subtokens, which is obtained by \subtokenizing the code or extracted
data's string format using the BPE (byte-pair encoding)
algorithm~\cite{radford2019language, SennrichETAL16Subword}.
\TecoTool combines the input pieces into a single input sequence by
concatenating them with a delimiter \aSep (the ordering of the
sequences is configurable):
\begin{flalign*}
\aInput =& 
\ \UseMacro{a-fields-notset}\ \aSep
\ \UseMacro{a-last-called-method}\ \aSep
\ \UseMacro{a-types-local}\ \aSep
\ \UseMacro{a-types-absent}\ \aSep
\ \UseMacro{a-similar-stmt}\ \aSep\\
&\ \UseMacro{a-setup-teardown}\ \aSep
\ \aMUT\ \aSep
\ \aSign\ \aSep
\ \aPriorStmts
\end{flalign*}

The maximum number of \subtokens that \TecoTool can accept, due to the
limitation of the underlying model, is \NumMaxSeqLen. If the input
sequence is longer than that, \TecoTool truncates the input sequence
from the beginning. In our experiments, the number of \subtokens in
the input sequence ranges from \UseMacro{corpus-all-seq_len-MIN} to
\UseMacro{corpus-all-seq_len-MAX} (average:
\UseMacro{corpus-all-seq_len-AVG}) and exceeds the limitation of
\NumMaxSeqLen \subtokens in \UseMacro{corpus-all-seq_len_gt_512}\% of
the cases.  As a result, more important information should be placed
at the end of the input sequence to avoid being truncated.
The ordering of \semdata is decided based on our domain knowledge of
which kind of data would contain more important information, and we
always put the \sufdata at the end (as they deliver the basic
information for the task).  Exploring all possible orderings may
discover better models but is too computationally expensive.  We plan
to investigate the impact of the orderings by designing experiments
with more affordable costs in the future.

Then, \TecoTool uses a transformer model~\cite{vaswani2017attention}
to learn the conditional probability distribution $P(\aOutput |
\aInput)$.  The model computes this by first using an encoder to
encode the input sequence into a deep representation, and then using a
decoder which reads the deep representation and generates the output
sequence one \subtoken at a time:
\begin{flalign*}
\aEncH =& \aEncoder(\aInput)\\
P(\aOutput[i] | \aOutput[:i], \aInput) =& \aDecoder(\aOutput[:i], \aEncH), \text{for each $i$}
\end{flalign*}

\subsection{\Finetuning}
\label{sec:model:finetuning}

Recent work shows that \pretraining a large-scale transformer model on
a large corpus of code and text and then \finetuning the model on
downstream tasks lead to better performance than training a model from
scratch~\cite{WangETAL21CodeT5, ChenETAL21Codex, feng2020codebert,
AhmadETAL21PLBART}.  However, \pretraining is only performed on
\sufdata in order to be generalizable to many downstream tasks with
different semantics.  Prior work shows that large-scale \pretrained
models may not perform well on simple tasks of executing the
code~\cite{NyeETAL21ScratchPad}. This indicates that they learned
little about the semantics of execution.

We believe that \finetuning on \semdata is vital for \pretrained
models to perform well on execution-related tasks such as \testcomp.
During \pretraining, the model mainly learns the syntax and grammar of
programming languages.  If only \sufdata is used during \finetuning,
the model would have to infer the semantics of execution, which can be
very inaccurate because the execution can be complicated and the
context provided by the \sufdata is limited.  By contrast, \semdata are
extracted using reliable \sanalysis algorithms in \TecoTool so that
the model can directly use such data instead of inferring.  Thus, during
\finetuning of \TecoTool, the model
learns how to understand
and process the additional \semdata in addition to the \sufdata.

For the \pretrained model, we use CodeT5~\cite{WangETAL21CodeT5}, which is
a large-scale encoder-decoder transformer model \pretrained on a
bimodal corpus of code in multiple programming languages (including
Java) and text.  We \finetune the model on a corpus for \testcomp
\aCorpus by minimizing the cross-entropy loss:
\begin{flalign*}
\aLoss =& \sum_{x, y \in \aCorpus} -\log P(\aOutput | \aInput) =
\sum_{\substack{x, y \in \aCorpus \\ i \in \left[0,
|\aOutput|\right)}} -\log P(\aOutput[i] | \aOutput[:i], \aInput)
\end{flalign*}

\subsection{Evaluation}
\label{sec:model:inference}

At evaluation time, \TecoTool uses the beam search algorithm with a
beam size of \NumBeamSize.  Specifically, starting from a special
begin-of-sequence \subtoken $\aOutput[0] = \aBos$, \TecoTool
iteratively runs \aDecoder to generate the most likely next \subtokens
which is appended to the output sequence; only the top \NumBeamSize
sequences with the highest total probability are kept at each step.
Each output sequence is completed upon generating a special
end-of-sequence \subtoken \aEos.  The beam search terminates after
generating \NumBeamSize completed output sequences.  %
As repeating the same \subtoken is unlikely, we apply a repetition
prevention mechanism that penalizes the probability of generating the
same \subtoken as the previous \subtoken~\cite{keskar2019ctrl}.

\subsection{Reranking by Execution}
\label{sec:model:reranking}

The model can generate plausible outputs by maximizing the generation
probability that it learnt during \finetuning.  However, there is no
guarantee for the generated \stmts---\subtoken sequences---to be
compilable and runnable code.  Arguably, generating compilable and
runnable code is more important than generating plausible but
non-executable code in the \testcomp task, because it is crucial for
developers to run the code and observe the runtime behavior, in order
to further improve the codebase.

We propose to use \emph{reranking by test execution} to improve the quality of the
generated \stmts.  Specifically, after collecting the top-\NumBeamSize
predictions from beam search \aOutputs ranked by their probabilities,
\TecoTool checks whether each of them is compilable and runnable.
Then, \TecoTool reranks the outputs into $\hat{\aOutputs}$ where one
output $\aOutput_i$ is ranked higher than another $\aOutput_j$ if:
(1)~$\aOutput_i$ is runnable and $\aOutput_j$ is not; or (2)~both are
not runnable, but $\aOutput_i$ is compilable and $\aOutput_j$ is not;
or (3)~both have the same runnable and compilable status, and
$P(\aOutput_i) > P(\aOutput_j)$.  In this way, generated \stmts that
are compilable and runnable are prioritized over the others.

\begin{figure}[t]
\centering
\includegraphics[width=\linewidth]{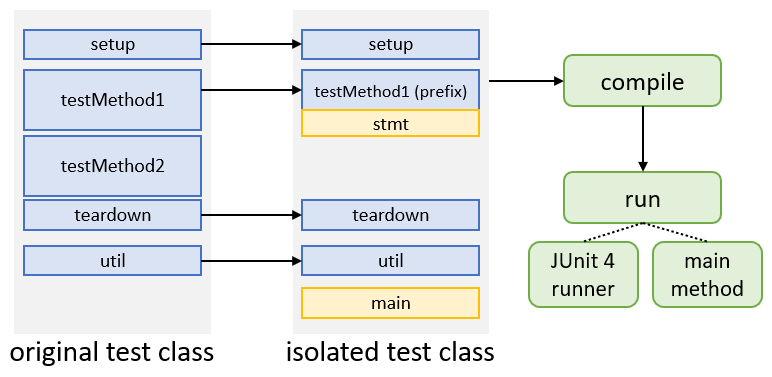}
\caption{Procedure of detecting whether a generated \stmt is
compilable and runnable.\label{fig:compilable-runnable}}
\end{figure}

\TecoTool detects whether a generated \stmt is compilable and runnable
by putting it in an ad-hoc \testc with the required context, isolated from
being affected by other \testms in the same project.  The procedure is
illustrated in \figurename~\ref{fig:compilable-runnable}, specifically:
\begin{enumerate}[topsep=3pt,itemsep=3pt,partopsep=0ex,parsep=0ex,leftmargin=*]
\item Create a class with a \testm using the
signature and \priorstmts, followed by the generated \stmt.
\item Extract the other non-test methods from the original \testc
(including setup, teardown, and utility methods) into the created
class.
\item Generate an ad-hoc main method which calls setup methods, the
\testm, and teardown methods in order.
\item Compile the generated class with all the dependencies specified
in the project's build configuration as well as all non-test classes
in the project.
\item If the compilation succeeds (at which point the \stmt is
considered to be compilable), execute the compiled class; the \stmt is
considered to be runnable only if there is no exception or assertion
failure during the execution. The execution of the class is performed
by one of the following:
\begin{enumerate}[topsep=3pt,itemsep=3pt,partopsep=0ex,parsep=0ex,leftmargin=*]
\item If the project is using JUnit 4~\cite{JUnit4} testing framework,
we try to use its command line runner to run the class with the
generated \stmt; the JUnit 4 runner is more robust because it properly
utilizes all JUnit features that our ad-hoc main method does not
handle, e.g., \CodeIn{@RunWith}.
\item If the project is not using JUnit 4 or running with JUnit 4 runner
failed, we run the ad-hoc main method.
\end{enumerate}
\end{enumerate}

\section{Corpus}
\label{sec:corpus}

As \testcomp is a new task, we construct a large-scale corpus that can
serve as a testbed for our work and future research.  We collected
data from the same subject projects used by
CodeSearchNet~\cite{HusainETAL20CodeSearchNet}, which is a large
corpus of code and comments that is frequently used in ML+code
research~\cite{WangETAL21CodeT5, LuETAL21CodeXGLUE, feng2020codebert}.  Out of the
\NumRepoCSNOrigin Java projects in
CodeSearchNet, we used the \NumRepoCSNFiltered projects
that: (1)~use the Maven build system (for the simplicity of data
collection; \TecoTool is not limited to any build system); (2)~compile
successfully, and (3)~have a license that permits the use of its data.
We collected the corpus in Spring 2022.  To ensure corpus quality, we
try to use the latest stable revision of each project by finding its
latest git-tag; but if it does not have any git-tag on or after Jan
1st, 2020, we use its latest revision.

To extract \testms from these projects, we first collected the set of
code elements from each project using the same toolchain for the
collection phase of \TecoTool's \sanalysis
(Section~\ref{sec:semdata}).  We identified the \testms written in
JUnit 4~\cite{JUnit4} and JUnit 5~\cite{JUnit5} testing frameworks,
which are the main frameworks used for writing tests in Java.
Specifically, we searched for methods with a test annotation
(\CodeIn{@org.junit.Test} or \CodeIn{@org.junit.jupiter.api.Test}) and
without an ignored-test annotation (\CodeIn{@org.junit.Ignore} or
\CodeIn{@org.junit.jupiter.api.Disabled}).  This initial search
resulted in \UseMacro{corpus-initial-num_test} \testms in all
projects.

\begin{table}[t]
\centering
\begin{small}
\caption{Statistics of our corpus. \UseMacro{TH-num_proj} = number of
projects; \UseMacro{TH-num_test} = number of \testm;
\UseMacro{TH-num_stmt} = number of \stmts; \UseMacro{TH-tok_test} =
average number of tokens in \testm; \UseMacro{TH-tok_mut} = average
number of tokens in \mut. \label{tab:corpus}}

\begin{tabular}{@{}l|r@{\hspace{6pt}}r@{\hspace{6pt}}r@{\hspace{6pt}}r@{\hspace{6pt}}r@{}}
\toprule
 & \textbf{\UseMacro{TH-num_proj}} & \textbf{\UseMacro{TH-num_test}} & \textbf{\UseMacro{TH-num_stmt}} & \textbf{\UseMacro{TH-tok_test}} & \textbf{\UseMacro{TH-tok_mut}} \\ 
\midrule
\UseMacro{TH-corpus-all}
 & \UseMacro{corpus-all-num_proj}
 & \UseMacro{corpus-all-num_test}
 & \UseMacro{corpus-all-num_stmt}
 & \UseMacro{corpus-all-tok_test}
 & \UseMacro{corpus-all-tok_mut}
\\
\UseMacro{TH-corpus-train}
 & \UseMacro{corpus-train-num_proj}
 & \UseMacro{corpus-train-num_test}
 & \UseMacro{corpus-train-num_stmt}
 & \UseMacro{corpus-train-tok_test}
 & \UseMacro{corpus-train-tok_mut}
\\
\UseMacro{TH-corpus-val}
 & \UseMacro{corpus-val-num_proj}
 & \UseMacro{corpus-val-num_test}
 & \UseMacro{corpus-val-num_stmt}
 & \UseMacro{corpus-val-tok_test}
 & \UseMacro{corpus-val-tok_mut}
\\
\UseMacro{TH-corpus-test}
 & \UseMacro{corpus-test-num_proj}
 & \UseMacro{corpus-test-num_test}
 & \UseMacro{corpus-test-num_stmt}
 & \UseMacro{corpus-test-tok_test}
 & \UseMacro{corpus-test-tok_mut}
\\
\bottomrule
\end{tabular}

\vspace{10pt}
\end{small}
\end{table}

Then, we further filtered the \testms to ensure corpus quality.  We
filtered \testms that are badly named (e.g., \CodeIn{test0};
\UseMacro{corpus-filter-remove_badly_named_tests} cases) or do not
follow the required signature of tests (e.g., parameter list is not
empty, return type is not void;
\UseMacro{corpus-filter-remove_irregular_tests} cases).  Then, we
tried to locate the \mut for each \testm, using the following enhanced
procedure originally proposed by Waston et
al.~\cite{WatsonETAL20ATLAS}:
\begin{enumerate}[topsep=2pt,itemsep=0pt,partopsep=0ex,parsep=0ex,leftmargin=*]
\item If there is only one call to a method, select it (as the \mut);
\item If a \clzut can be found by removing ``Test'' from the \testc's name:
\begin{enumerate}[topsep=0pt,itemsep=0pt,partopsep=0ex,parsep=0ex,leftmargin=*]
    \item If there is only one call to a method declared in \clzut, select it;
    \item Select the last method declared in \clzut called before the first assertion \stmt, if any;
\end{enumerate}
\item Select the last method called before the first assertion \stmt, if any;
\item Select the last method called, if any.
\end{enumerate}
\noindent
We removed \UseMacro{corpus-filter-cannot_locate_focalm} \testms for
which we could not locate the \mut after this procedure.

We used the line number table to find the bytecode instructions
corresponding to each \stmt, and we removed
\UseMacro{corpus-filter-code_mapping_error} cases where we could not
do this because of multiple \stmts on the same line.  After that, we
set size constraints on the data: the \testm should have at least
\UseMacro{corpus-limit-min_test_stmt} \stmt (filtered
\UseMacro{corpus-filter-min_test_stmt} cases) and at most
\UseMacro{corpus-limit-max_test_stmt} \stmts (filtered
\UseMacro{corpus-filter-max_test_stmt} cases); the \mut should have at
most \UseMacro{corpus-limit-max_focalm_tok} tokens (filtered
\UseMacro{corpus-filter-max_focalm_tok} cases); the \mut and the
\testm together should have at most
\UseMacro{corpus-limit-max_test_focalm_tok} tokens (filtered
\UseMacro{corpus-filter-max_test_focalm_tok} cases); each \stmt in the
\testm should have at most \UseMacro{corpus-limit-max_tok_per_stmt}
tokens (filtered \UseMacro{corpus-filter-max_tok_per_stmt} cases).

We also removed several cases that introduce extra overhead during
analysis:
\testms with if statements, loops, and try blocks, because they entail
non-sequential control flow which is not suitable to be modeled by
predicting the next \stmt given \priorstmts
(\UseMacro{corpus-filter-has_control_flow} cases); and \testms using
lambda expressions~\cite{LambdaExpressions}, because they prevent many
\sanalysis algorithms from working
(\UseMacro{corpus-filter-has_invokedynamic} cases).
We plan to lift these limitations in future work.

Lastly, we mask the string literals in the data by replacing them with
a common token ``STR'', similar to prior work on code
completion~\cite{SvyatkovskiyETAL20IntelliCode}.  Although string
literals are frequently used in \testms, for example as logging
messages, test inputs, or expected outputs, they pose challenges for a
pure-deep-learning solution to generate because they have a different
style than other parts of the code and can sometimes be very long.
Thus, we focus on predicting the next \stmt with masked string
literals, and leave predicting the content of the string literals as
future work.

After filtering, we obtained a corpus with \UseMacro{corpus-all-num_proj}
projects (removed \UseMacro{corpus-filter-no_data} projects because no
data was left after filtering), \UseMacro{corpus-all-num_test} \testms,
and \UseMacro{corpus-all-num_stmt} \stmts.

We follow the same project-level \train/\val/\test split as
CodeSearchNet.  Because CodeT5, the \pretrained model that \TecoTool
uses in our experiments, also followed the same project-level split,
our experiments will not have data leakage issues of evaluating on the
data that the model was \pretrained on.  Table~\ref{tab:corpus} shows
the statistics of our corpus, where the first row is for the entire
corpus, and the other three rows are for each set after the split.
Out of the \UseMacro{corpus-all-num_test} \testms,
\UseMacro{corpus-all-runnable-any-num_test}
(\UseMacro{corpus-all-runnable-any-pct}\%) are runnable following our
procedure described in Section~\ref{sec:model:reranking}.
Note that with the masking of string literals, some \testms that would
originally pass may be considered as ``not runnable'' in our current
corpus (e.g., when the \testm compares a variable with a string
literal).

\section{Experiments Setup}
\label{sec:exp}

We assess the performance of \TecoTool by answering the following research
questions:

\DefMacro{rq-overall}{RQ1}
\DefMacro{rq-runnable}{RQ2}
\DefMacro{rq-oracle}{RQ3}
\DefMacro{rq-ablation}{RQ4}
\DefMacro{rq-semdata}{RQ5}

\vspace{4pt}
\noindent\textbf{\UseMacro{rq-overall}}: What is the performance of
\TecoTool on the \testcomp task and how does it compare to baselines?

\vspace{4pt}
\noindent\textbf{\UseMacro{rq-runnable}}: On the runnable subset of
\test set, how frequently can \TecoTool predict a compilable and runnable next \stmt?

\vspace{4pt}
\noindent\textbf{\UseMacro{rq-oracle}}: What is the performance of
\TecoTool on \oraclegen, which is a downstream application of the
\testcomp task, and how does it compare to prior work?

\vspace{4pt}
\noindent\textbf{\UseMacro{rq-ablation}}: How does reranking by
execution help with more accurately predicting the next \stmt?

\vspace{4pt}
\noindent\textbf{\UseMacro{rq-semdata}}: How does each kind of
\semdata help with more accurately predicting the next \stmt, and how
complementary are different kinds of \semdata?

\vspace{4pt}

To answer these questions, we setup an experiment to evaluate
\TecoTool and baseline models on our \testcomp corpus.  We train each
model on the \train and \val sets (\val set is used for tuning
hyper-parameters and early stopping), apply the model to predict each
\stmt of each \testm in the \test set (or subsets of the \test set), 
and measure the quality of the prediction via a number of evaluation
metrics, both intrinsically and extrinsically.

All models are trained and evaluated on machines equipped with 4
NVidia 1080-TI GPUs and Intel(R) Xeon(R) CPU E5-2620 v4 @ 2.10GHz.  We
ran each experiment \NumTrials times with different random seeds and
report average values.  When comparing models, we conducted
statistical significance tests using bootstrap
tests~\cite{Berg-KirkpatrickETAL12Empirical} with a 95\% confidence level.

We next describe the \TecoTool models
(Section~\ref{sec:eval:our-models}) and baseline models
(Section~\ref{sec:eval:baseline-models}) used in the experiments, the
subsets of the \test set for
computing compilable and runnable metrics and evaluating on the
\oraclegen task (Section~\ref{sec:eval:subsets}), and the evaluation
metrics (Section~\ref{sec:eval:metrics}).

\subsection{\TecoTool Models}
\label{sec:eval:our-models}

We run a \UseMacro{model-teco-icse23} model that uses all \NumSemData
kinds of \semdata and with reranking by test execution.  
To study \UseMacro{rq-ablation}, we run a
\UseMacro{model-teco-icse23-norr} model that uses the same \semdata
but does not use reranking.
To study \UseMacro{rq-semdata}, we run \NumSemData \TecoTool models
with only one kind of \semdata at a time, which we call
\UseMacro{model-teco-id} (e.g., \UseMacro{model-teco-s1} only uses
\UseMacro{no-types-local}).

\subsection{Baseline Models}
\label{sec:eval:baseline-models}

We compare our \TecoTool models to the following baseline models that only use \sufdata.

\MyParaOnly{CodeT5}~\cite{WangETAL21CodeT5} is a \pretrained
encoder-decoder transformer model for code-related tasks, and is built
on top of Google's popular T5 framework~\cite{raffel2020exploring}.
CodeT5 was \pretrained on eight commonly used
programming
languages (including Java) using both mask language modeling and
identifier name recovering tasks.  We \finetune \TecoTool models based
on CodeT5.  As such, we compare to a baseline \UseMacro{model-codet5}
model that is finetuned on \sufdata.  For completeness, we also
compare to a \UseMacro{model-codet5-noft} that is only \pretrained
and not \finetuned.

\MyParaOnly{\UseMacro{model-codegpt}}~\cite{LuETAL21CodeXGLUE} is a
decoder-only transformer model built on
GPT-2~\cite{radford2019language}.  We used the \CodeIn{java\text{-}adapted}
version of it, which is initialized from GPT-2 \pretrained on natural
language, and then further \pretrained on a corpus of Java code.
Svyatkovskiy et al.~\cite{SvyatkovskiyETAL20IntelliCode} used a very
similar model (which is not publicly available) for code completion.
As \UseMacro{model-codegpt} tends to generate longer code than a \stmt
(without generating the \aEos \subtoken to stop the generation), we
slightly modify its decoding algorithm to terminate upon generating
the first `;' \subtoken for the \testcomp task.

\Oraclegen is the task of generating the assertion statement given the
\codeut (including the \mut), \testsign, and \priorstmts
before the assertion statement.  When studying this task, we additionally compare to the
following two deep learning baseline models for \oraclegen developed
in prior work, both of which only use \sufdata.  Following the prior
works, we only consider generating the first assertion statement in
each \testm.

\MyParaOnly{\UseMacro{model-atlas}}~\cite{WatsonETAL20ATLAS} is a RNN
encoder-decoder model for \oraclegen.  We used the ``raw model''
version of it, i.e., that does not abstract out the identifiers in
code.

\MyParaOnly{\UseMacro{model-toga}}~\cite{DinellaETAL22TOGA} is a
transformer encoder-only model for classifying the suitability of an
assertion statement for an incomplete \testm without assertions.  It
can be used for \oraclegen by first generating a set of assertion
statements and then using the model to rank them and select the best
one.  The model is initialized from CodeBERT~\cite{feng2020codebert},
which is also \pretrained on the CodeSearchNet
corpus~\cite{HusainETAL20CodeSearchNet}.

For all baseline models, we use the default hyper-parameters and
training configurations recommended by the authors.  We train
\UseMacro{model-codet5} and \UseMacro{model-codegpt} on the entire
\train and \val set of our corpus.  We train \UseMacro{model-atlas}
and \UseMacro{model-toga} on a subset of our \train and \val set that
only predicts the first assertion \stmt in each \testm, which contains
\UseMacro{corpus-train-assert-num_stmt} \stmts and
\UseMacro{corpus-val-assert-num_stmt} \stmts, respectively.

\subsection{Subsets of the \Test Set}
\label{sec:eval:subsets}

To study the ability of models in predicting a compilable and runnable
next statement, we evaluate models on the \runnablesubset. That is,
the subset of the \test set where the
gold  (i.e., developer-written) \stmt is runnable.
We follow the same procedure to check if the gold \stmt is runnable as
described in Section~\ref{sec:model:reranking}.  Not all gold \stmts
can be successfully executed because of the difficulties in setting up
the proper runtime environment, such as missing resources (that may
need to be downloaded or generated via other commands), requiring
other runtime environments than Java, etc.  Our \runnablesubset
contains \UseMacro{corpus-test-runnable-any-num_stmt} \stmts
(\UseMacro{corpus-test-runnable-any-pct}\% of all \stmts in the \test
set) from \UseMacro{corpus-test-runnable-any-num_test} \testms.

To study the \oraclegen task, we evaluate models on the \oraclesubset:
the subset of the \test set where the \stmt to generate is the first
assertion \stmt in the \testm, which contains
\UseMacro{corpus-test-assert-num_stmt} \stmts.  To compute compilable
and runnable metrics on the \oraclegen task, we evaluate models
on the \oraclerunnablesubset: the subset of the \oraclesubset where
the gold \stmt is runnable, which contains
\UseMacro{corpus-test-runnable-assert-num_stmt} \stmts.

\begin{table*}[t]
\centering
\begin{small}
\caption{Results for \TecoTool and baseline models. \UseMacro{TC-notice-bold} \UseMacro{TC-notice-sign-subtable}\label{tab:results-main}}
\begin{minipage}[t]{\textwidth}
  \centering
  \subcaption{On the \test set.\label{tab:results-main-any-CSNm}}
  \vspace{-5pt}
  
%% Automatically generated by pyutil.latex 

\begin{tabular}{@{}l|@{\hspace{2pt}}r@{\hspace{2pt}}r@{\hspace{2pt}}r@{\hspace{2pt}}r@{\hspace{2pt}}r@{\hspace{2pt}}r@{}}
\toprule
\textbf{\UseMacro{TH-model}}
 & \textbf{\UseMacro{TH-xmatch}}
 & \textbf{\UseMacro{TH-xmatch-top10}}
 & \textbf{\UseMacro{TH-bleu}}
 & \textbf{\UseMacro{TH-code-bleu}}
 & \textbf{\UseMacro{TH-edit-sim}}
 & \textbf{\UseMacro{TH-rouge}}
\\
\midrule
\UseMacro{model-codet5}
 & \UseMacro{res-codet5-CSNm-eval-any-stmt/test-bs10-last-xmatch}
 & \UseMacro{res-codet5-CSNm-eval-any-stmt/test-bs10-last-xmatch-top10}
 & \UseMacro{res-codet5-CSNm-eval-any-stmt/test-bs10-last-bleu}
 & \UseMacro{res-codet5-CSNm-eval-any-stmt/test-bs10-last-code-bleu}
 & \UseMacro{res-codet5-CSNm-eval-any-stmt/test-bs10-last-edit-sim}
 & \UseMacro{res-codet5-CSNm-eval-any-stmt/test-bs10-last-rouge-f}
\\
\UseMacro{model-codet5-noft}
 & \UseMacro{res-codet5-noft-CSNm-eval-any-stmt/test-bs10-last-xmatch}
 & \UseMacro{res-codet5-noft-CSNm-eval-any-stmt/test-bs10-last-xmatch-top10}
 & \UseMacro{res-codet5-noft-CSNm-eval-any-stmt/test-bs10-last-bleu}
 & \UseMacro{res-codet5-noft-CSNm-eval-any-stmt/test-bs10-last-code-bleu}
 & \UseMacro{res-codet5-noft-CSNm-eval-any-stmt/test-bs10-last-edit-sim}
 & \UseMacro{res-codet5-noft-CSNm-eval-any-stmt/test-bs10-last-rouge-f}
\\
\UseMacro{model-codegpt}
 & \UseMacro{res-codegpt-CSNm-eval-any-stmt/test-bs10-last-xmatch}
 & \UseMacro{res-codegpt-CSNm-eval-any-stmt/test-bs10-last-xmatch-top10}
 & \UseMacro{res-codegpt-CSNm-eval-any-stmt/test-bs10-last-bleu}
 & \UseMacro{res-codegpt-CSNm-eval-any-stmt/test-bs10-last-code-bleu}
 & \UseMacro{res-codegpt-CSNm-eval-any-stmt/test-bs10-last-edit-sim}
 & \UseMacro{res-codegpt-CSNm-eval-any-stmt/test-bs10-last-rouge-f}
\\
\midrule
\UseMacro{model-teco-icse23}
 & \textbf{\UseMacro{res-teco-icse23-CSNm-eval-any-stmt/test-bs10-last-xmatch}}
 & \textbf{\UseMacro{res-teco-icse23-CSNm-eval-any-stmt/test-bs10-last-xmatch-top10}}
 & \textbf{\UseMacro{res-teco-icse23-CSNm-eval-any-stmt/test-bs10-last-bleu}}
 & \textbf{\UseMacro{res-teco-icse23-CSNm-eval-any-stmt/test-bs10-last-code-bleu}}
 & \textbf{\UseMacro{res-teco-icse23-CSNm-eval-any-stmt/test-bs10-last-edit-sim}}
 & \textbf{\UseMacro{res-teco-icse23-CSNm-eval-any-stmt/test-bs10-last-rouge-f}}
\\
\bottomrule
\end{tabular}

  \vspace{5pt}
\end{minipage}
\begin{minipage}[t]{\textwidth}
  \centering
  \subcaption{On the \runnablesubset.\label{tab:results-main-runnable-any-CSNm}}
  \vspace{-5pt}
  
%% Automatically generated by pyutil.latex 

\begin{tabular}{@{}l|@{\hspace{2pt}}r@{\hspace{2pt}}r@{\hspace{2pt}}r@{\hspace{2pt}}r@{\hspace{2pt}}r@{\hspace{2pt}}r@{\hspace{2pt}}r@{\hspace{2pt}}r@{}}
\toprule
\textbf{\UseMacro{TH-model}}
 & \textbf{\UseMacro{TH-compilable}}
 & \textbf{\UseMacro{TH-runnable}}
 & \textbf{\UseMacro{TH-xmatch}}
 & \textbf{\UseMacro{TH-xmatch-top10}}
 & \textbf{\UseMacro{TH-bleu}}
 & \textbf{\UseMacro{TH-code-bleu}}
 & \textbf{\UseMacro{TH-edit-sim}}
 & \textbf{\UseMacro{TH-rouge}}
\\
\midrule
\UseMacro{model-codet5}
 & \UseMacro{res-codet5-CSNm-eval-runnable-any-stmt/test-bs10-last-compilable}
 & \UseMacro{res-codet5-CSNm-eval-runnable-any-stmt/test-bs10-last-runnable}
 & \UseMacro{res-codet5-CSNm-eval-runnable-any-stmt/test-bs10-last-xmatch}
 & \UseMacro{res-codet5-CSNm-eval-runnable-any-stmt/test-bs10-last-xmatch-top10}
 & \UseMacro{res-codet5-CSNm-eval-runnable-any-stmt/test-bs10-last-bleu}
 & \UseMacro{res-codet5-CSNm-eval-runnable-any-stmt/test-bs10-last-code-bleu}
 & \UseMacro{res-codet5-CSNm-eval-runnable-any-stmt/test-bs10-last-edit-sim}
 & \UseMacro{res-codet5-CSNm-eval-runnable-any-stmt/test-bs10-last-rouge-f}
\\
\UseMacro{model-codet5-noft}
 & \UseMacro{res-codet5-noft-CSNm-eval-runnable-any-stmt/test-bs10-last-compilable}
 & \UseMacro{res-codet5-noft-CSNm-eval-runnable-any-stmt/test-bs10-last-runnable}
 & \UseMacro{res-codet5-noft-CSNm-eval-runnable-any-stmt/test-bs10-last-xmatch}
 & \UseMacro{res-codet5-noft-CSNm-eval-runnable-any-stmt/test-bs10-last-xmatch-top10}
 & \UseMacro{res-codet5-noft-CSNm-eval-runnable-any-stmt/test-bs10-last-bleu}
 & \UseMacro{res-codet5-noft-CSNm-eval-runnable-any-stmt/test-bs10-last-code-bleu}
 & \UseMacro{res-codet5-noft-CSNm-eval-runnable-any-stmt/test-bs10-last-edit-sim}
 & \UseMacro{res-codet5-noft-CSNm-eval-runnable-any-stmt/test-bs10-last-rouge-f}
\\
\UseMacro{model-codegpt}
 & \UseMacro{res-codegpt-CSNm-eval-runnable-any-stmt/test-bs10-last-compilable}
 & \UseMacro{res-codegpt-CSNm-eval-runnable-any-stmt/test-bs10-last-runnable}
 & \UseMacro{res-codegpt-CSNm-eval-runnable-any-stmt/test-bs10-last-xmatch}
 & \UseMacro{res-codegpt-CSNm-eval-runnable-any-stmt/test-bs10-last-xmatch-top10}
 & \UseMacro{res-codegpt-CSNm-eval-runnable-any-stmt/test-bs10-last-bleu}
 & \UseMacro{res-codegpt-CSNm-eval-runnable-any-stmt/test-bs10-last-code-bleu}
 & \UseMacro{res-codegpt-CSNm-eval-runnable-any-stmt/test-bs10-last-edit-sim}
 & \UseMacro{res-codegpt-CSNm-eval-runnable-any-stmt/test-bs10-last-rouge-f}
\\
\midrule
\UseMacro{model-teco-icse23}
 & \textbf{\UseMacro{res-teco-icse23-CSNm-eval-runnable-any-stmt/test-bs10-last-compilable}}
 & \textbf{\UseMacro{res-teco-icse23-CSNm-eval-runnable-any-stmt/test-bs10-last-runnable}}
 & \textbf{\UseMacro{res-teco-icse23-CSNm-eval-runnable-any-stmt/test-bs10-last-xmatch}}
 & \textbf{\UseMacro{res-teco-icse23-CSNm-eval-runnable-any-stmt/test-bs10-last-xmatch-top10}}
 & \textbf{\UseMacro{res-teco-icse23-CSNm-eval-runnable-any-stmt/test-bs10-last-bleu}}
 & \textbf{\UseMacro{res-teco-icse23-CSNm-eval-runnable-any-stmt/test-bs10-last-code-bleu}}
 & \textbf{\UseMacro{res-teco-icse23-CSNm-eval-runnable-any-stmt/test-bs10-last-edit-sim}}
 & \textbf{\UseMacro{res-teco-icse23-CSNm-eval-runnable-any-stmt/test-bs10-last-rouge-f}}
\\
\bottomrule
\end{tabular}

  \vspace{5pt}
\end{minipage}
\begin{minipage}[t]{\textwidth}
  \centering
  \subcaption{On the \oraclesubset.\label{tab:results-main-assert-CSNm}}
  \vspace{-5pt}
  
%% Automatically generated by pyutil.latex 

\begin{tabular}{@{}l|@{\hspace{2pt}}r@{\hspace{2pt}}r@{\hspace{2pt}}r@{\hspace{2pt}}r@{\hspace{2pt}}r@{\hspace{2pt}}r@{}}
\toprule
\textbf{\UseMacro{TH-model}}
 & \textbf{\UseMacro{TH-xmatch}}
 & \textbf{\UseMacro{TH-xmatch-top10}}
 & \textbf{\UseMacro{TH-bleu}}
 & \textbf{\UseMacro{TH-code-bleu}}
 & \textbf{\UseMacro{TH-edit-sim}}
 & \textbf{\UseMacro{TH-rouge}}
\\
\midrule
\UseMacro{model-codet5}
 & \UseMacro{res-codet5-CSNm-eval-assert-stmt/test-bs10-last-xmatch}
 & \UseMacro{res-codet5-CSNm-eval-assert-stmt/test-bs10-last-xmatch-top10}
 & \UseMacro{res-codet5-CSNm-eval-assert-stmt/test-bs10-last-bleu}
 & \UseMacro{res-codet5-CSNm-eval-assert-stmt/test-bs10-last-code-bleu}
 & \UseMacro{res-codet5-CSNm-eval-assert-stmt/test-bs10-last-edit-sim}
 & \UseMacro{res-codet5-CSNm-eval-assert-stmt/test-bs10-last-rouge-f}
\\
\UseMacro{model-codet5-noft}
 & $^{\alpha}$\UseMacro{res-codet5-noft-CSNm-eval-assert-stmt/test-bs10-last-xmatch}
 & \UseMacro{res-codet5-noft-CSNm-eval-assert-stmt/test-bs10-last-xmatch-top10}
 & \UseMacro{res-codet5-noft-CSNm-eval-assert-stmt/test-bs10-last-bleu}
 & \UseMacro{res-codet5-noft-CSNm-eval-assert-stmt/test-bs10-last-code-bleu}
 & \UseMacro{res-codet5-noft-CSNm-eval-assert-stmt/test-bs10-last-edit-sim}
 & \UseMacro{res-codet5-noft-CSNm-eval-assert-stmt/test-bs10-last-rouge-f}
\\
\UseMacro{model-codegpt}
 & \UseMacro{res-codegpt-CSNm-eval-assert-stmt/test-bs10-last-xmatch}
 & \UseMacro{res-codegpt-CSNm-eval-assert-stmt/test-bs10-last-xmatch-top10}
 & \UseMacro{res-codegpt-CSNm-eval-assert-stmt/test-bs10-last-bleu}
 & \UseMacro{res-codegpt-CSNm-eval-assert-stmt/test-bs10-last-code-bleu}
 & \UseMacro{res-codegpt-CSNm-eval-assert-stmt/test-bs10-last-edit-sim}
 & \UseMacro{res-codegpt-CSNm-eval-assert-stmt/test-bs10-last-rouge-f}
\\
\UseMacro{model-atlas}
 & $^{\alpha}$\UseMacro{res-atlas-CSNm-eval-assert-stmt/test-bs10-last-xmatch}
 & \UseMacro{res-atlas-CSNm-eval-assert-stmt/test-bs10-last-xmatch-top10}
 & \UseMacro{res-atlas-CSNm-eval-assert-stmt/test-bs10-last-bleu}
 & \UseMacro{res-atlas-CSNm-eval-assert-stmt/test-bs10-last-code-bleu}
 & \UseMacro{res-atlas-CSNm-eval-assert-stmt/test-bs10-last-edit-sim}
 & \UseMacro{res-atlas-CSNm-eval-assert-stmt/test-bs10-last-rouge-f}
\\
\UseMacro{model-toga}
 & \UseMacro{res-toga-CSNm-eval-assert-stmt/test-bs10-last-xmatch}
 & \UseMacro{res-toga-CSNm-eval-assert-stmt/test-bs10-last-xmatch-top10}
 & \UseMacro{res-toga-CSNm-eval-assert-stmt/test-bs10-last-bleu}
 & \UseMacro{res-toga-CSNm-eval-assert-stmt/test-bs10-last-code-bleu}
 & \UseMacro{res-toga-CSNm-eval-assert-stmt/test-bs10-last-edit-sim}
 & \UseMacro{res-toga-CSNm-eval-assert-stmt/test-bs10-last-rouge-f}
\\
\midrule
\UseMacro{model-teco-icse23}
 & \textbf{\UseMacro{res-teco-icse23-CSNm-eval-assert-stmt/test-bs10-last-xmatch}}
 & \textbf{\UseMacro{res-teco-icse23-CSNm-eval-assert-stmt/test-bs10-last-xmatch-top10}}
 & \textbf{\UseMacro{res-teco-icse23-CSNm-eval-assert-stmt/test-bs10-last-bleu}}
 & \textbf{\UseMacro{res-teco-icse23-CSNm-eval-assert-stmt/test-bs10-last-code-bleu}}
 & \textbf{\UseMacro{res-teco-icse23-CSNm-eval-assert-stmt/test-bs10-last-edit-sim}}
 & \textbf{\UseMacro{res-teco-icse23-CSNm-eval-assert-stmt/test-bs10-last-rouge-f}}
\\
\bottomrule
\end{tabular}

  \vspace{5pt}
\end{minipage}
\begin{minipage}[t]{\textwidth}
  \centering
  \subcaption{On the \oraclerunnablesubset.\label{tab:results-main-runnable-assert-CSNm}}
  \vspace{-5pt}
  
%% Automatically generated by pyutil.latex 

\begin{tabular}{@{}l|@{\hspace{2pt}}r@{\hspace{2pt}}r@{\hspace{2pt}}r@{\hspace{2pt}}r@{\hspace{2pt}}r@{\hspace{2pt}}r@{\hspace{2pt}}r@{\hspace{2pt}}r@{}}
\toprule
\textbf{\UseMacro{TH-model}}
 & \textbf{\UseMacro{TH-compilable}}
 & \textbf{\UseMacro{TH-runnable}}
 & \textbf{\UseMacro{TH-xmatch}}
 & \textbf{\UseMacro{TH-xmatch-top10}}
 & \textbf{\UseMacro{TH-bleu}}
 & \textbf{\UseMacro{TH-code-bleu}}
 & \textbf{\UseMacro{TH-edit-sim}}
 & \textbf{\UseMacro{TH-rouge}}
\\
\midrule
\UseMacro{model-codet5}
 & \UseMacro{res-codet5-CSNm-eval-runnable-assert-stmt/test-bs10-last-compilable}
 & \UseMacro{res-codet5-CSNm-eval-runnable-assert-stmt/test-bs10-last-runnable}
 & \UseMacro{res-codet5-CSNm-eval-runnable-assert-stmt/test-bs10-last-xmatch}
 & \UseMacro{res-codet5-CSNm-eval-runnable-assert-stmt/test-bs10-last-xmatch-top10}
 & \UseMacro{res-codet5-CSNm-eval-runnable-assert-stmt/test-bs10-last-bleu}
 & \UseMacro{res-codet5-CSNm-eval-runnable-assert-stmt/test-bs10-last-code-bleu}
 & \UseMacro{res-codet5-CSNm-eval-runnable-assert-stmt/test-bs10-last-edit-sim}
 & \UseMacro{res-codet5-CSNm-eval-runnable-assert-stmt/test-bs10-last-rouge-f}
\\
\UseMacro{model-codet5-noft}
 & \UseMacro{res-codet5-noft-CSNm-eval-runnable-assert-stmt/test-bs10-last-compilable}
 & \UseMacro{res-codet5-noft-CSNm-eval-runnable-assert-stmt/test-bs10-last-runnable}
 & $^{\alpha}$\UseMacro{res-codet5-noft-CSNm-eval-runnable-assert-stmt/test-bs10-last-xmatch}
 & \UseMacro{res-codet5-noft-CSNm-eval-runnable-assert-stmt/test-bs10-last-xmatch-top10}
 & \UseMacro{res-codet5-noft-CSNm-eval-runnable-assert-stmt/test-bs10-last-bleu}
 & \UseMacro{res-codet5-noft-CSNm-eval-runnable-assert-stmt/test-bs10-last-code-bleu}
 & \UseMacro{res-codet5-noft-CSNm-eval-runnable-assert-stmt/test-bs10-last-edit-sim}
 & \UseMacro{res-codet5-noft-CSNm-eval-runnable-assert-stmt/test-bs10-last-rouge-f}
\\
\UseMacro{model-codegpt}
 & \UseMacro{res-codegpt-CSNm-eval-runnable-assert-stmt/test-bs10-last-compilable}
 & \UseMacro{res-codegpt-CSNm-eval-runnable-assert-stmt/test-bs10-last-runnable}
 & \UseMacro{res-codegpt-CSNm-eval-runnable-assert-stmt/test-bs10-last-xmatch}
 & \textbf{\UseMacro{res-codegpt-CSNm-eval-runnable-assert-stmt/test-bs10-last-xmatch-top10}}
 & \UseMacro{res-codegpt-CSNm-eval-runnable-assert-stmt/test-bs10-last-bleu}
 & \UseMacro{res-codegpt-CSNm-eval-runnable-assert-stmt/test-bs10-last-code-bleu}
 & \UseMacro{res-codegpt-CSNm-eval-runnable-assert-stmt/test-bs10-last-edit-sim}
 & \UseMacro{res-codegpt-CSNm-eval-runnable-assert-stmt/test-bs10-last-rouge-f}
\\
\UseMacro{model-atlas}
 & \UseMacro{res-atlas-CSNm-eval-runnable-assert-stmt/test-bs10-last-compilable}
 & \UseMacro{res-atlas-CSNm-eval-runnable-assert-stmt/test-bs10-last-runnable}
 & $^{\alpha}$\UseMacro{res-atlas-CSNm-eval-runnable-assert-stmt/test-bs10-last-xmatch}
 & \UseMacro{res-atlas-CSNm-eval-runnable-assert-stmt/test-bs10-last-xmatch-top10}
 & \UseMacro{res-atlas-CSNm-eval-runnable-assert-stmt/test-bs10-last-bleu}
 & \UseMacro{res-atlas-CSNm-eval-runnable-assert-stmt/test-bs10-last-code-bleu}
 & \UseMacro{res-atlas-CSNm-eval-runnable-assert-stmt/test-bs10-last-edit-sim}
 & \UseMacro{res-atlas-CSNm-eval-runnable-assert-stmt/test-bs10-last-rouge-f}
\\
\UseMacro{model-toga}
 & \UseMacro{res-toga-CSNm-eval-runnable-assert-stmt/test-bs10-last-compilable}
 & \UseMacro{res-toga-CSNm-eval-runnable-assert-stmt/test-bs10-last-runnable}
 & \UseMacro{res-toga-CSNm-eval-runnable-assert-stmt/test-bs10-last-xmatch}
 & \UseMacro{res-toga-CSNm-eval-runnable-assert-stmt/test-bs10-last-xmatch-top10}
 & \UseMacro{res-toga-CSNm-eval-runnable-assert-stmt/test-bs10-last-bleu}
 & \UseMacro{res-toga-CSNm-eval-runnable-assert-stmt/test-bs10-last-code-bleu}
 & \UseMacro{res-toga-CSNm-eval-runnable-assert-stmt/test-bs10-last-edit-sim}
 & \UseMacro{res-toga-CSNm-eval-runnable-assert-stmt/test-bs10-last-rouge-f}
\\
\midrule
\UseMacro{model-teco-icse23}
 & \textbf{\UseMacro{res-teco-icse23-CSNm-eval-runnable-assert-stmt/test-bs10-last-compilable}}
 & \textbf{\UseMacro{res-teco-icse23-CSNm-eval-runnable-assert-stmt/test-bs10-last-runnable}}
 & \textbf{\UseMacro{res-teco-icse23-CSNm-eval-runnable-assert-stmt/test-bs10-last-xmatch}}
 & \UseMacro{res-teco-icse23-CSNm-eval-runnable-assert-stmt/test-bs10-last-xmatch-top10}
 & \textbf{\UseMacro{res-teco-icse23-CSNm-eval-runnable-assert-stmt/test-bs10-last-bleu}}
 & \textbf{\UseMacro{res-teco-icse23-CSNm-eval-runnable-assert-stmt/test-bs10-last-code-bleu}}
 & \textbf{\UseMacro{res-teco-icse23-CSNm-eval-runnable-assert-stmt/test-bs10-last-edit-sim}}
 & \textbf{\UseMacro{res-teco-icse23-CSNm-eval-runnable-assert-stmt/test-bs10-last-rouge-f}}
\\
\bottomrule
\end{tabular}

\end{minipage}
\end{small}
\end{table*}

\subsection{Evaluation Metrics}
\label{sec:eval:metrics}

\textit{(1) Lexical-level metrics}: We use the following automatic
metrics to measure how close the predicted \stmts are to the gold
\stmts; these metrics have been frequently used in prior work on code
generation and comment generation~\cite{IyerETAL16Summarizing,
HuETAL19Deep, LiangAndZhu18Automatic, LeClairETAL19Neural}:

\MyParaOnly{\Xmatch} (\xmatchAcro) is the percentage of predicted
\stmts matches exactly with the gold. This metric is the most strict
one; each point of improvement directly entails a larger portion of
code that is both syntactically and semantically correct, yet it does
not take into account paraphrases or give any partial credit.

\MyParaOnly{\Xmatchk} (\xmatchkAcro) is the percentage of any top-10
predicted \stmts matches exactly with the gold. This metric evaluates
the use case where the developer can see and select from the top-10
predictions of the model.

\MyParaOnly{\BLEU}~\cite{PapineniETAL02BLEU} calculates the
number of n-grams (consecutive n \subtokens) in the prediction that
also appear in the gold; specifically, we compute the
$1\!\sim\!4$-grams overlap between the \subtokens in the prediction
and the \subtokens in the gold, averaged between $1\!\sim\!4$-grams
with smoothing method proposed by Lin and
Och~\cite{LinAndOch04ORANGE}.

\MyParaOnly{\CodeBLEU}~\cite{RenETAL20CodeBLEU} is an improved
version of \BLEU adapted for code.  It is a combination of the
traditional \BLEU, the \BLEU if only considering keywords, syntactical
AST match, and semantic data-flow match.

\MyParaOnly{\Editsim} (\editsimAcro) = 1 - \Leditdis, where the
\Leditdis measures the amount of single-character edits (including
insertion, substitution, or deletion) that need to be made to
transform the prediction to the gold, normalized by the maximum number
of characters in the prediction and the gold.  This metric was
proposed and used in prior work on code
completion~\cite{SvyatkovskiyETAL20IntelliCode}.

\MyParaOnly{\ROUGEL}~\cite{LinAndOch04Automatic} measures the overlap
between the prediction \subtokens and the gold \subtokens based on the
Longest Common Subsequence statistics, using F1 score.

\begin{table*}[t]
\centering
\begin{small}
\caption{Results for \TecoTool without and with reranking by
execution. \UseMacro{TC-notice-bold} \UseMacro{TC-notice-allsign}\label{tab:results-ablation}}
\begin{minipage}[t]{\textwidth}
  \centering
  \subcaption{On the \test set.\label{tab:results-ablation-any-CSNm}}
  \vspace{-5pt}
  
%% Automatically generated by pyutil.latex 

\begin{tabular}{@{}l|@{\hspace{2pt}}r@{\hspace{2pt}}r@{\hspace{2pt}}r@{\hspace{2pt}}r@{\hspace{2pt}}r@{}}
\toprule
\textbf{\UseMacro{TH-model}}
 & \textbf{\UseMacro{TH-xmatch}}
 & \textbf{\UseMacro{TH-bleu}}
 & \textbf{\UseMacro{TH-code-bleu}}
 & \textbf{\UseMacro{TH-edit-sim}}
 & \textbf{\UseMacro{TH-rouge}}
\\
\midrule
\UseMacro{model-codet5}
 & \UseMacro{res-codet5-CSNm-eval-any-stmt/test-bs10-last-xmatch}
 & \UseMacro{res-codet5-CSNm-eval-any-stmt/test-bs10-last-bleu}
 & \UseMacro{res-codet5-CSNm-eval-any-stmt/test-bs10-last-code-bleu}
 & \UseMacro{res-codet5-CSNm-eval-any-stmt/test-bs10-last-edit-sim}
 & \UseMacro{res-codet5-CSNm-eval-any-stmt/test-bs10-last-rouge-f}
\\
\UseMacro{model-teco-icse23-norr}
 & \UseMacro{res-teco-icse23-norr-CSNm-eval-any-stmt/test-bs10-last-xmatch}
 & \UseMacro{res-teco-icse23-norr-CSNm-eval-any-stmt/test-bs10-last-bleu}
 & \UseMacro{res-teco-icse23-norr-CSNm-eval-any-stmt/test-bs10-last-code-bleu}
 & \UseMacro{res-teco-icse23-norr-CSNm-eval-any-stmt/test-bs10-last-edit-sim}
 & \UseMacro{res-teco-icse23-norr-CSNm-eval-any-stmt/test-bs10-last-rouge-f}
\\
\UseMacro{model-teco-icse23}
 & \textbf{\UseMacro{res-teco-icse23-CSNm-eval-any-stmt/test-bs10-last-xmatch}}
 & \textbf{\UseMacro{res-teco-icse23-CSNm-eval-any-stmt/test-bs10-last-bleu}}
 & \textbf{\UseMacro{res-teco-icse23-CSNm-eval-any-stmt/test-bs10-last-code-bleu}}
 & \textbf{\UseMacro{res-teco-icse23-CSNm-eval-any-stmt/test-bs10-last-edit-sim}}
 & \textbf{\UseMacro{res-teco-icse23-CSNm-eval-any-stmt/test-bs10-last-rouge-f}}
\\
\bottomrule
\end{tabular}

  \vspace{5pt}
\end{minipage}
\begin{minipage}[t]{\textwidth}
  \centering
  \subcaption{On the \runnablesubset.\label{tab:results-ablation-runnable-any-CSNm}}
  \vspace{-5pt}
  
%% Automatically generated by pyutil.latex 

\begin{tabular}{@{}l|@{\hspace{2pt}}r@{\hspace{2pt}}r@{\hspace{2pt}}r@{\hspace{2pt}}r@{\hspace{2pt}}r@{\hspace{2pt}}r@{\hspace{2pt}}r@{}}
\toprule
\textbf{\UseMacro{TH-model}}
 & \textbf{\UseMacro{TH-compilable}}
 & \textbf{\UseMacro{TH-runnable}}
 & \textbf{\UseMacro{TH-xmatch}}
 & \textbf{\UseMacro{TH-bleu}}
 & \textbf{\UseMacro{TH-code-bleu}}
 & \textbf{\UseMacro{TH-edit-sim}}
 & \textbf{\UseMacro{TH-rouge}}
\\
\midrule
\UseMacro{model-codet5}
 & \UseMacro{res-codet5-CSNm-eval-runnable-any-stmt/test-bs10-last-compilable}
 & \UseMacro{res-codet5-CSNm-eval-runnable-any-stmt/test-bs10-last-runnable}
 & \UseMacro{res-codet5-CSNm-eval-runnable-any-stmt/test-bs10-last-xmatch}
 & \UseMacro{res-codet5-CSNm-eval-runnable-any-stmt/test-bs10-last-bleu}
 & \UseMacro{res-codet5-CSNm-eval-runnable-any-stmt/test-bs10-last-code-bleu}
 & \UseMacro{res-codet5-CSNm-eval-runnable-any-stmt/test-bs10-last-edit-sim}
 & \UseMacro{res-codet5-CSNm-eval-runnable-any-stmt/test-bs10-last-rouge-f}
\\
\UseMacro{model-teco-icse23-norr}
 & \UseMacro{res-teco-icse23-norr-CSNm-eval-runnable-any-stmt/test-bs10-last-compilable}
 & \UseMacro{res-teco-icse23-norr-CSNm-eval-runnable-any-stmt/test-bs10-last-runnable}
 & \UseMacro{res-teco-icse23-norr-CSNm-eval-runnable-any-stmt/test-bs10-last-xmatch}
 & \UseMacro{res-teco-icse23-norr-CSNm-eval-runnable-any-stmt/test-bs10-last-bleu}
 & \UseMacro{res-teco-icse23-norr-CSNm-eval-runnable-any-stmt/test-bs10-last-code-bleu}
 & \UseMacro{res-teco-icse23-norr-CSNm-eval-runnable-any-stmt/test-bs10-last-edit-sim}
 & \UseMacro{res-teco-icse23-norr-CSNm-eval-runnable-any-stmt/test-bs10-last-rouge-f}
\\
\UseMacro{model-teco-icse23}
 & \textbf{\UseMacro{res-teco-icse23-CSNm-eval-runnable-any-stmt/test-bs10-last-compilable}}
 & \textbf{\UseMacro{res-teco-icse23-CSNm-eval-runnable-any-stmt/test-bs10-last-runnable}}
 & \textbf{\UseMacro{res-teco-icse23-CSNm-eval-runnable-any-stmt/test-bs10-last-xmatch}}
 & \textbf{\UseMacro{res-teco-icse23-CSNm-eval-runnable-any-stmt/test-bs10-last-bleu}}
 & \textbf{\UseMacro{res-teco-icse23-CSNm-eval-runnable-any-stmt/test-bs10-last-code-bleu}}
 & \textbf{\UseMacro{res-teco-icse23-CSNm-eval-runnable-any-stmt/test-bs10-last-edit-sim}}
 & \textbf{\UseMacro{res-teco-icse23-CSNm-eval-runnable-any-stmt/test-bs10-last-rouge-f}}
\\
\bottomrule
\end{tabular}

  \vspace{5pt}
\end{minipage}
\begin{minipage}[t]{\textwidth}
  \centering
  \subcaption{On the \oraclesubset.\label{tab:results-ablation-assert-CSNm}}
  \vspace{-5pt}
  
%% Automatically generated by pyutil.latex 

\begin{tabular}{@{}l|@{\hspace{2pt}}r@{\hspace{2pt}}r@{\hspace{2pt}}r@{\hspace{2pt}}r@{\hspace{2pt}}r@{}}
\toprule
\textbf{\UseMacro{TH-model}}
 & \textbf{\UseMacro{TH-xmatch}}
 & \textbf{\UseMacro{TH-bleu}}
 & \textbf{\UseMacro{TH-code-bleu}}
 & \textbf{\UseMacro{TH-edit-sim}}
 & \textbf{\UseMacro{TH-rouge}}
\\
\midrule
\UseMacro{model-codet5}
 & \UseMacro{res-codet5-CSNm-eval-assert-stmt/test-bs10-last-xmatch}
 & \UseMacro{res-codet5-CSNm-eval-assert-stmt/test-bs10-last-bleu}
 & \UseMacro{res-codet5-CSNm-eval-assert-stmt/test-bs10-last-code-bleu}
 & \UseMacro{res-codet5-CSNm-eval-assert-stmt/test-bs10-last-edit-sim}
 & \UseMacro{res-codet5-CSNm-eval-assert-stmt/test-bs10-last-rouge-f}
\\
\UseMacro{model-teco-icse23-norr}
 & \UseMacro{res-teco-icse23-norr-CSNm-eval-assert-stmt/test-bs10-last-xmatch}
 & \UseMacro{res-teco-icse23-norr-CSNm-eval-assert-stmt/test-bs10-last-bleu}
 & \UseMacro{res-teco-icse23-norr-CSNm-eval-assert-stmt/test-bs10-last-code-bleu}
 & \UseMacro{res-teco-icse23-norr-CSNm-eval-assert-stmt/test-bs10-last-edit-sim}
 & \UseMacro{res-teco-icse23-norr-CSNm-eval-assert-stmt/test-bs10-last-rouge-f}
\\
\UseMacro{model-teco-icse23}
 & \textbf{\UseMacro{res-teco-icse23-CSNm-eval-assert-stmt/test-bs10-last-xmatch}}
 & \textbf{\UseMacro{res-teco-icse23-CSNm-eval-assert-stmt/test-bs10-last-bleu}}
 & \textbf{\UseMacro{res-teco-icse23-CSNm-eval-assert-stmt/test-bs10-last-code-bleu}}
 & \textbf{\UseMacro{res-teco-icse23-CSNm-eval-assert-stmt/test-bs10-last-edit-sim}}
 & \textbf{\UseMacro{res-teco-icse23-CSNm-eval-assert-stmt/test-bs10-last-rouge-f}}
\\
\bottomrule
\end{tabular}

  \vspace{5pt}
\end{minipage}
\begin{minipage}[t]{\textwidth}
  \centering
  \subcaption{On the \oraclerunnablesubset.\label{tab:results-ablation-runnable-assert-CSNm}}
  \vspace{-5pt}
  
%% Automatically generated by pyutil.latex 

\begin{tabular}{@{}l|@{\hspace{2pt}}r@{\hspace{2pt}}r@{\hspace{2pt}}r@{\hspace{2pt}}r@{\hspace{2pt}}r@{\hspace{2pt}}r@{\hspace{2pt}}r@{}}
\toprule
\textbf{\UseMacro{TH-model}}
 & \textbf{\UseMacro{TH-compilable}}
 & \textbf{\UseMacro{TH-runnable}}
 & \textbf{\UseMacro{TH-xmatch}}
 & \textbf{\UseMacro{TH-bleu}}
 & \textbf{\UseMacro{TH-code-bleu}}
 & \textbf{\UseMacro{TH-edit-sim}}
 & \textbf{\UseMacro{TH-rouge}}
\\
\midrule
\UseMacro{model-codet5}
 & \UseMacro{res-codet5-CSNm-eval-runnable-assert-stmt/test-bs10-last-compilable}
 & \UseMacro{res-codet5-CSNm-eval-runnable-assert-stmt/test-bs10-last-runnable}
 & \UseMacro{res-codet5-CSNm-eval-runnable-assert-stmt/test-bs10-last-xmatch}
 & \UseMacro{res-codet5-CSNm-eval-runnable-assert-stmt/test-bs10-last-bleu}
 & \UseMacro{res-codet5-CSNm-eval-runnable-assert-stmt/test-bs10-last-code-bleu}
 & \UseMacro{res-codet5-CSNm-eval-runnable-assert-stmt/test-bs10-last-edit-sim}
 & \UseMacro{res-codet5-CSNm-eval-runnable-assert-stmt/test-bs10-last-rouge-f}
\\
\UseMacro{model-teco-icse23-norr}
 & \UseMacro{res-teco-icse23-norr-CSNm-eval-runnable-assert-stmt/test-bs10-last-compilable}
 & \UseMacro{res-teco-icse23-norr-CSNm-eval-runnable-assert-stmt/test-bs10-last-runnable}
 & \UseMacro{res-teco-icse23-norr-CSNm-eval-runnable-assert-stmt/test-bs10-last-xmatch}
 & \UseMacro{res-teco-icse23-norr-CSNm-eval-runnable-assert-stmt/test-bs10-last-bleu}
 & \UseMacro{res-teco-icse23-norr-CSNm-eval-runnable-assert-stmt/test-bs10-last-code-bleu}
 & \UseMacro{res-teco-icse23-norr-CSNm-eval-runnable-assert-stmt/test-bs10-last-edit-sim}
 & \UseMacro{res-teco-icse23-norr-CSNm-eval-runnable-assert-stmt/test-bs10-last-rouge-f}
\\
\UseMacro{model-teco-icse23}
 & \textbf{\UseMacro{res-teco-icse23-CSNm-eval-runnable-assert-stmt/test-bs10-last-compilable}}
 & \textbf{\UseMacro{res-teco-icse23-CSNm-eval-runnable-assert-stmt/test-bs10-last-runnable}}
 & \textbf{\UseMacro{res-teco-icse23-CSNm-eval-runnable-assert-stmt/test-bs10-last-xmatch}}
 & \textbf{\UseMacro{res-teco-icse23-CSNm-eval-runnable-assert-stmt/test-bs10-last-bleu}}
 & \textbf{\UseMacro{res-teco-icse23-CSNm-eval-runnable-assert-stmt/test-bs10-last-code-bleu}}
 & \textbf{\UseMacro{res-teco-icse23-CSNm-eval-runnable-assert-stmt/test-bs10-last-edit-sim}}
 & \textbf{\UseMacro{res-teco-icse23-CSNm-eval-runnable-assert-stmt/test-bs10-last-rouge-f}}
\\
\bottomrule
\end{tabular}

\end{minipage}
\end{small}
\end{table*}

\textit{(2) Functional correctness}: The aforementioned metrics only
capture the lexical similarity between the prediction against the
gold, but the gold \stmt may not be the only correct solution for
competing the next \stmt.  Namely, the prediction can be functionally
correct despite being different from the gold \stmt.  To measure the
functional correctness, we additionally use the following automatic
metrics:

\MyParaOnly{\PctCompilable} is the percentage of the predicted
\stmts that are compilable when appended to the incomplete test.

\MyParaOnly{\PctRunnable} is the percentage of the predicted
\stmts that are compilable and runnable when appended to the
incomplete test, without incurring assertion failures or runtime
errors.

Note that \PctCompilable and \PctRunnable are over-estimations of the
functional correctness, as they do not consider whether the underlying
logic of the code is meaningful.  That said, most functional
correctness errors relevant to tests, such as generating the wrong
expected outputs, can be captured by the \PctRunnable metric.  Prior
work has used a similar methodology to evaluate the functional
correctness of text-to-code transduction by running generated code
with test cases~\cite{ChenETAL21Codex}, which was performed on a
rather small dataset because of the difficulty in collecting manual
labelled data.  Thanks to the executable nature of tests, we are able
to design the two automatic functional correctness metrics for a large
corpus.

\section{Results}
\label{sec:results}

\subsection{\UseMacro{rq-overall}: Performance of \TecoTool vs. Baseline Models}

Table~\ref{tab:results-main-any-CSNm} shows the results of \TecoTool
and baseline models on solving the \testcomp task.  Our model
\UseMacro{model-teco-icse23} significantly outperforms all baseline
models on all automatic metrics.  \UseMacro{model-teco-icse23}
achieves \UseMacro{res-ours-CSNm-eval-any-stmt/test-bs10-last-xmatch}
\xmatch, which is
\UseMacro{res-improv-pct-CSNm-eval-any-stmt/test-bs10-last-xmatch}\%
higher than the best baseline model, \UseMacro{model-codet5}'s
\UseMacro{res-baseline-CSNm-eval-any-stmt/test-bs10-last-xmatch}.
This indicates that using \semdata and reranking by execution can
greatly improve deep learning model's performance on \testcomp.

The non-\finetuned baseline model, \UseMacro{model-codet5-noft}, is
not capable of solving \testcomp task.  This is because the
model is optimized to solve different tasks during \pretraining and
does not have the domain knowledge of the input-output format of the
\testcomp task.

\UseMacro{model-codegpt} has shown to be effective on the task of code
completion~\cite{LuETAL21CodeXGLUE,SvyatkovskiyETAL20IntelliCode},
where the primary goal is to continue generating code similar to the
context code.  However, 
it performs slightly worse than the encoder-decoder baseline
\UseMacro{model-codet5} on \testcomp, because the task requires
generating \stmt in the \testm which has different style than the \mut
in the provided context.

\subsection{\UseMacro{rq-runnable}: Functional Correctness}

Table~\ref{tab:results-main-runnable-any-CSNm} shows the results of
\TecoTool and baseline models on the \runnablesubset, with
\PctCompilable and \PctRunnable metrics that measure the functional
correctness of the generated \stmts.  Our model,
\UseMacro{model-teco-icse23} can generate runnable \stmts for
\UseMacro{res-teco-icse23-CSNm-eval-runnable-any-stmt/test-bs10-last-runnable}\%
of the time, and compilable \stmts for
\UseMacro{res-teco-icse23-CSNm-eval-runnable-any-stmt/test-bs10-last-compilable}\%
of the time, much higher than the best baseline model's
\UseMacro{res-codet5-CSNm-eval-runnable-any-stmt/test-bs10-last-runnable}\%
and
\UseMacro{res-codet5-CSNm-eval-runnable-any-stmt/test-bs10-last-compilable}\%.
On this \runnablesubset, \TecoTool also outperforms all baseline
models on other
metrics measuring lexical similarity.

The other two baseline models, \UseMacro{model-codet5-noft} and
\UseMacro{model-codegpt}, fail to generate any compilable or runnable
\stmts.  After closer inspection, we found that
\UseMacro{model-codet5-noft} always generate broken non-code
outputs, as it is not \finetuned to process the inputs; and
\UseMacro{model-codegpt} always generate code that is not a valid
\stmt in Java, e.g., code that starts with a method signature.

\subsection{\UseMacro{rq-oracle}: Performance on \OracleGen}

Tables~\ref{tab:results-main-assert-CSNm} and
\ref{tab:results-main-runnable-assert-CSNm} show the results of the
downstream application of \oraclegen, on the \oraclesubset and the
\oraclerunnablesubset, respectively.  \TecoTool significantly improves
the \xmatch on this task by a large margin (by
\UseMacro{res-improv-pct-CSNm-eval-assert-stmt/test-bs10-last-xmatch}\%),
from \UseMacro{res-toga-CSNm-eval-assert-stmt/test-bs10-last-xmatch}
for the prior state-of-the-art, \UseMacro{model-toga}, to ours
\UseMacro{res-teco-icse23-CSNm-eval-assert-stmt/test-bs10-last-xmatch}.
Note that \UseMacro{model-toga}'s \xmatch is on-par with
\UseMacro{model-codet5}, the model that \TecoTool is \finetuned from,
which confirms that \TecoTool's improvements primarily come from
using \semdata and reranking by execution.

\UseMacro{model-toga} is the strongest prior model on this task in
terms of \xmatch.  However, it is worse than the
\UseMacro{model-codet5} baseline model on other metrics that consider
partial matches.  This is because \UseMacro{model-toga} is a
classification model that ranks a set of assertion statement
candidates generated using heuristics, and when the gold \stmt is not
in the set, the model fails to correctly rank a sub-optimal candidate.

\subsection{\UseMacro{rq-ablation}: Improvements from Reranking by Execution}

Table~\ref{tab:results-ablation} shows the results of
\UseMacro{model-teco-icse23-norr} (the \xmatchk for
\UseMacro{model-teco-icse23-norr} is always the same as
\UseMacro{model-teco-icse23}, because the reranking is performed on
top-10 predictions, thus we did not include this metric in the table).

Comparing \UseMacro{model-teco-icse23} with
\UseMacro{model-teco-icse23-norr} on the \test set
(Table~\ref{tab:results-ablation-any-CSNm}), reranking by execution
alone contributes to 2 points in \xmatch.  However, the improvements
over other similarity metrics, which take into account partial
matches, are smaller.  This indicates that reranking by execution is
effective in prioritizing the exact correct generated \stmt than other
non-runnable candidates most of the times, but in a few cases it may
prioritize runnable candidates that are less similar to the gold \stmt
than the original top-1.  \UseMacro{model-teco-icse23-norr} still
significantly outperforms \UseMacro{model-codet5} on all metrics.  On
the \runnablesubset
(Table~\ref{tab:results-ablation-runnable-any-CSNm}),
\UseMacro{model-teco-icse23} improves both \PctCompilable and
\PctRunnable over \UseMacro{model-teco-icse23-norr} by large margins,
which shows that reranking by execution is an effective strategy for
improving the quality of generated \stmts.

Reranking by execution ended up being very important for improving
performance on the task of \oraclegen, as shown on the \oraclesubset
(Table~\ref{tab:results-ablation-assert-CSNm}) and the
\oraclerunnablesubset
(Table~\ref{tab:results-ablation-runnable-assert-CSNm}).  For example,
\UseMacro{model-teco-icse23} outperforms
\UseMacro{model-teco-icse23-norr} by 6--8 points in \xmatch and 12
points in \PctRunnable.  This is because logical errors in assertion
statements can be easily found by execution (e.g., generating the
wrong expected value will cause an assertion to fail).

\subsection{\UseMacro{rq-semdata}: Comparisons of \SemData}

\begin{table}[t]
\centering
\begin{small}
\caption{Results for \TecoTool models with only one kind of \semdata
on the \test set. \UseMacro{TC-notice-bold}\label{tab:results-single}}
\begin{minipage}[t]{.5\textwidth}
  \centering
  \subcaption{On the \test set.\label{tab:results-single-any-CSNm}}
  \vspace{-5pt}
  
%% Automatically generated by pyutil.latex 

\begin{tabular}{@{}l|@{\hspace{2pt}}r@{\hspace{2pt}}r@{\hspace{2pt}}r@{\hspace{2pt}}r@{\hspace{2pt}}r@{\hspace{2pt}}r@{}}
\toprule
\textbf{\UseMacro{TH-model}}
 & \textbf{\UseMacro{TH-xmatch}}
 & \textbf{\UseMacro{TH-xmatch-top10}}
 & \textbf{\UseMacro{TH-bleu}}
 & \textbf{\UseMacro{TH-code-bleu}}
 & \textbf{\UseMacro{TH-edit-sim}}
 & \textbf{\UseMacro{TH-rouge}}
\\
\midrule
\UseMacro{model-codet5}
 & \UseMacro{res-codet5-CSNm-eval-any-stmt/test-bs10-last-xmatch}
 & \UseMacro{res-codet5-CSNm-eval-any-stmt/test-bs10-last-xmatch-top10}
 & \UseMacro{res-codet5-CSNm-eval-any-stmt/test-bs10-last-bleu}
 & \UseMacro{res-codet5-CSNm-eval-any-stmt/test-bs10-last-code-bleu}
 & \UseMacro{res-codet5-CSNm-eval-any-stmt/test-bs10-last-edit-sim}
 & \UseMacro{res-codet5-CSNm-eval-any-stmt/test-bs10-last-rouge-f}
\\
\midrule
\UseMacro{model-teco-s1}
 & \UseMacro{res-teco-s1-CSNm-eval-any-stmt/test-bs10-last-xmatch}
 & \UseMacro{res-teco-s1-CSNm-eval-any-stmt/test-bs10-last-xmatch-top10}
 & \UseMacro{res-teco-s1-CSNm-eval-any-stmt/test-bs10-last-bleu}
 & \UseMacro{res-teco-s1-CSNm-eval-any-stmt/test-bs10-last-code-bleu}
 & \UseMacro{res-teco-s1-CSNm-eval-any-stmt/test-bs10-last-edit-sim}
 & \UseMacro{res-teco-s1-CSNm-eval-any-stmt/test-bs10-last-rouge-f}
\\
\UseMacro{model-teco-s2}
 & \UseMacro{res-teco-s2-CSNm-eval-any-stmt/test-bs10-last-xmatch}
 & \UseMacro{res-teco-s2-CSNm-eval-any-stmt/test-bs10-last-xmatch-top10}
 & \textbf{\UseMacro{res-teco-s2-CSNm-eval-any-stmt/test-bs10-last-bleu}}
 & \textbf{\UseMacro{res-teco-s2-CSNm-eval-any-stmt/test-bs10-last-code-bleu}}
 & \textbf{\UseMacro{res-teco-s2-CSNm-eval-any-stmt/test-bs10-last-edit-sim}}
 & \textbf{\UseMacro{res-teco-s2-CSNm-eval-any-stmt/test-bs10-last-rouge-f}}
\\
\UseMacro{model-teco-s3}
 & \UseMacro{res-teco-s3-CSNm-eval-any-stmt/test-bs10-last-xmatch}
 & \UseMacro{res-teco-s3-CSNm-eval-any-stmt/test-bs10-last-xmatch-top10}
 & \UseMacro{res-teco-s3-CSNm-eval-any-stmt/test-bs10-last-bleu}
 & \UseMacro{res-teco-s3-CSNm-eval-any-stmt/test-bs10-last-code-bleu}
 & \UseMacro{res-teco-s3-CSNm-eval-any-stmt/test-bs10-last-edit-sim}
 & \UseMacro{res-teco-s3-CSNm-eval-any-stmt/test-bs10-last-rouge-f}
\\
\UseMacro{model-teco-s4}
 & \textbf{\UseMacro{res-teco-s4-CSNm-eval-any-stmt/test-bs10-last-xmatch}}
 & \textbf{\UseMacro{res-teco-s4-CSNm-eval-any-stmt/test-bs10-last-xmatch-top10}}
 & \UseMacro{res-teco-s4-CSNm-eval-any-stmt/test-bs10-last-bleu}
 & \UseMacro{res-teco-s4-CSNm-eval-any-stmt/test-bs10-last-code-bleu}
 & \UseMacro{res-teco-s4-CSNm-eval-any-stmt/test-bs10-last-edit-sim}
 & \UseMacro{res-teco-s4-CSNm-eval-any-stmt/test-bs10-last-rouge-f}
\\
\UseMacro{model-teco-s5}
 & \UseMacro{res-teco-s5-CSNm-eval-any-stmt/test-bs10-last-xmatch}
 & \UseMacro{res-teco-s5-CSNm-eval-any-stmt/test-bs10-last-xmatch-top10}
 & \UseMacro{res-teco-s5-CSNm-eval-any-stmt/test-bs10-last-bleu}
 & \UseMacro{res-teco-s5-CSNm-eval-any-stmt/test-bs10-last-code-bleu}
 & \UseMacro{res-teco-s5-CSNm-eval-any-stmt/test-bs10-last-edit-sim}
 & \UseMacro{res-teco-s5-CSNm-eval-any-stmt/test-bs10-last-rouge-f}
\\
\UseMacro{model-teco-s6}
 & \UseMacro{res-teco-s6-CSNm-eval-any-stmt/test-bs10-last-xmatch}
 & \UseMacro{res-teco-s6-CSNm-eval-any-stmt/test-bs10-last-xmatch-top10}
 & \UseMacro{res-teco-s6-CSNm-eval-any-stmt/test-bs10-last-bleu}
 & \UseMacro{res-teco-s6-CSNm-eval-any-stmt/test-bs10-last-code-bleu}
 & \UseMacro{res-teco-s6-CSNm-eval-any-stmt/test-bs10-last-edit-sim}
 & \UseMacro{res-teco-s6-CSNm-eval-any-stmt/test-bs10-last-rouge-f}
\\
\bottomrule
\end{tabular}

  \vspace{5pt}
\end{minipage}
\begin{minipage}[t]{.5\textwidth}
  \centering
  \subcaption{On the \oraclesubset.\label{tab:results-single-assert-CSNm}}
  \vspace{-5pt}
  
%% Automatically generated by pyutil.latex 

\begin{tabular}{@{}l|@{\hspace{2pt}}r@{\hspace{2pt}}r@{\hspace{2pt}}r@{\hspace{2pt}}r@{\hspace{2pt}}r@{\hspace{2pt}}r@{}}
\toprule
\textbf{\UseMacro{TH-model}}
 & \textbf{\UseMacro{TH-xmatch}}
 & \textbf{\UseMacro{TH-xmatch-top10}}
 & \textbf{\UseMacro{TH-bleu}}
 & \textbf{\UseMacro{TH-code-bleu}}
 & \textbf{\UseMacro{TH-edit-sim}}
 & \textbf{\UseMacro{TH-rouge}}
\\
\midrule
\UseMacro{model-codet5}
 & \UseMacro{res-codet5-CSNm-eval-assert-stmt/test-bs10-last-xmatch}
 & \UseMacro{res-codet5-CSNm-eval-assert-stmt/test-bs10-last-xmatch-top10}
 & \UseMacro{res-codet5-CSNm-eval-assert-stmt/test-bs10-last-bleu}
 & \UseMacro{res-codet5-CSNm-eval-assert-stmt/test-bs10-last-code-bleu}
 & \UseMacro{res-codet5-CSNm-eval-assert-stmt/test-bs10-last-edit-sim}
 & \UseMacro{res-codet5-CSNm-eval-assert-stmt/test-bs10-last-rouge-f}
\\
\midrule
\UseMacro{model-teco-s1}
 & \UseMacro{res-teco-s1-CSNm-eval-assert-stmt/test-bs10-last-xmatch}
 & \UseMacro{res-teco-s1-CSNm-eval-assert-stmt/test-bs10-last-xmatch-top10}
 & \UseMacro{res-teco-s1-CSNm-eval-assert-stmt/test-bs10-last-bleu}
 & \UseMacro{res-teco-s1-CSNm-eval-assert-stmt/test-bs10-last-code-bleu}
 & \UseMacro{res-teco-s1-CSNm-eval-assert-stmt/test-bs10-last-edit-sim}
 & \UseMacro{res-teco-s1-CSNm-eval-assert-stmt/test-bs10-last-rouge-f}
\\
\UseMacro{model-teco-s2}
 & \UseMacro{res-teco-s2-CSNm-eval-assert-stmt/test-bs10-last-xmatch}
 & \UseMacro{res-teco-s2-CSNm-eval-assert-stmt/test-bs10-last-xmatch-top10}
 & \UseMacro{res-teco-s2-CSNm-eval-assert-stmt/test-bs10-last-bleu}
 & \UseMacro{res-teco-s2-CSNm-eval-assert-stmt/test-bs10-last-code-bleu}
 & \UseMacro{res-teco-s2-CSNm-eval-assert-stmt/test-bs10-last-edit-sim}
 & \UseMacro{res-teco-s2-CSNm-eval-assert-stmt/test-bs10-last-rouge-f}
\\
\UseMacro{model-teco-s3}
 & \UseMacro{res-teco-s3-CSNm-eval-assert-stmt/test-bs10-last-xmatch}
 & \UseMacro{res-teco-s3-CSNm-eval-assert-stmt/test-bs10-last-xmatch-top10}
 & \textbf{\UseMacro{res-teco-s3-CSNm-eval-assert-stmt/test-bs10-last-bleu}}
 & \UseMacro{res-teco-s3-CSNm-eval-assert-stmt/test-bs10-last-code-bleu}
 & \textbf{\UseMacro{res-teco-s3-CSNm-eval-assert-stmt/test-bs10-last-edit-sim}}
 & \textbf{\UseMacro{res-teco-s3-CSNm-eval-assert-stmt/test-bs10-last-rouge-f}}
\\
\UseMacro{model-teco-s4}
 & \UseMacro{res-teco-s4-CSNm-eval-assert-stmt/test-bs10-last-xmatch}
 & \UseMacro{res-teco-s4-CSNm-eval-assert-stmt/test-bs10-last-xmatch-top10}
 & \UseMacro{res-teco-s4-CSNm-eval-assert-stmt/test-bs10-last-bleu}
 & \UseMacro{res-teco-s4-CSNm-eval-assert-stmt/test-bs10-last-code-bleu}
 & \UseMacro{res-teco-s4-CSNm-eval-assert-stmt/test-bs10-last-edit-sim}
 & \UseMacro{res-teco-s4-CSNm-eval-assert-stmt/test-bs10-last-rouge-f}
\\
\UseMacro{model-teco-s5}
 & \UseMacro{res-teco-s5-CSNm-eval-assert-stmt/test-bs10-last-xmatch}
 & \UseMacro{res-teco-s5-CSNm-eval-assert-stmt/test-bs10-last-xmatch-top10}
 & \UseMacro{res-teco-s5-CSNm-eval-assert-stmt/test-bs10-last-bleu}
 & \UseMacro{res-teco-s5-CSNm-eval-assert-stmt/test-bs10-last-code-bleu}
 & \UseMacro{res-teco-s5-CSNm-eval-assert-stmt/test-bs10-last-edit-sim}
 & \UseMacro{res-teco-s5-CSNm-eval-assert-stmt/test-bs10-last-rouge-f}
\\
\UseMacro{model-teco-s6}
 & \textbf{\UseMacro{res-teco-s6-CSNm-eval-assert-stmt/test-bs10-last-xmatch}}
 & \textbf{\UseMacro{res-teco-s6-CSNm-eval-assert-stmt/test-bs10-last-xmatch-top10}}
 & \UseMacro{res-teco-s6-CSNm-eval-assert-stmt/test-bs10-last-bleu}
 & \textbf{\UseMacro{res-teco-s6-CSNm-eval-assert-stmt/test-bs10-last-code-bleu}}
 & \UseMacro{res-teco-s6-CSNm-eval-assert-stmt/test-bs10-last-edit-sim}
 & \UseMacro{res-teco-s6-CSNm-eval-assert-stmt/test-bs10-last-rouge-f}
\\
\bottomrule
\end{tabular}

\end{minipage}
\end{small}
\end{table}

Tables~\ref{tab:results-single-any-CSNm} and
\ref{tab:results-single-assert-CSNm} show the results of the \TecoTool
models with only one kind of \semdata, comparing with the strongest
baseline model \UseMacro{model-codet5}, on the full \test set and the
\oraclesubset, respectively.  We did not perform statistical
significance tests for the results here as the performances of the
models are too close.
Each model outperforms \UseMacro{model-codet5} on at least one metric,
meaning that each \semdata provides some information useful for
\testcomp.
In Table~\ref{tab:results-single-any-CSNm}, \UseMacro{model-teco-s2}
(\UseMacro{name-types-absent}) is the best model in terms of \BLEU,
\CodeBLEU, \editsimAcro and \ROUGEL metrics, and
\UseMacro{model-teco-s4} (\UseMacro{name-setup-teardown}) is the best
model in terms of \xmatch and \xmatchk, which indicates that these two
kinds of \semdata are relatively more important than others.
Interestingly, in Table~\ref{tab:results-single-assert-CSNm}, the
models that achieved the best performance among single-data models
changed: \UseMacro{model-teco-s3} (\UseMacro{name-fields-notset}) is
the best model in terms of \BLEU, \editsimAcro, and \ROUGEL, and
\UseMacro{model-teco-s6} (\UseMacro{name-similar-stmt}) is the best
model in terms of \xmatch, \xmatchk, and \CodeBLEU.  Thus, different
kinds of \semdata provide complementary information for \testcomp.

\section{Limitations and Future Work}
\label{sec:limitations}

We discuss several limitations of our work and the future work
inspired by those limitations.

\MyPara{Usability} We envision our models being integrated into an
IDE.  At any point, a user would be able to see top-k results from our
models and potentially decide to use one of the suggestions.
This is similar to email completion that has recently been integrated
into several popular web-based email clients, e.g., GMail.

\MyPara{Structured representation} Currently we do not considering
using any structured representation of code, e.g., abstract syntax
trees (ASTs).  Such a representation could enhance performance of our
models and enable a quick check of validity of generated code.  We
leave this for future work.

\MyPara{Test-Driven Development (TDD)} We assume that \codeut is
written before tests when defining the \testcomp task, which is the
opposite order of TDD.  Future work could explore the mirror task of
code completion with a \testm context that is applicable to projects
adopting TDD. 

\MyPara{Testing frameworks} We focused on tests written in the JUnit
style.  Although other testing frameworks are available (e.g.,
TestNG), JUnit is the most popular among Java projects.

\MyPara{Large language models for code}  Recent large language models
that scale up to billions of parameters create new state-of-the-art
for many code-related
tasks~\cite{ChenETAL21Codex,chowdhery2022palm,li2022competition}.
However, these models do not perform well on simple code execution
tasks~\cite{NyeETAL21ScratchPad}.  
Incorporating \semdata into large language models for code is a
promising direction which we leave as future work.

\section{Related Work}
\label{sec:related}

\MyPara{Automated test generation} Existing automatic test generation
work includes fuzz/random testing~\cite{PachecoETAL07Randoop,
ZellerETAL19Fuzzing, ZangETAL22JAttack}, property-based
testing~\cite{ClaessenAndHughes00QuickCheck, BoyapatiETAL02Korat,
GligoricETAL10Test, IvanETAL15Programming, CelikETAL17intKorat,
HolmesETAL20TestGeneration, AlAwarETAL21Tempo}, search-based
testing~\cite{HarmanAndMcMinn10Theoretical,
FraserAndArcuri11EvoSuite}, and combinatorial
testing~\cite{CohenETAL06Testing}.  The typical goal in automated test
generation techniques, e.g., Randoop~\cite{PachecoETAL07Randoop} and
EvoSuite~\cite{FraserAndArcuri11EvoSuite}, is to achieve high code
coverage of the code under test by generating a large amount of tests,
either randomly or systematically. However, the generated tests would
not be added to the manually written tests in the code repository due
to their low quality and the excessive amount.  Some prior work
explored improving the quality of the generated tests, for example:
Helmes et al.~\cite{HolmesETAL20TestGeneration} proposed to use
relative LOC to guide the choosing of test generation targets; Reddy
et al.~\cite{ReddyETAL20RLCheck} proposed to use reinforcement
learning to guide the random input generator in property-based test
generation.  So far, these automated techniques are used only in
addition to manually written tests.  In contrast, we focus on
improving developers' productivity when writing manual tests.

Another disadvantage of the automated test generation approaches is
the lack of test oracles.  To remedy that, prior work explored
extracting test oracles from code comments, focusing on test oracles
related to exceptional behaviors, null pointer checks, and boundary
conditions~\cite{TanETAL12TComment,GoffiETAL16Automatic,BlasiETAL18Translating,MotwaniETAL19Automatically}.  
These techniques target generating/completing test oracles, but we
target completing any part of the tests, including test oracles.

Prior work also explored using deep learning models for \oraclegen
without the use of comments, including ATLAS~\cite{WatsonETAL20ATLAS}
and TOGA~\cite{DinellaETAL22TOGA}.  We have described both models in
Section~\ref{sec:eval:baseline-models} and compared \TecoTool with
them on the task of \oraclegen, which can be considered as downstream
application of our \testcomp task.

Tufano et al.~\cite{TufanoETAL20TestGeneration} developed a code
generation technique for tests based on a BART architecture
\pretrained on English and code corpora.  While they target to
generate the entire \testm as a whole, we target to complete one
statement at a time, which allows the developer to observe and control
the process of writing a \testm.

\MyPara{Test recommendation} Prior work also explored improving
developers' productivity in testing by test recommendation: given a
\mut, suggest relevant \testms from the existing test suite using a
recommendation
system~\cite{JanjicAndAtkinson13Utilizing,PhamETAL15Automatically,QianETAL20Test,ZhuETAL20HomoTR}.
These techniques rely on having a set of relevant existing tests to
recommend tests from, which is usually not the case when developers
are starting a new project or adding tests to a project without tests.
Our technique helps developers by providing completions while they are
writing tests and does not have this limitation.

\MyPara{ML for SE} The applications of ML models on SE tasks is an
active research area in recent years.  One of the most studied task is
code completion, which improves developers' productivity by suggesting
next tokens or \stmts as developers are writing
code~\cite{RobbesAndLanza10Improving,BruchETAL09Learning,HanETAL09Code,NguyenETAL12Graphbased,SvyatkovskiyETAL19Pythia,RaychevETAL14Code,LiETAL18Code,SvyatkovskiyETAL20IntelliCode,WangETAL21CodeT5,LuETAL21CodeXGLUE,LeeETAL20Naturalness}.  %
Researchers have also studied developing ML models for other SE tasks,
including code
summarization~\cite{ZhangETAL20Leveraging,NieETAL22EvalMethodologies,HuETAL19Deep,LiangAndZhu18Automatic,IyerETAL16Summarizing,AhmadETAL20Transformer-based,LeClairETAL19Neural,WeiETAL19CodeGen},
code and comment
maintenance~\cite{PanthaplackelETAL20Learning,ZhangETAL22CoditT5,PanthaplackelETAL21InconsistencyDetection,PanthaplackelETAL20AssociatingCommentAndCode},
bug
fixing~\cite{PanthaplackelETAL22DeveloperDiscussionBugFixes,PanthaplackelETAL22LearningToDescribeSolutionsForBugReports,ChakrabortyAndRay21MultiModal},
etc.
In this work, we propose the novel task of \testcomp, which brings
several unique features (e.g., \mut) and necessitates reasoning about
code execution.  We also compared \TecoTool to recent work on code
completion~\cite{WangETAL21CodeT5,LuETAL21CodeXGLUE}.

Prior work explored the use of code execution data in ML for SE.
Wang et al.~\cite{WangETAL18Semantic} proposed to train semantic code
embeddings from execution traces, which can be used to improve the
performance of program repair models.
Wang and Su~\cite{WangAndSu20Semantic} blended syntactical and
semantic code embeddings and applied them in a method naming model.
Nie et al.~\cite{NieETAL20Roosterize} developed Roosterize, a model
for suggesting lemma names in verification projects which is trained
using the runtime representations of lemmas.
Pei et al.~\cite{PeiETAL22Trex} developed a transfer learning
framework called TREX that learns execution semantics from
forced-execution traces to detect similar binary functions.
Pi et al.~\cite{PiETAL22PoEt} proposed PoEt that improves the
reasoning capabilities of language models by \pretraining on
code execution data.
Shi et al.~\cite{ShiETAL22nl2code} proposed to improve code generation
models' outputs using a minimum Bayes risk decoding algorithm based on
execution results.
\TecoTool is the first model designed with code execution in the
testing domain, specifically on the \testcomp task, where reasoning
about the execution of the \codeut is needed. Moreover, \TecoTool
integrates execution to improve both training (using \semdata) and
inference (using reranking via execution) of the model.

\section{Conclusion}

We introduced an idea of designing ML models for code-related tasks
with \semdata inputs and reranking based on test execution outcomes.
Based on this idea, we developed a concrete model, named \TecoTool,
targeting a novel task: test completion.
We evaluated \TecoTool on a new corpus, containing
\UseMacro{corpus-all-num_test} methods and
\UseMacro{corpus-all-runnable-any-num_test} executable methods.  Our
results show that \TecoTool
significantly outperforms the state-of-the-art on code completion and
oracle generation tasks, across a number of evaluation metrics.
We believe that \TecoTool is only a starting point in the exciting
area of ML for code with \semdata and execution data.

\section*{Acknowledgments}

We thank Nader Al Awar, Alex Dimakis, Greg Durrett, Kush Jain, Yu Liu,
Sheena Panthaplackel, August Shi, Aditya Thimmaiah, Zhiqiang Zang,
Jiyang Zhang, and the anonymous reviewers for their comments and
feedback.
The authors acknowledge the Texas Advanced Computing Center (TACC) at
The University of Texas at Austin for providing HPC resources that
have contributed to the research results reported within this paper.
This work is partially supported by
the US National Science Foundation under Grant Nos.~CCF-1652517,
CCF-2107291, IIS-2145479, and CCF-2217696.

\bibliography{bib}
\balance

\clearpage\newpage
\appendices

\section{Comparisons with Large Language Models}

\begin{table*}[!h]
\centering
\begin{small}
\caption{Results for Codex, \TecoTool, and other baseline models on
\testcomp. \UseMacro{TC-notice-bold}
\UseMacro{TC-notice-sign-subtable}\label{tab:results-latest}}
\begin{minipage}[t]{\textwidth}
  \centering
  \subcaption{On the \test set.\label{tab:results-latest-any-CSNm}}
  \vspace{-5pt}
  
%% Automatically generated by pyutil.latex 

\begin{tabular}{@{}l|@{\hspace{2pt}}r@{\hspace{2pt}}r@{\hspace{2pt}}r@{\hspace{2pt}}r@{\hspace{2pt}}r@{\hspace{2pt}}r@{}}
\toprule
\textbf{\UseMacro{TH-model}}
 & \textbf{\UseMacro{TH-xmatch}}
 & \textbf{\UseMacro{TH-xmatch-top10}}
 & \textbf{\UseMacro{TH-bleu}}
 & \textbf{\UseMacro{TH-code-bleu}}
 & \textbf{\UseMacro{TH-edit-sim}}
 & \textbf{\UseMacro{TH-rouge}}
\\
\midrule
\UseMacro{model-codex}
 & \UseMacro{res-codex-CSNm-eval-any-stmt/test-bs10-last-xmatch}
 & \THNA 
 & \UseMacro{res-codex-CSNm-eval-any-stmt/test-bs10-last-bleu}
 & \UseMacro{res-codex-CSNm-eval-any-stmt/test-bs10-last-code-bleu}
 & \UseMacro{res-codex-CSNm-eval-any-stmt/test-bs10-last-edit-sim}
 & \UseMacro{res-codex-CSNm-eval-any-stmt/test-bs10-last-rouge-f}
\\
\UseMacro{model-codet5}
 & \UseMacro{res-codet5-CSNm-eval-any-stmt/test-bs10-last-xmatch}
 & \UseMacro{res-codet5-CSNm-eval-any-stmt/test-bs10-last-xmatch-top10}
 & \UseMacro{res-codet5-CSNm-eval-any-stmt/test-bs10-last-bleu}
 & \UseMacro{res-codet5-CSNm-eval-any-stmt/test-bs10-last-code-bleu}
 & \UseMacro{res-codet5-CSNm-eval-any-stmt/test-bs10-last-edit-sim}
 & \UseMacro{res-codet5-CSNm-eval-any-stmt/test-bs10-last-rouge-f}
\\
\UseMacro{model-codet5-noft}
 & \UseMacro{res-codet5-noft-CSNm-eval-any-stmt/test-bs10-last-xmatch}
 & \UseMacro{res-codet5-noft-CSNm-eval-any-stmt/test-bs10-last-xmatch-top10}
 & \UseMacro{res-codet5-noft-CSNm-eval-any-stmt/test-bs10-last-bleu}
 & \UseMacro{res-codet5-noft-CSNm-eval-any-stmt/test-bs10-last-code-bleu}
 & \UseMacro{res-codet5-noft-CSNm-eval-any-stmt/test-bs10-last-edit-sim}
 & \UseMacro{res-codet5-noft-CSNm-eval-any-stmt/test-bs10-last-rouge-f}
\\
\UseMacro{model-codegpt}
 & \UseMacro{res-codegpt-CSNm-eval-any-stmt/test-bs10-last-xmatch}
 & \UseMacro{res-codegpt-CSNm-eval-any-stmt/test-bs10-last-xmatch-top10}
 & \UseMacro{res-codegpt-CSNm-eval-any-stmt/test-bs10-last-bleu}
 & \UseMacro{res-codegpt-CSNm-eval-any-stmt/test-bs10-last-code-bleu}
 & \UseMacro{res-codegpt-CSNm-eval-any-stmt/test-bs10-last-edit-sim}
 & \UseMacro{res-codegpt-CSNm-eval-any-stmt/test-bs10-last-rouge-f}
\\
\midrule
\UseMacro{model-teco-icse23}
 & \textbf{\UseMacro{res-teco-icse23-CSNm-eval-any-stmt/test-bs10-last-xmatch}}
 & \textbf{\UseMacro{res-teco-icse23-CSNm-eval-any-stmt/test-bs10-last-xmatch-top10}}
 & \textbf{\UseMacro{res-teco-icse23-CSNm-eval-any-stmt/test-bs10-last-bleu}}
 & \textbf{\UseMacro{res-teco-icse23-CSNm-eval-any-stmt/test-bs10-last-code-bleu}}
 & \textbf{\UseMacro{res-teco-icse23-CSNm-eval-any-stmt/test-bs10-last-edit-sim}}
 & \textbf{\UseMacro{res-teco-icse23-CSNm-eval-any-stmt/test-bs10-last-rouge-f}}
\\
\bottomrule
\end{tabular}

  \vspace{5pt}
\end{minipage}
\begin{minipage}[t]{\textwidth}
  \centering
  \subcaption{On the \runnablesubset.\label{tab:results-latest-runnable-any-CSNm}}
  \vspace{-5pt}
  
%% Automatically generated by pyutil.latex 

\begin{tabular}{@{}l|@{\hspace{2pt}}r@{\hspace{2pt}}r@{\hspace{2pt}}r@{\hspace{2pt}}r@{\hspace{2pt}}r@{\hspace{2pt}}r@{\hspace{2pt}}r@{\hspace{2pt}}r@{}}
\toprule
\textbf{\UseMacro{TH-model}}
 & \textbf{\UseMacro{TH-compilable}}
 & \textbf{\UseMacro{TH-runnable}}
 & \textbf{\UseMacro{TH-xmatch}}
 & \textbf{\UseMacro{TH-xmatch-top10}}
 & \textbf{\UseMacro{TH-bleu}}
 & \textbf{\UseMacro{TH-code-bleu}}
 & \textbf{\UseMacro{TH-edit-sim}}
 & \textbf{\UseMacro{TH-rouge}}
\\
\midrule
\UseMacro{model-codex}
 & \UseMacro{res-codex-CSNm-eval-runnable-any-stmt/test-bs10-last-compilable}
 & \UseMacro{res-codex-CSNm-eval-runnable-any-stmt/test-bs10-last-runnable}
 & \UseMacro{res-codex-CSNm-eval-runnable-any-stmt/test-bs10-last-xmatch}
 & \THNA 
 & \UseMacro{res-codex-CSNm-eval-runnable-any-stmt/test-bs10-last-bleu}
 & \UseMacro{res-codex-CSNm-eval-runnable-any-stmt/test-bs10-last-code-bleu}
 & \UseMacro{res-codex-CSNm-eval-runnable-any-stmt/test-bs10-last-edit-sim}
 & \UseMacro{res-codex-CSNm-eval-runnable-any-stmt/test-bs10-last-rouge-f}
\\
\UseMacro{model-codet5}
 & \UseMacro{res-codet5-CSNm-eval-runnable-any-stmt/test-bs10-last-compilable}
 & \UseMacro{res-codet5-CSNm-eval-runnable-any-stmt/test-bs10-last-runnable}
 & \UseMacro{res-codet5-CSNm-eval-runnable-any-stmt/test-bs10-last-xmatch}
 & \UseMacro{res-codet5-CSNm-eval-runnable-any-stmt/test-bs10-last-xmatch-top10}
 & \UseMacro{res-codet5-CSNm-eval-runnable-any-stmt/test-bs10-last-bleu}
 & \UseMacro{res-codet5-CSNm-eval-runnable-any-stmt/test-bs10-last-code-bleu}
 & \UseMacro{res-codet5-CSNm-eval-runnable-any-stmt/test-bs10-last-edit-sim}
 & \UseMacro{res-codet5-CSNm-eval-runnable-any-stmt/test-bs10-last-rouge-f}
\\
\UseMacro{model-codet5-noft}
 & \UseMacro{res-codet5-noft-CSNm-eval-runnable-any-stmt/test-bs10-last-compilable}
 & \UseMacro{res-codet5-noft-CSNm-eval-runnable-any-stmt/test-bs10-last-runnable}
 & \UseMacro{res-codet5-noft-CSNm-eval-runnable-any-stmt/test-bs10-last-xmatch}
 & \UseMacro{res-codet5-noft-CSNm-eval-runnable-any-stmt/test-bs10-last-xmatch-top10}
 & \UseMacro{res-codet5-noft-CSNm-eval-runnable-any-stmt/test-bs10-last-bleu}
 & \UseMacro{res-codet5-noft-CSNm-eval-runnable-any-stmt/test-bs10-last-code-bleu}
 & \UseMacro{res-codet5-noft-CSNm-eval-runnable-any-stmt/test-bs10-last-edit-sim}
 & \UseMacro{res-codet5-noft-CSNm-eval-runnable-any-stmt/test-bs10-last-rouge-f}
\\
\UseMacro{model-codegpt}
 & \UseMacro{res-codegpt-CSNm-eval-runnable-any-stmt/test-bs10-last-compilable}
 & \UseMacro{res-codegpt-CSNm-eval-runnable-any-stmt/test-bs10-last-runnable}
 & \UseMacro{res-codegpt-CSNm-eval-runnable-any-stmt/test-bs10-last-xmatch}
 & \UseMacro{res-codegpt-CSNm-eval-runnable-any-stmt/test-bs10-last-xmatch-top10}
 & \UseMacro{res-codegpt-CSNm-eval-runnable-any-stmt/test-bs10-last-bleu}
 & \UseMacro{res-codegpt-CSNm-eval-runnable-any-stmt/test-bs10-last-code-bleu}
 & \UseMacro{res-codegpt-CSNm-eval-runnable-any-stmt/test-bs10-last-edit-sim}
 & \UseMacro{res-codegpt-CSNm-eval-runnable-any-stmt/test-bs10-last-rouge-f}
\\
\midrule
\UseMacro{model-teco-icse23}
 & \textbf{\UseMacro{res-teco-icse23-CSNm-eval-runnable-any-stmt/test-bs10-last-compilable}}
 & \textbf{\UseMacro{res-teco-icse23-CSNm-eval-runnable-any-stmt/test-bs10-last-runnable}}
 & \textbf{\UseMacro{res-teco-icse23-CSNm-eval-runnable-any-stmt/test-bs10-last-xmatch}}
 & \textbf{\UseMacro{res-teco-icse23-CSNm-eval-runnable-any-stmt/test-bs10-last-xmatch-top10}}
 & \textbf{\UseMacro{res-teco-icse23-CSNm-eval-runnable-any-stmt/test-bs10-last-bleu}}
 & \textbf{\UseMacro{res-teco-icse23-CSNm-eval-runnable-any-stmt/test-bs10-last-code-bleu}}
 & \textbf{\UseMacro{res-teco-icse23-CSNm-eval-runnable-any-stmt/test-bs10-last-edit-sim}}
 & \textbf{\UseMacro{res-teco-icse23-CSNm-eval-runnable-any-stmt/test-bs10-last-rouge-f}}
\\
\bottomrule
\end{tabular}

  \vspace{5pt}
\end{minipage}
\begin{minipage}[t]{\textwidth}
  \centering
  \subcaption{On the \oraclesubset.\label{tab:results-latest-assert-CSNm}}
  \vspace{-5pt}
  
%% Automatically generated by pyutil.latex 

\begin{tabular}{@{}l|@{\hspace{2pt}}r@{\hspace{2pt}}r@{\hspace{2pt}}r@{\hspace{2pt}}r@{\hspace{2pt}}r@{\hspace{2pt}}r@{}}
\toprule
\textbf{\UseMacro{TH-model}}
 & \textbf{\UseMacro{TH-xmatch}}
 & \textbf{\UseMacro{TH-xmatch-top10}}
 & \textbf{\UseMacro{TH-bleu}}
 & \textbf{\UseMacro{TH-code-bleu}}
 & \textbf{\UseMacro{TH-edit-sim}}
 & \textbf{\UseMacro{TH-rouge}}
\\
\midrule
\UseMacro{model-codex}
 & \UseMacro{res-codex-CSNm-eval-assert-stmt/test-bs10-last-xmatch}
 & \THNA 
 & \UseMacro{res-codex-CSNm-eval-assert-stmt/test-bs10-last-bleu}
 & \UseMacro{res-codex-CSNm-eval-assert-stmt/test-bs10-last-code-bleu}
 & \UseMacro{res-codex-CSNm-eval-assert-stmt/test-bs10-last-edit-sim}
 & \UseMacro{res-codex-CSNm-eval-assert-stmt/test-bs10-last-rouge-f}
\\
\UseMacro{model-codet5}
 & \UseMacro{res-codet5-CSNm-eval-assert-stmt/test-bs10-last-xmatch}
 & \UseMacro{res-codet5-CSNm-eval-assert-stmt/test-bs10-last-xmatch-top10}
 & \UseMacro{res-codet5-CSNm-eval-assert-stmt/test-bs10-last-bleu}
 & \UseMacro{res-codet5-CSNm-eval-assert-stmt/test-bs10-last-code-bleu}
 & \UseMacro{res-codet5-CSNm-eval-assert-stmt/test-bs10-last-edit-sim}
 & \UseMacro{res-codet5-CSNm-eval-assert-stmt/test-bs10-last-rouge-f}
\\
\UseMacro{model-codet5-noft}
 & $^{\alpha}$\UseMacro{res-codet5-noft-CSNm-eval-assert-stmt/test-bs10-last-xmatch}
 & \UseMacro{res-codet5-noft-CSNm-eval-assert-stmt/test-bs10-last-xmatch-top10}
 & \UseMacro{res-codet5-noft-CSNm-eval-assert-stmt/test-bs10-last-bleu}
 & \UseMacro{res-codet5-noft-CSNm-eval-assert-stmt/test-bs10-last-code-bleu}
 & \UseMacro{res-codet5-noft-CSNm-eval-assert-stmt/test-bs10-last-edit-sim}
 & \UseMacro{res-codet5-noft-CSNm-eval-assert-stmt/test-bs10-last-rouge-f}
\\
\UseMacro{model-codegpt}
 & \UseMacro{res-codegpt-CSNm-eval-assert-stmt/test-bs10-last-xmatch}
 & \UseMacro{res-codegpt-CSNm-eval-assert-stmt/test-bs10-last-xmatch-top10}
 & \UseMacro{res-codegpt-CSNm-eval-assert-stmt/test-bs10-last-bleu}
 & \UseMacro{res-codegpt-CSNm-eval-assert-stmt/test-bs10-last-code-bleu}
 & \UseMacro{res-codegpt-CSNm-eval-assert-stmt/test-bs10-last-edit-sim}
 & \UseMacro{res-codegpt-CSNm-eval-assert-stmt/test-bs10-last-rouge-f}
\\
\UseMacro{model-atlas}
 & $^{\alpha}$\UseMacro{res-atlas-CSNm-eval-assert-stmt/test-bs10-last-xmatch}
 & \UseMacro{res-atlas-CSNm-eval-assert-stmt/test-bs10-last-xmatch-top10}
 & \UseMacro{res-atlas-CSNm-eval-assert-stmt/test-bs10-last-bleu}
 & \UseMacro{res-atlas-CSNm-eval-assert-stmt/test-bs10-last-code-bleu}
 & \UseMacro{res-atlas-CSNm-eval-assert-stmt/test-bs10-last-edit-sim}
 & \UseMacro{res-atlas-CSNm-eval-assert-stmt/test-bs10-last-rouge-f}
\\
\UseMacro{model-toga}
 & \UseMacro{res-toga-CSNm-eval-assert-stmt/test-bs10-last-xmatch}
 & \UseMacro{res-toga-CSNm-eval-assert-stmt/test-bs10-last-xmatch-top10}
 & \UseMacro{res-toga-CSNm-eval-assert-stmt/test-bs10-last-bleu}
 & \UseMacro{res-toga-CSNm-eval-assert-stmt/test-bs10-last-code-bleu}
 & \UseMacro{res-toga-CSNm-eval-assert-stmt/test-bs10-last-edit-sim}
 & \UseMacro{res-toga-CSNm-eval-assert-stmt/test-bs10-last-rouge-f}
\\
\midrule
\UseMacro{model-teco-icse23}
 & \textbf{\UseMacro{res-teco-icse23-CSNm-eval-assert-stmt/test-bs10-last-xmatch}}
 & \textbf{\UseMacro{res-teco-icse23-CSNm-eval-assert-stmt/test-bs10-last-xmatch-top10}}
 & \textbf{\UseMacro{res-teco-icse23-CSNm-eval-assert-stmt/test-bs10-last-bleu}}
 & \textbf{\UseMacro{res-teco-icse23-CSNm-eval-assert-stmt/test-bs10-last-code-bleu}}
 & \textbf{\UseMacro{res-teco-icse23-CSNm-eval-assert-stmt/test-bs10-last-edit-sim}}
 & \textbf{\UseMacro{res-teco-icse23-CSNm-eval-assert-stmt/test-bs10-last-rouge-f}}
\\
\bottomrule
\end{tabular}

  \vspace{5pt}
\end{minipage}
\begin{minipage}[t]{\textwidth}
  \centering
  \subcaption{On the \oraclerunnablesubset.\label{tab:results-latest-runnable-assert-CSNm}}
  \vspace{-5pt}
  
%% Automatically generated by pyutil.latex 

\begin{tabular}{@{}l|@{\hspace{2pt}}r@{\hspace{2pt}}r@{\hspace{2pt}}r@{\hspace{2pt}}r@{\hspace{2pt}}r@{\hspace{2pt}}r@{\hspace{2pt}}r@{\hspace{2pt}}r@{}}
\toprule
\textbf{\UseMacro{TH-model}}
 & \textbf{\UseMacro{TH-compilable}}
 & \textbf{\UseMacro{TH-runnable}}
 & \textbf{\UseMacro{TH-xmatch}}
 & \textbf{\UseMacro{TH-xmatch-top10}}
 & \textbf{\UseMacro{TH-bleu}}
 & \textbf{\UseMacro{TH-code-bleu}}
 & \textbf{\UseMacro{TH-edit-sim}}
 & \textbf{\UseMacro{TH-rouge}}
\\
\midrule
\UseMacro{model-codex}
 & \UseMacro{res-codex-CSNm-eval-runnable-assert-stmt/test-bs10-last-compilable}
 & \UseMacro{res-codex-CSNm-eval-runnable-assert-stmt/test-bs10-last-runnable}
 & \UseMacro{res-codex-CSNm-eval-runnable-assert-stmt/test-bs10-last-xmatch}
 & \THNA 
 & \UseMacro{res-codex-CSNm-eval-runnable-assert-stmt/test-bs10-last-bleu}
 & \UseMacro{res-codex-CSNm-eval-runnable-assert-stmt/test-bs10-last-code-bleu}
 & \UseMacro{res-codex-CSNm-eval-runnable-assert-stmt/test-bs10-last-edit-sim}
 & \UseMacro{res-codex-CSNm-eval-runnable-assert-stmt/test-bs10-last-rouge-f}
\\
\UseMacro{model-codet5}
 & \UseMacro{res-codet5-CSNm-eval-runnable-assert-stmt/test-bs10-last-compilable}
 & \UseMacro{res-codet5-CSNm-eval-runnable-assert-stmt/test-bs10-last-runnable}
 & \UseMacro{res-codet5-CSNm-eval-runnable-assert-stmt/test-bs10-last-xmatch}
 & \UseMacro{res-codet5-CSNm-eval-runnable-assert-stmt/test-bs10-last-xmatch-top10}
 & \UseMacro{res-codet5-CSNm-eval-runnable-assert-stmt/test-bs10-last-bleu}
 & \UseMacro{res-codet5-CSNm-eval-runnable-assert-stmt/test-bs10-last-code-bleu}
 & \UseMacro{res-codet5-CSNm-eval-runnable-assert-stmt/test-bs10-last-edit-sim}
 & \UseMacro{res-codet5-CSNm-eval-runnable-assert-stmt/test-bs10-last-rouge-f}
\\
\UseMacro{model-codet5-noft}
 & \UseMacro{res-codet5-noft-CSNm-eval-runnable-assert-stmt/test-bs10-last-compilable}
 & \UseMacro{res-codet5-noft-CSNm-eval-runnable-assert-stmt/test-bs10-last-runnable}
 & $^{\alpha}$\UseMacro{res-codet5-noft-CSNm-eval-runnable-assert-stmt/test-bs10-last-xmatch}
 & \UseMacro{res-codet5-noft-CSNm-eval-runnable-assert-stmt/test-bs10-last-xmatch-top10}
 & \UseMacro{res-codet5-noft-CSNm-eval-runnable-assert-stmt/test-bs10-last-bleu}
 & \UseMacro{res-codet5-noft-CSNm-eval-runnable-assert-stmt/test-bs10-last-code-bleu}
 & \UseMacro{res-codet5-noft-CSNm-eval-runnable-assert-stmt/test-bs10-last-edit-sim}
 & \UseMacro{res-codet5-noft-CSNm-eval-runnable-assert-stmt/test-bs10-last-rouge-f}
\\
\UseMacro{model-codegpt}
 & \UseMacro{res-codegpt-CSNm-eval-runnable-assert-stmt/test-bs10-last-compilable}
 & \UseMacro{res-codegpt-CSNm-eval-runnable-assert-stmt/test-bs10-last-runnable}
 & \UseMacro{res-codegpt-CSNm-eval-runnable-assert-stmt/test-bs10-last-xmatch}
 & \textbf{\UseMacro{res-codegpt-CSNm-eval-runnable-assert-stmt/test-bs10-last-xmatch-top10}}
 & \UseMacro{res-codegpt-CSNm-eval-runnable-assert-stmt/test-bs10-last-bleu}
 & \UseMacro{res-codegpt-CSNm-eval-runnable-assert-stmt/test-bs10-last-code-bleu}
 & \UseMacro{res-codegpt-CSNm-eval-runnable-assert-stmt/test-bs10-last-edit-sim}
 & \UseMacro{res-codegpt-CSNm-eval-runnable-assert-stmt/test-bs10-last-rouge-f}
\\
\UseMacro{model-atlas}
 & \UseMacro{res-atlas-CSNm-eval-runnable-assert-stmt/test-bs10-last-compilable}
 & \UseMacro{res-atlas-CSNm-eval-runnable-assert-stmt/test-bs10-last-runnable}
 & $^{\alpha}$\UseMacro{res-atlas-CSNm-eval-runnable-assert-stmt/test-bs10-last-xmatch}
 & \UseMacro{res-atlas-CSNm-eval-runnable-assert-stmt/test-bs10-last-xmatch-top10}
 & \UseMacro{res-atlas-CSNm-eval-runnable-assert-stmt/test-bs10-last-bleu}
 & \UseMacro{res-atlas-CSNm-eval-runnable-assert-stmt/test-bs10-last-code-bleu}
 & \UseMacro{res-atlas-CSNm-eval-runnable-assert-stmt/test-bs10-last-edit-sim}
 & \UseMacro{res-atlas-CSNm-eval-runnable-assert-stmt/test-bs10-last-rouge-f}
\\
\UseMacro{model-toga}
 & \UseMacro{res-toga-CSNm-eval-runnable-assert-stmt/test-bs10-last-compilable}
 & \UseMacro{res-toga-CSNm-eval-runnable-assert-stmt/test-bs10-last-runnable}
 & \UseMacro{res-toga-CSNm-eval-runnable-assert-stmt/test-bs10-last-xmatch}
 & \UseMacro{res-toga-CSNm-eval-runnable-assert-stmt/test-bs10-last-xmatch-top10}
 & \UseMacro{res-toga-CSNm-eval-runnable-assert-stmt/test-bs10-last-bleu}
 & \UseMacro{res-toga-CSNm-eval-runnable-assert-stmt/test-bs10-last-code-bleu}
 & \UseMacro{res-toga-CSNm-eval-runnable-assert-stmt/test-bs10-last-edit-sim}
 & \UseMacro{res-toga-CSNm-eval-runnable-assert-stmt/test-bs10-last-rouge-f}
\\
\midrule
\UseMacro{model-teco-icse23}
 & \textbf{\UseMacro{res-teco-icse23-CSNm-eval-runnable-assert-stmt/test-bs10-last-compilable}}
 & \textbf{\UseMacro{res-teco-icse23-CSNm-eval-runnable-assert-stmt/test-bs10-last-runnable}}
 & \textbf{\UseMacro{res-teco-icse23-CSNm-eval-runnable-assert-stmt/test-bs10-last-xmatch}}
 & \UseMacro{res-teco-icse23-CSNm-eval-runnable-assert-stmt/test-bs10-last-xmatch-top10}
 & \textbf{\UseMacro{res-teco-icse23-CSNm-eval-runnable-assert-stmt/test-bs10-last-bleu}}
 & \textbf{\UseMacro{res-teco-icse23-CSNm-eval-runnable-assert-stmt/test-bs10-last-code-bleu}}
 & \textbf{\UseMacro{res-teco-icse23-CSNm-eval-runnable-assert-stmt/test-bs10-last-edit-sim}}
 & \textbf{\UseMacro{res-teco-icse23-CSNm-eval-runnable-assert-stmt/test-bs10-last-rouge-f}}
\\
\bottomrule
\end{tabular}

\end{minipage}
\end{small}
\end{table*}

Recent large language models for code that scale up to billions of
parameters, such as Codex~\cite{ChenETAL21Codex}, have been shown to
be promising for many code-related tasks.  In parallel with our work,
researchers applied large language models to generate
tests~\cite{BareissETAL22CodeGen,SchaferETAL23TestPilot,LemieuxETAL23CodaMosa}.
In this appendix, we perform an additional experiment to evaluate the
performance of large language models on \testcomp and compare with
\TecoTool.

The large language model we used is Codex~\cite{ChenETAL21Codex},
which is the state-of-the-art specialized large language model for
code.  Following the contemporary work on using large language models
for test
generation~\cite{BareissETAL22CodeGen,SchaferETAL23TestPilot,LemieuxETAL23CodaMosa},
we used Codex to perform \testcomp in the zero-shot learning
setup~\cite{KojimaETAL22ZeroShot}, i.e., providing Codex with a prompt
that contains the \mut, \testsign, and \priorstmts, and letting it
generate the next \stmt.  Because Codex is \pretrained to complete
code, the prompt needs to be carefully designed as a code fragment to
be completed. \figurename~\ref{fig:example-prompt} illustrates the
prompt format we used.  We configured Codex to generate until seeing
the first `;', similar to the way we used \UseMacro{model-codegpt}.
Because the current generation speed of Codex is quite slow, we
configured Codex to only generate the top-1 next statement using the
greedy decoding algorithm.  We used the
\CodeIn{code\text{-}davinci\text{-}002} version of the Codex model.
Running Codex on our \test set (with \UseMacro{corpus-test-num_stmt}
\stmts) took 18 hours.

Table~\ref{tab:results-latest} shows the results of Codex for the
\testcomp task with comparisons to \TecoTool and the other baseline
models; the results are organized into four parts---on the full \test
set, \runnablesubset, \oraclesubset, and \oraclerunnablesubset---as
explained in Section~\ref{sec:exp}.  \TecoTool statistically
significantly outperforms Codex on all metrics, which confirms the
importance of using code semantics and code execution together with
ML.  Compared with the other baseline models
(\UseMacro{model-codet5}/\UseMacro{model-codegpt}), Codex has better
performance on some metrics (e.g., \PctRunnable on the \runnablesubset
and \oraclerunnablesubset; \xmatch on the \oraclesubset and
\oraclerunnablesubset) but has slightly worse performance on others.
Although Codex is expected to be much more powerful than
\UseMacro{model-codet5}/\UseMacro{model-codegpt} due to the larger
scale (billions of parameters vs. millions of parameters) and more
\pretraining data, we hypothesize that \finetuning
\UseMacro{model-codet5}/\UseMacro{model-codegpt} on our large
\testcomp corpus helped with improving their performance. Codex
performs better on the \oraclegen task than the \testcomp task, which
may be because of the more \priorstmts context available when
performing \oraclegen.

\begin{figure}[t]
  \centering
\begin{lstlisting}[language=java-pretty]
public GMOperation addImage(final File file) {
  if (file == null) {
    throw new IllegalArgumentException("file must be defined"); 
  }
  getCmdArgs().add(file.getPath());
  return this; 
}

// Here is a test for the above method. Please complete the next statement of the test
@Test public void addImage_ThrowsException_WhenFileIsNull() throws Exception {
exception.except(IllegalArgumentException.class);
// Please compete the next statement:
\end{lstlisting}
\caption{The prompt to Codex for the example \testcomp task in \figurename~\ref{fig:example-data}. \label{fig:example-prompt}}
\end{figure}

\balance

\end{document}